\documentclass[11pt]{article}

\usepackage{amssymb,amsmath,amsfonts,eurosym,geometry,ulem,graphicx,caption,color,setspace,sectsty,comment,footmisc,caption,pdflscape,array}
\usepackage[hidelinks]{hyperref}
\usepackage{authblk}
\usepackage{mathtools}
\usepackage{tabularx}
\usepackage{longtable}
\usepackage{threeparttable}
\usepackage{multirow}
\usepackage{tabu}
\usepackage{bigints}
\usepackage{enumerate}
\usepackage{subcaption}  
\usepackage{booktabs}
\usepackage{natbib}

\def\mathclap#1{\text{\hbox to 0pt{\hss$\mathsurround=0pt#1$\hss}}}

\newtheorem{theorem}{Theorem}

\newtheorem{proposition}{Proposition}
\newtheorem{assumption}{Assumption}
\newtheorem{lemma}{Lemma}

\numberwithin{equation}{section}

\renewcommand{\theequation}{\thesection.\arabic{equation}}

\newcounter{step}[section]
\newenvironment{step}[1][]{\refstepcounter{step}\par\medskip\noindent%
   \textit{\textbf{Step~\thestep. #1}} \rmfamily}{\bigskip}
\newcolumntype{L}[1]{>{\raggedright\let\newline\\arraybackslash\hspace{0pt}}m{#1}}
\newcolumntype{C}[1]{>{\centering\let\newline\\arraybackslash\hspace{0pt}}m{#1}}
\newcolumntype{R}[1]{>{\raggedleft\let\newline\\arraybackslash\hspace{0pt}}m{#1}}

\allowdisplaybreaks

\begin{document}

\begin{titlepage}
\title{Quantile regression with generated dependent variable and covariates }
\author{Jayeeta Bhattacharya\thanks{Lecturer, Department of Economics, University of Southampton. Correspondence email: j.bhattacharya@soton.ac.uk. This work was completed during my PhD at Queen Mary University of London (QMUL) and I am deeply thankful to Emmanuel Guerre for his valuable comments and supervision. I also thank participants at various conferences for insightful comments. Generous funding from the School of Economics and Finance, QMUL, is also gratefully acknowledged.}}
\affil{December 2020}
\date{{ }}
\maketitle
\begin{abstract}
\noindent We study linear quantile regression models when regressors and/or dependent variable are not directly observed but estimated in an initial first step and used in the second step quantile regression for estimating the quantile parameters. This general class of generated quantile regression (GQR) covers various statistical applications, for instance, estimation of endogenous quantile regression models and triangular structural equation models, and some new relevant applications are discussed. We study the asymptotic distribution of the two-step estimator, which is challenging because of the presence of generated covariates and/or dependent variable in the non-smooth quantile regression estimator. We employ techniques from empirical process theory to find uniform Bahadur expansion for the two step estimator, which is used to establish the asymptotic results. We illustrate the performance of the GQR estimator through simulations and an empirical application based on auctions.  \\
\vspace{0in}
\vspace{0in}\\

\bigskip
\end{abstract}
\setcounter{page}{0}
\thispagestyle{empty}
\end{titlepage}
\pagebreak \newpage

\section{Introduction} \label{sec:introduction}

Econometric analysis often requires the use of regressors that are not directly observed but have been estimated in a preliminary first step. A rich literature exists on estimation and inference in models with generated regressors. \cite{P84} and \cite{OM93} provide surveys for parametric models with generated regressors while Mammen, Rothe \& Schienle (2012)\nocite{MRS12} study and illustrate various examples for non-parametric regression with generated covariates. Studying the asymptotic properties of two step estimators in a parametric context, \cite{MT85} points out that ignoring the effect of first step estimation leads to incorrect asymptotic standard errors. 

While these models are concerned with the characterization of the conditional mean, a more complete picture of the conditional distribution of a dependent variable is provided by quantile regression (QR) models. Since the seminal work of \cite{KB78}, quantile regression is widely used in both empirical studies and theoretical statistics for analysing conditional quantile functions in linear and nonlinear response models. Quantile regression applications using generated regressors abound in literature, most prominently related to models with endogenous covariates. \cite{CH05, CH06, CH08} develop identification and estimation for QR models in the presence of endogeneity. Another popular approach to deal with endogeneity uses the estimated reduced form residuals as control variables in quantile regression. This technique has been applied in endogenous censored quantile regression models by \cite{BP07} and Chernozhukov, Fern{\'a}ndez-Val \& Kowalski (2015)\nocite{CFK15}. Estimation of quantile treatment effects or quantile parameters in triangular simultaneous equation models using the control variable approach have been considered in \cite{C03}, \cite{KM06}, \cite{L07}, \cite{IN09}, and Chernozhukov, Fern{\'a}ndez-Val, Newey, Stouli \& Vella (2017)\nocite{CFNSV17}. There are, however, few references that develop a general theory for quantile models with generated covariates and systematically study its statistical properties. The only related work seems to be Chen, Galvao \& Song (2018)\nocite{CGS18}, who consider estimation and inference of quantile regression when regressors are generated. However, they study two step quantile estimation when the second step estimator is differentiable with
respect to the first stage, which may not hold true for some relevant
applications since the quantile regression objective function is not smooth. They also do not consider transformation for the dependent variable as permitted here. 

This paper considers the general framework of QR models when either the regressors or the dependent variable (or both) are generated and studies the asymptotic behaviour of the two-step QR estimator, called the generated quantile regression (GQR) estimator, without being tailored to any specific application. An example giving rise to generated dependent variable is quantile specifications with some constant slope parameters, as in the setup of \cite{ZY08}. Their composite quantile regression (CQR) method can be used to estimate the constant and quantile-varying parameters together, but its asymptotic properties have been studied for the estimation of constant parameters only. To focus on quantile estimation, the constant slope parameters can first be estimated by linear regression (or any other suitable method) in a first step. Estimation of the quantile-varying slope parameters, thereafter, involves quantile regression with the dependent variable generated as a function of the constant slope parameters and the corresponding covariates. Removing some parameters through the first step estimation may alleviate the computational burden of the CQR method caused by a large number of variables. Moreover, covariates make the QR estimators non-monotonic, even if the quantile function is increasing. So, the expectation is that reducing the dimensionality of the regressors in the QR stage by removing some covariates through the first stage of estimation allows to get closer to monotonicity, a desirable property for quantile estimation. The two-step procedure can also simplify estimation of complex models like random coefficient models, popularised in demand analysis by Berry, Levinsohn \& Pakes (1995)\nocite{BLP95}. \nocite{HKM10} Hoderlein, Klemel{\"a} \& Mammen (2010) propose non-parametric estimation of the distribution of the random coefficients. Studying the econometrics of auctions, \cite{GG20} consider quantile specifications arising from elliptically distributed random coefficients, which includes multivariate normal, lognormal or Student distribution. Its two-step estimation involves a median normalisation to identify and estimate the elliptical distribution location and dispersion parameters in the first step, which can generate the dependent variable for the second stage quantile regression for estimating sample quantiles. Another example for generated dependent variable arises in quantile models where the dependent variable is transformed based on some transformation parameter, like Box-Cox transformation, to induce some desirable properties for statistical inference. The joint estimation of quantile varying transformation and slope parameters through non-linear quantile regression is computationally difficult, in addition to a numerical problem being that the objective function is not defined for all parameter values and observations (meaning estimation occurs by omitting such values). Estimating the transformation parameter in a first step will avoid such numerical problem and involve a linear quantile regression, ensuring a better performance of the numerical algorithm used to compute the estimator. 

It is well known that the first step estimates impact the overall asymptotic behaviour of the final estimator, understanding which is crucial for obtaining consistent standard errors which can be used for constructing correct confidence intervals. The wide range of quantile regression applications that give rise to generated regressors or dependent variable obtained from estimation in a preliminary step suggest the need for a systematic analysis of their impact on the statistical properties of the QR estimator. The classical way in which asymptotic analysis is carried out for two step estimators with smooth objective functions relies on a Taylor expansion based technique for the second stage estimates, as applied in \cite{MT85}. However, such methods are not applicable for the QR estimator, since it is difficult to differentiate the QR estimator\footnote{This could be done in principle by applying the Implicit Function Theorem to the first-order condition that defines the estimator. However, the  QR estimator is not always unique and the QR objective function is not twice differentiable, preventing the use of this approach.}.
Finding the asymptotic variance of the non-smooth two-step GQR estimator is not a trivial task and requires alternative techniques. 
Chen, Linton \& Van Keilegom (2003) develop the asymptotic theory for semi-parametric GMM-type estimators with non-smooth criterion function and non-parametric first-stage; closely related papers are \cite{IL10}, \cite{HR13}, and Mammen, Rothe \& Schienle (2016)\nocite{MRS16} (the latter two also involve generated regressors in the non-parametric component, such that estimation occurs in three steps). As a consequence of generality, they have a standard two step proof approach, where they first give conditions for consistency and then establish asymptotic normality. However, since the QR objective function is convex, it allows bypassing the tedious task of checking conditions for consistency as in \cite{CLV03} and to establish asymptotic normality in one step instead, by applying the convexity trick of \cite{HP11}. Also, a Bahadur representation of the non-smooth two-step estimator with its rate is not present in these works. 

This paper systematically handles the associated issues for the asymptotic analysis of the generalised two-step GQR estimator using techniques from  asymptotic analysis for quantile regression and empirical process theory. We derive the Bahadur expansion of the GQR estimator, with precise stochastic order of the remainder term, which holds uniformly with respect to the first step parameter and the quantile levels. This involves establishing a stochastic equicontinuity result that allows approximating the score evaluated at the estimated first stage parameter by that taken at the true parameter, which is interesting in evaluating the effect of the first stage. Using the Bahadur expansion approach, under the assumption that the first stage estimation is asymptotically normal and some other regularity conditions, we establish asymptotic normality and obtain explicit expression for the asymptotic variance of the GQR estimator. 

Several applications fit the generated QR framework and four motivating examples are discussed - quantile regression involving constant slope parameters, an ellipticly distributed random coefficient model, a Box-Cox power transformed quantile regression, and a variant for endogenous quantile regression model. 
The example of QR model with constant slope also forms the basis for simulation experiments and an empirical application; the analysis suggests potential benefits of the two
stage GQR procedure over standard QR. The simulation exercises illustrate the validity of the GQR asymptotic normality result and the effect of the first stage estimation; further analysis of the asymptotic variance suggests that the GQR estimator produces efficiency improvements over standard QR estimator for central quantiles.
Finally, an empirical application based on auction models in quantile framework confirms that the GQR estimator improves the monotonicity and accuracy of quantile slopes as compared to an unconstrained estimation using standard quantile regression. 

The rest of the paper is organised as follows. Section \ref{sec:model} introduces the baseline model and the GQR estimator, and presents four applications to motivate the framework. Section \ref{sec:asymptotics} carries out the asymptotic analysis and presents the Bahadur expansion results and the central limit theorem for the GQR estimator. The asymptotic results are applied to the motivating examples in Section \ref{sec:examples2}. Section \ref{sec:sims} presents simulation results while Section \ref{sec:empirical} reports results of the empirical application to first price auctions. Proofs of the main results are given in the Appendices.

\section{Quantile regression with generated variables} \label{sec:model}
We consider the following linear quantile specification.
\begin{equation}\label{M:eq1}
Y(\theta) = X(\theta)^{\prime}\beta(U);\  \ U \vert X(\theta) \sim \mathcal{U}[0,1],
\end{equation}
where, provided that $\tau \mapsto X(\theta)^{\prime}\beta(\tau)$ is strictly increasing and continuous in $\tau$, $X(\theta)' \beta (\tau)$ is the $\tau$-quantile of $Y(\theta)$ conditional on $X(\theta)$. Here, $Y(\theta)$ and $X(\theta)$ are functions of a vector of parameters $\theta$, which includes elements that generate the dependent variable $Y$, or the regressor $X$, or both. The true value of the parameter $\theta$ in \eqref{M:eq1}, denoted by $\theta_0$, is not known but estimated. Hence, estimation proceeds in two steps.

\paragraph{First step: Estimation of $\theta_0$.} It is assumed that a consistent estimator $\widehat{\theta}$ is available. For the sake of generality, any estimation method is allowed at this stage, provided it satisfies an expansion typical of regular estimators, see for example \cite{NM94}. As discussed for the examples, a suitable choice of $\widehat{\theta}$ can be done on a case-by-case basis.

\paragraph{Second step: Estimation of quantile parameter.} The quantile parameter estimate $\widehat{\beta}(\tau)$ in \eqref{M:eq1} is given by
\begin{equation}\label{M:eq2}
\begin{aligned}
\widehat{\beta}(\tau) = \widehat{\beta}(\tau;\widehat{\theta}) &= \arg\min_{\beta
}\frac{1}{n}\sum_{i=1}^{n}\rho_{\tau  }\left(  Y_i(\widehat{\theta})-X_i(\widehat{\theta})^{\prime}\beta\right),
\end{aligned}
\end{equation}
where $\rho_\tau(u) = \left(\tau - \mathbb{I}\left(u < 0\right)\right)u$ is the check function of \cite{KB78}. 

\subsection{Motivating examples} \label{sec:examples1}
The general framework of quantile regression with dependent variable and/or covariates obtained as a function of parameters estimated in a first step finds wide application in economics and statistics. We present four applications here, which we revisit later to derive their asymptotic results.

\subsubsection{Quantile regression with constant slope}\label{subsec:regandQR1}

Consider the quantile regression (QR) model%
\begin{equation}\label{eq:ols1}
Q_{Y}\left( \tau |X\right) =\beta _{0}\left( \tau \right) +\beta
_{1}\left( \tau \right) X_{1}+\beta _{2}\left( \tau \right) X_{2}
\end{equation}
and assume that $\beta _{1}\left( \tau \right) =\beta _{1}$ for all $\tau $, ie $\beta
_{1}\left( \cdot \right) $ is constant. This model can be estimated using \cite{ZY08}'s composite quantile regression (CQR) method as follows:
\[
\left(\widehat{\beta}_1, \widehat{\beta}_0(\tau_1), \widehat{\beta}_2(\tau_1), \cdots, \widehat{\beta}_0(\tau_K), \widehat{\beta}_2(\tau_K) \right) = \arg \min_{\substack{b_1,b_{0k},b_{2k}; \\ k=1,\cdots,K
}}\sum_{k=1}^{K} \sum_{i=1}^{n}\rho _{\tau_k }\left( Y_{i}-X_{1i}b_1 - b_{0k}- X_{2i}b_{2k}  
\right),
\]
for $0<\tau_1<\tau_2<\cdots<\tau_K<1$. This could lead to an intractable system due to very large number of variables, especially with more quantile parameters and quantile levels. Moreover, \cite{ZY08} studies the asymptotic properties of the CQR estimator for estimation of constant slope parameters and compares efficiency with least squares, while the asymptotic behaviour for quantile varying slope parameters remains unstudied. As an alternative to \cite{ZY08}, consider a two step estimation of this model as described below. 

As there exist uniform variables $U_{i}$ independent of $X_{i}$ such that $%
Y_{i}=$ $Q_{Y}\left( U_{i}|X_{i}\right) $, it holds
\begin{equation}\nonumber
Y_{i}=\overline{\beta }_{0}+\overline{\beta }_{1}X_{1i}+\overline{\beta }%
_{2}X_{2i}+\varepsilon _{i}
\end{equation}
where $\overline{\beta }_{k}=\mathbb{E}\left[ \beta _{k}\left( U_{i}\right) %
\right] $, $k=\{0,1,2\}$, and $\varepsilon _{i}=\beta _{0}\left( U_{i}\right) -\overline{\beta }_{0}+\left(
\beta _{2}\left( U_{i}\right) -\overline{\beta }_{2}\right) X_{2i}$ (since $\overline{\beta }_{1} = {\beta }_{1} = \beta _{1}\left( U_{i}\right)$).
It follows that the $\overline{\beta }_{k}$'s can be estimated using OLS,
that is,%
\begin{equation}\label{eq:ols4}
\left( \widehat{\overline{\beta }}_{0},\widehat{\overline{\beta }}_{1},%
\widehat{\overline{\beta }}_{2}\right) =\arg
\min_{b_{0},b_{1},b_{2}}\sum_{i=1}^{n}\left(
Y_{i}-b_{0}-b_{1}X_{1i}-b_{2}X_{2i}\right) ^{2}.
\end{equation}
Set $\widehat{\overline{\beta }}_{1}=\widehat{\beta }_{1}$. A two step estimator of $\left( \beta _{0}\left( \tau \right) ,\beta
_{2}\left( \tau \right) \right) $ is then%
\begin{equation}\label{eq:ols5}
\left( \widehat{\beta }_{0}\left( \tau \right) ,\widehat{\beta }_{2}\left(
\tau \right) \right) =\arg \min_{b_{0},b_{2}}\sum_{i=1}^{n}\rho _{\tau
}\left( Y_{i}-\widehat{\beta }_{1}X_{1i}-b_{0}-b_{2}X_{2i}\right).
\end{equation}
Hence, in this example, the first step parameter is $\theta \equiv \beta_1$, and the dependent variable is generated as $Y_i(\beta_1) = Y_i - \beta_1X_{1i}$.

\subsubsection{Random coefficient model}\label{subsec:randomcoeff1}
Consider the model \vspace{-3mm}
\begin{align}\label{eq:random1}
Y_i = X_i^\prime \beta_i, \quad i=1,\hdots,n
\end{align} 
where, $\beta_i$ is a $(K+1)$-dimensional vector of random coefficients, independent from the $(K+1)$-vector of covariates $X_i$ whose first element is $1$ (such that the first element of $\beta_i$ represents the error in this model). Note that linear regression is a special case of \eqref{eq:random1} where $\beta_i=\beta$ for all $i$. 

Suppose $\beta_i$ is drawn from elliptical distribution with location parameter $\mu$ and symmetric nonnegative dispersion matrix $\Sigma$, which includes distributions like multivariate normal, log-normal and t-distribution, as considered in \cite{GG20}'s auctions based application. Let $\mathcal{R}_i$ denote a random vector distributed uniformly on the unit sphere in $\mathbb{R}^{(K+1)}$ and consider the Euclidean norm ${E}_i = \left\vert\left\vert \Sigma^{-1/2}(\beta_i - \mu) \right\vert\right\vert$, independent from $\mathcal{R}_i$. Then, following Fang, Kotz \& Ng (1990, pg-32)\nocite{FKN90}, $\beta_i$ has the same distribution as $\mu + E_i\Sigma^{1/2} \mathcal{R}_i$. Let ${r}_i$ denote the first coordinate of $\mathcal{R}_i$ such that $t^\prime \mathcal{R}_i = \left\vert\left\vert  t \right\vert\right\vert r_i$ (see \cite{FKN90}, Theorem 2.4). Hence, using the symbol $\stackrel{d}{=}$ for denoting identical distribution of random variables, we have from \eqref{eq:random1}

\begin{align*}
Y_i \stackrel{d}{=} X_i^\prime \mu + \left(\Sigma^{1/2} X_i\right)^\prime E_i\mathcal{R}_i \stackrel{d}{=} X_i^\prime \mu + \left\vert\left\vert\Sigma^{1/2} X_i\right\vert\right\vert E_i r_i.
\end{align*}
Hence, the quantile specification for \eqref{eq:random1} is given by 
\begin{align}\label{eq:randomquant}
Q_Y(\tau \vert X)= X^\prime \mu + \left\vert\left\vert \Sigma^{1/2}X \right\vert\right\vert \xi(\tau)
\end{align} 
where $\xi(\tau)$ is the $\tau$-th quantile of $E_i r_i$. The above model can be estimated in two steps as follows. Under the normalisation $\xi(1/2)=1$, the parameters $\mu$ and $\Sigma$ are identified by conditional median regression:
\begin{align}\label{eq:randomfirststep}
(\widehat{\mu},\widehat{\Sigma})=\arg \min_{\mu, \Sigma} \sum_{i=1}^{n}\left\vert Y_i - X_i^\prime \mu - \left\vert\left\vert \Sigma^{1/2}X_i \right\vert\right\vert \right\vert.
\end{align}
The second step involves quantile regression using the generated dependent variable $Y_i(\widehat{\mu},\widehat{\Sigma})=\frac{Y_i - X_i^\prime \widehat{\mu}}{ \left\vert\left\vert \widehat{\Sigma}^{1/2}X \right\vert\right\vert }$ to obtain the sample quantiles:
\begin{align}\label{eq:randomsecondstep}
\widehat{\xi}(\tau)=\arg \min_{\xi} \sum_{i=1}^{n}\rho_\tau\left(Y_i(\widehat{\mu},\widehat{\Sigma}) - \xi \right).
\end{align}

\subsubsection{Box-Cox power transformation} \label{subsec:boxcox1} \cite{BC64} proposes finding a transformation parameter $\lambda$ such that with the following transformation on the original observations $Y$,
\begin{equation}\label{eq:bc1}
\begin{aligned}
Y(\lambda)  &= 
\begin{cases}
\frac{Y^\lambda - 1}{\lambda},& \text{if } \lambda \neq 0,\\
\log Y,												& \text{if } \lambda = 0,
\end{cases} 
\end{aligned}
\end{equation}
$Y(\lambda)$ is normally distributed with conditional variance $\sigma^2$, and $\mathbb{E}[Y(\lambda) \vert X]= X^\prime \beta$. The desirous property for quantile regression is linearity, that is, 
\begin{equation}\nonumber
Q_{Y(\lambda)}(\tau \vert X) = X^\prime \beta(\tau). 
\end{equation}

The Box-Cox quantile regression literature has mostly focussed on finding a quantile dependent transformation parameter (see, for instance,  \cite{P91}, \cite{C94}, \cite{B95}, \cite{MM00} and Fitzenberger, Wilke \& Zhang (2009)\nocite{F09}). Owing to the equivariance property of quantiles, this leads to minimization of the non-linear function $\sum_{i=1}^{n}\rho _{\tau
}\left( Y_{i}- \left(\lambda  X_i^\prime \beta + 1 \right)^{1/\lambda}\right)$. Quantile varying $\lambda$ adds flexibility to the model, but joint estimation of $\left(\lambda(\tau), \beta(\tau) \right)$ requires effort, see \cite{K17}. Also, a basic numerical problem is that $\left(\lambda  X_i^\prime \beta + 1 \right)$ needs to be positive for all $\lambda$ and all observations. 

A constrained estimation with a constant $\lambda$ has obvious computational and numerical benefits. \cite{MH07} considers constancy of $\lambda(\tau)$. In the empirical application of \cite{B95} studying transformation of log wages over $25$ years, $\lambda(\tau)$ seems to be constant for all quantiles except the highest. 
A simpler approach would, therefore, involve estimating $\widehat{\lambda}$ separately in a first step and thereafter performing linear quantile regression using the transformed $Y$ for estimating $\beta(\tau)$.    
 $\widehat{\lambda}$ can be estimated from the linear regression $Y(\lambda)=X^\prime \beta + \varepsilon$. A consistent estimator for $\widehat{\lambda}$ is \cite{A74}'s nonlinear IV (NIV) estimator,   
\begin{equation}\label{eq:bc3}
\left( \widehat{\lambda}_{NIV},\widehat{\beta}_{NIV}\right) =\arg
\min_{\ell,b} \left(\sum_{i=1}^{n}\left(Y_i\left(\ell \right) - X_i^\prime b \right) W_i^\prime\right) \Omega \left(\sum_{i=1}^{n} W_i\left(Y_i\left(\ell \right) - X_i^\prime b \right)\right),
\end{equation}
where $W_i$ always contains $X_i$ as well as additional instruments (\cite{AP81} recommends using squares and cross-products of $X_i$'s). Set $\widehat{\lambda} = \widehat{\lambda}_{NIV}$.  
The dependent variables $Y_i(\widehat{\lambda})$ is, then, generated using equation \eqref{eq:bc1}. $\beta(\tau)$ is estimated from quantile regression of $Y(\widehat{\lambda})$ on $X$,
\begin{equation}\label{eq:bc4}
 \widehat{\beta }\left( \tau \right)  =\arg \min_{b}\sum_{i=1}^{n}\rho _{\tau
}\left( Y_{i}(\widehat{\lambda})-X_{i}^{\prime}b\right). 
\end{equation}

\subsubsection{Endogeneity in quantile regression - control variable approach} \label{subsec:IV}
Control variable approach views endogeneity bias as an omitted variable bias and proceeds by estimating the `control variable', which is the residual of the regression of the endogenous regressor on the instruments, conditional on which error becomes independent of the regressors (see \cite{BP03}). 	

Consider the following system of equations%
\begin{align}
Y &=W^{\prime}\alpha+ X\beta + \varepsilon,\label{eq:endo}  \\ 
X &=Z^{\prime }\gamma +\eta \label{eq:ivfirst} 
\end{align}
where $W$ is a vector of exogenous covariates and $X$ is the endogenous regressor of interest generated by \eqref{eq:ivfirst} in which $Z$ is the vector of instruments uncorrelated with $\eta$ and $\varepsilon$, $\eta $ being centered with a finite variance. Hence,  endogeneity in $X$ arises due to the unobserved latent variable $\eta$, adding which as regressor in the first equation `corrects' for endogeneity, as in the following quantile specification:
\begin{align} \label{eq:iv1}
Q_{Y|W,X,\eta }\left( \tau |W,X,\eta \right) =W^{\prime}\alpha(\tau)+ X\beta \left( \tau\right) +\eta\lambda \left( \tau \right) .
\end{align}
The above model can be estimated in two steps as follows. The first stage
least squares estimates the control variable $\eta $,%
\begin{eqnarray}\label{eq:iv2}
\widehat{\eta }_{i}=X_{i}-Z_{i}^{\prime }\widehat{\gamma },\quad \widehat{%
\gamma }=\left( \sum_{i=1}^{N}Z_{i}Z_{i}^{\prime }\right)
^{-1}\sum_{i=1}^{N}Z_{i}X_{i}.
\end{eqnarray}%
The second stage estimator for the quantile coefficients is%
\begin{align}
\left[  \widehat{\alpha }^{\prime }\left( \tau \right) ,\widehat{\beta }\left( \tau \right) ,\widehat{\lambda }%
\left( \tau \right) \right] ^{\prime }&=\arg \min_{\alpha,\beta ,\lambda
}\sum_{i=1}^{N}\rho _{\tau }\left( Y_{i}-W_{i}^{\prime }\alpha-X_{i}\beta - \widehat{\eta } \lambda 
\right) \nonumber \\
& = \arg \min_{\alpha, \beta ,\lambda
}\sum_{i=1}^{N}\rho _{\tau }\left( Y_{i}-W_{i}^{\prime }\alpha-X_{i}\beta - (X_{i} - Z_i^\prime \widehat{\gamma }) \lambda 
\right).\label{eq:iv3}
\end{align}
Hence, in this example, the first step estimator is $\theta \equiv \gamma$, and the second stage involves quantile regression of $Y_i$ on generated regressors, $X_i(\theta) \equiv \left[W_i^\prime, X_i, \left(X_i-Z_i^\prime \gamma\right) \right]^{\prime}$.

\section{Asymptotic analysis} \label{sec:asymptotics} 

Our main assumptions are as follows:

\begin{assumption}[First step estimator] \label{A3}
There exists a function $\psi(z)$ such that the estimator of the true $\theta_0$ is asymptotically linear:
\[
\sqrt{n}\left(\widehat{\theta} - \theta_0\right)=\frac{1}{\sqrt{n}}\sum_{i=1}^{n}\psi(z_i) + o_{\mathbb{P}}(1),\ \ \mathbb{E}\left[\psi(z)\right]=0,\ \ \mathbb{E}\left[\psi(z)\psi(z)^\prime\right] < \infty.
\]
\end{assumption}

\begin{assumption}[Model]  \label{A1}
$(X_i,Y_i)$ are i.i.d. There exists a compact set $\Theta $ with a non empty interior containing $\theta _{0}$ such that $X_i\left( \theta \right) =h\left(X_i,\theta \right) $ and $Y_i\left( \theta \right) =g\left( Y_i,X_i,\theta \right) $ are continuous and differentiable with respect to $\theta $ in $\Theta $ for all $\left( Y_i,X_i\right) $. Denoting $\left\Vert \cdot \right\Vert$ as the Euclidean norm, it holds moreover that 
\[
\sup_{\theta \in \Theta }\left\Vert \frac{\partial g\left( Y,X,\theta
\right) }{\partial \theta }\right\Vert <\infty .
\]
\end{assumption}

In the next Assumption $F\left(y|x,\theta \right) $ and $f\left( y|x,\theta \right) $ stands for the c.d.f. and p.d.f. of $Y\left( \theta \right) $ given $X\left( \theta \right) $, $f_{X}\left( \cdot |\theta \right) $ being the p.d.f. of $X\left( \theta \right) $. The set $\mathcal{X}\left( \theta \right) $ is the support of $X\left( \theta \right) $. All p.d.f. are defined with respect to the Lebesgue measure. The set $\Theta $ is as in Assumption \ref{A1}.

\begin{assumption}[Smoothness] \label{A2}
(i) $X\left( \theta \right) $ lies in $\mathbb{R}^{d}$ for each $\theta $
and $\mathcal{X}\left( \theta \right) $ is a compact subset of $\mathbb{R}%
^{d}$ with non empty interior. $f_X\left( x|\theta \right) >0$ over the
interior of $\mathcal{X}\left( \theta \right) $ and vanishes at its
boundaries. $f_X\left( x|\theta \right) $ is continuously differentiable with
respect to $\theta $. (ii) the p.d.f. $f\left(
y|x,\theta \right)$ of $Y\left( \theta \right) $ given $X$
is continuously differentiable in $\left( y,x,\theta \right) $ with $f\left(
y|x,\theta \right) >0$ for all $\left( y,x,\theta \right) $ such that $%
\left( x,\theta \right) \in \bigcup\limits_{\theta \in \Theta }\left\{ 
\mathcal{X}\left( \theta \right) \mathcal{\times \theta }\right\} $ and $y$
is in the interior of the support of $F\left( \cdot |x,\theta \right) $.
\end{assumption}

Asymptotically linear estimators in Assumption \ref{A3} refer to the class of extremum estimators as considered in \cite{NM94}.  Examples include MLE, NLS, and the GMM class. It implies $\sqrt{n}$-consistency of the first step estimator and is key to the derivation of the asymptotic normality result for the second-step estimator. The triangular structure imposed by Assumption \ref{A1} ensures that $X(\theta)$ is not a function of $Y$ and therefore remains exogenous; it is useful in the example of Section \ref{subsec:regandQR1}. Assumption \ref{A2}-(ii) is a high level assumption that can be derived from
Assumption \ref{A1} and the quantile regression slope $\beta \left( \cdot \right) $
since $g\left( Y,X,\theta _{0}\right) =X\left( \theta _{0}\right) ^{\prime
}\beta \left( U\right) $. It implicitly requests a monotone $%
g\left( \cdot ,X,\theta \right) $ with non zero derivatives, as $f\left(
\cdot |x,\theta \right) $ may diverge otherwise. Indeed, if $\partial
g\left( y,x,\theta \right) /\partial y>0$ and $f\left( y|x\right) $ is the
p.d.f. of $Y$ given $X$ (assuming $X\left( \theta \right) =X$ for the sake
of the brevity of this discussion), it holds%
\[
f\left( y|x,\theta \right) =\frac{1}{\frac{\partial g}{\partial y}\left[
g^{-1}\left( y,x,\theta \right) ,x,\theta \right] }f\left[ g^{-1}\left(
y,x,\theta \right) |x\right] 
\]%
which may not be bounded if $\partial g\left( y,x,\theta \right) /\partial y$
vanishes. Assumption \ref{A2}-(ii) then holds if $f\left( y|x\right) $ is
continuously differentiable in $\left( x,y\right) $ and $g\left( y,x,\theta
\right) $ twice differentiable with respect to $y$ and $\theta $ with
bounded partial derivatives. Assumption \ref{A2}-(i) is similar, but note that the
transformation $X\left( \theta \right) =h\left( X,\theta \right) $ does not
need to be one to one, as $X\left( \theta \right) $ may have a smaller
dimension than $X$.

The QR estimator of the slope coefficient is
an estimator of $\beta \left( \tau ;\widehat{\theta }\right) $ where%
\begin{align}\label{eq:QRdef}
\beta \left( \tau ;\theta \right) =\arg \min_{\beta }\mathbb{E}\left[ \rho
_{\tau }\left( Y\left( \theta \right) -X^{\prime }\left( \theta \right)
\beta \right) \right] .
\end{align}

Assumption \ref{A2} ensures that the objective function is strictly convex
for all $\theta $, so that $\beta \left( \tau ;\theta \right) $ is the
unique solution of the first order condition%
\[
0=\mathbb{E}\left[ \left\{ \mathbb{I}\left( Y\left( \theta \right) \leq
X^{\prime }\left( \theta \right) \beta \right) -\tau \right\} X\left( \theta
\right) \right] =\mathbb{E}\left[ \left\{ F\left( \left. X^{\prime }\left(
\theta \right) \beta \right\vert X,\theta \right) -\tau \right\} X\left(
\theta \right) \right] 
\]%
This together with the Implicit Function Theorem implies that $\beta \left(
\tau ;\theta \right) $ is differentiable with respect to $\theta $, as
established in the following Proposition.

\begin{proposition}\label{P1}
Under Assumptions 2 and 3, $\beta \left( \tau ;\theta \right) $ is
continuously differentiable with respect to $\theta $ for any $\theta \in \Theta$ and $0<\tau <1$.
It holds moreover%
\[
\frac{\partial \beta \left( \tau ;\theta _{0}\right) }{\partial \theta }%
=H\left( \tau ;\theta _{0}\right) ^{-1}D\left( \tau ;\theta _{0}\right) 
\]%
where
\begin{eqnarray*}
H\left( \tau ;\theta _{0}\right)  &=&\mathbb{E}\left[ f\left( \left.
X^{\prime }\left( \theta _{0}\right) \beta \left(\tau; \theta _{0}\right)
\right\vert X,\theta _{0}\right) X\left( \theta _{0}\right) X^{\prime
}\left( \theta _{0}\right) \right]  \\
D\left( \tau ;\theta _{0}\right)  &=&-\left. \frac{\partial }{\partial
\theta }\left[ \mathbb{E}\left[ \left\{ F\left( \left. X^{\prime }\left(
\theta \right) \beta \right\vert X,\theta \right) -\tau \right\} X\left(
\theta \right) \right] \right] \right\vert _{\theta =\theta _{0},\beta
=\beta \left( \tau ;\theta _{0}\right) }.
\end{eqnarray*}
\end{proposition}

\textbf{Proof of Proposition \ref{P1}}: See proof section in Appendix \hyperref[sec:Ap1]{1}.

\medskip

\noindent The matrix $H\left( \tau ;\theta _{0}\right) $ plays an important role in
the asymptotic distribution of standard QR estimators, see below and \cite{K05}. The existence of its inverse is established in Lemma \ref{AN:L2} of the proof section in Appendix \hyperref[sec:Ap1]{1}.
The matrix $D\left( \tau ;\theta _{0}\right) $ is specific to two stage
estimation. With known $\theta_0$, a linear representation for $\sqrt{n}\left( \widehat{\beta }\left( \tau ;%
\theta_0\right) -\beta \left( \tau ,\theta_0\right)
\right) $ can be found in \cite{K05} Section $4.3$, among others, from which asymptotic normality easily follows. But estimating the parameter $\theta$ induces some important changes compared to a known $\theta _{0}$ and requires finding an approximation for $\sqrt{n}\left( \widehat{\beta }\left( \tau ;%
\widehat{\theta }\right) -\beta \left( \tau ,\widehat{\theta }\right)
\right) $. The approach used here builds on a Bahadur expansion which holds uniformly in $\theta $ and $\tau $.  While detailed proofs are in Appendix \hyperref[sec:Ap1]{1}-\hyperref[sec:Ap2]{2}, a heuristic description of the Bahadur expansion proof is as below. 

\paragraph{Heuristics.}  
Define%
\begin{eqnarray}
\widehat{S}\left( \tau ;\theta \right)  &=&\frac{1}{\sqrt{n}}\sum_{i=1}^{n}%
\left[ \mathbb{I}\left( Y_{i}\left( \theta \right) \leq X_{i}^{\prime
}\left( \theta \right) \beta \left( \tau ;\theta \right) \right) -\tau %
\right] X_{i}\left( \theta \right)\label{eq:def1} , \\
J\left( \tau ;\theta \right)  &=&\tau \left( 1-\tau \right) \mathbb{E}\left[
X_{i}\left( \theta \right) X_{i}^{\prime }\left( \theta \right) \right]\label{eq:def2}  \\
\widehat{\mathcal{E}}\left( \tau ;\theta \right)  &=&\sqrt{n}\left( \widehat{%
\beta }\left( \tau ;\theta \right) -\beta \left( \tau ;\theta \right)
\right) -\left(-H^{-1}\left( \tau ;\theta \right) \widehat{S}\left( \tau ;\theta
\right) \right)\label{eq:def3}.
\end{eqnarray}%
Note that for a given $\theta$, $\widehat{S}\left( \tau ;\theta \right) / \sqrt{n}$ is the score of the
objective function in \eqref{eq:QRdef} and $\widehat{S}\left( \tau ;\theta
\right) $ is centered for $0<\tau <1$ with variance $J\left( \tau ;\theta
\right) $. The basic idea is to approximate  $\sqrt{n}\left( \widehat{\beta }\left( \tau ;%
\widehat{\theta }\right) -\beta \left( \tau ,\widehat{\theta }\right)
\right) $ with $-H^{-1}\left( \tau ;\theta_0 \right) \widehat{S}\left( \tau ;\theta_0
\right)$, assuming that the approximation error is of the right order.  If so, the asymptotic normality of the two-step QR estimator follows from that of its score evaluated at the true first stage. 

Crucially, it needs to be shown that the approximation error is small. The approach here is based on two main results: showing that the Bahadur error term $\widehat{\mathcal{E}}\left( \tau ;\theta \right)$ is small by finding its uniform bound for all $\theta$ and $\tau$, and proving the stochastic equicontinuity of $H^{-1}( \tau ;\theta)\hat{S}( \tau ;\theta)$ at the true $\theta$. The outline of the proof is as follows. Let $\widehat{\mathcal{E}}\left( \tau ;\theta \right) = \arg \min_\epsilon \mathbb{L}_n \left( \epsilon, \tau ;\theta \right)$, where $\mathbb{L}_n$ is so defined to be a linear combination of $\rho_{\tau}(\cdot)$ and, thus, is convex. Consider the decomposition $\mathbb{L}_n \left(\epsilon, \tau ;\theta \right) = \mathbb{L}^0_n \left(\epsilon, \tau ;\theta \right) + \mathbb{R}_n \left( \epsilon, \tau ;\theta \right)$, where $\mathbb{L}^0_n$ is the quadratic approximation of $\mathbb{L}_n$ and $\mathbb{R}_n$ is the remainder term. Finding uniform order for $\widehat{\mathcal{E}}$ means finding bounds for the probability of $\vert \vert \widehat{\mathcal{E}} \vert\vert \geq t_n$, for a small number $t_n$ such that $\lim_{n\rightarrow \infty} t_n \rightarrow 0$, for all $\theta$ and $\tau$. This involves finding bounds on $\inf_{\vert \vert \epsilon \vert\vert \geq t_n} \mathbb{L}_n \left( \epsilon, \tau ;\theta \right)$, which, in turn, requests placing bounds on $\inf_{\vert \vert \epsilon \vert\vert = t_n} \mathbb{L}^0_n \left( \epsilon, \tau ;\theta \right)$ and $\sup_{\vert \vert \epsilon \vert\vert = t_n} \mathbb{R}_n \left( \epsilon, \tau ;\theta \right)$. Note that convexity allows us to make inference for the non-compact set $\vert \vert \epsilon \vert\vert \geq t_n$ by considering a compact set $\vert \vert \epsilon \vert\vert = t_n$ (as detailed under heading `Uniform order for $\widehat{\mathcal{E}}(\tau;\theta)$' in Appendix \hyperref[sec:Ap1]{1}). Obtaining bounds for $\mathbb{L}^0_n$ is straightforward. Uniform order of $\mathbb{R}_n$ for all $\theta$ and $\tau$, as obtained in Appendix \hyperref[sec:Ap1]{1} equation \eqref{AN:eq15}, relies on establishing Bernstein type maximal inequality for the empirical process $\mathbb{R}_n$ using Theorem $6.8$ of \cite{M07}; see Lemmas \ref{AN:L4}-\ref{AN:L5} in Appendix \hyperref[sec:Ap1]{1}. The derivation of the maximal inequality Lemmas is given in Appendix \hyperref[sec:Ap2]{2} in two steps; this requires finding sets of all functions that cover the empirical process (called brackets) (see Step \hyperref[L2.1S2]{1} in Appendix \hyperref[sec:Ap2]{2}), such that the minimum number of such brackets of a given metric can be used to derive a maximum limit on the uniform order for the remainder term (see Step \hyperref[L2.1S3]{2} in Appendix \hyperref[sec:Ap2]{2}). 
Stochastic equicontinuity of $H^{-1}\hat{S}$ also follows from similar arguments of maximal inequality under bracketing entropy.  The next Proposition presents the Bahadur error bound and stochastic equicontinuity result to establish linearisation of the GQR estimator, uniformly in $\tau$ and $\theta$. 

\begin{proposition}\label{P2}
Under Assumptions \ref{A3}-\ref{A2} it holds for any compact parameter set $\Theta $ and $C>0$%
\begin{align}
&\text{\textit{(i)} } \sup_{\left( \tau ,\theta \right) \in \left[ \underline{\tau },\overline{%
\tau }\right] \times \Theta }\left\Vert \widehat{\mathcal{E}}\left( \tau
;\theta \right) \right\Vert  = O_{\mathbb{P}}\left( \frac{\log ^{3/4}n}{%
n^{1/4}}\right) ,  \label{Baha} \\
&\text{\textit{(ii)} } \sup_{\left( \tau ,\theta \right) \in \left[ \underline{\tau },\overline{%
\tau }\right] \times \mathcal{B}\left( \theta _{0},Cn^{-1/2}\right)
}\left\Vert H^{-1}\left( \tau ;\theta \right) \widehat{S}\left( \tau ;\theta
\right) -H^{-1}\left( \tau ;\theta _{0}\right) \widehat{S}\left( \tau
;\theta _{0}\right) \right\Vert  =O_{\mathbb{P}}\left( \frac{\log ^{1/2}n}{%
n^{1/4}}\right)  \label{Stocheq}
\end{align}%
where $0<\underline{\tau }\leq \overline{\tau }<1$ and $\mathcal{B}\left(
\theta _{0},\varrho \right) =\left\{ \theta ;\left\Vert \theta -\theta
_{0}\right\Vert \leq \varrho \right\} $.
\end{proposition}

\textbf{Proof of Proposition \ref{P2}}: See proof section in Appendix \hyperref[sec:Ap1]{1}.

\medskip

\noindent Propositions \ref{P1} and \ref{P2} give the next Theorem, which
states a Central Limit Theorem for the two step estimator of the slope
coefficient. Note that in absence of the parameter $\theta$, the asymptotic normality result is the same as that for usual quantile regression estimator, as derived in Theorem $4.1$ of \cite{K05}.  

\begin{theorem}\label{AN:th1}
Under Assumptions \ref{A3}-\ref{A2}, it holds for any $\tau $ in $\left( 0,1\right) $%
\[
\sqrt{n}\left( \widehat{\beta }\left( \tau \right) -\beta \left( \tau
\right) \right) \overset{d}{\rightarrow }\mathcal{N}\left( 0,V\left( \tau
\right) \right) 
\]%
where%
\begin{eqnarray*}
&V\left( \tau \right)  =H\left( \tau ;\theta _{0}\right) ^{-1}\left[
J\left( \tau ;\theta _{0}\right) +D\left( \tau ;\theta _{0}\right) C_{\Psi
S}\left( \tau \right) +C_{\Psi S}^{\prime }\left( \tau \right) D^{\prime
}\left( \tau ;\theta _{0}\right) \right.\\
&\left. +D\left( \tau ;\theta _{0}\right) C_{\Psi
\Psi }D^{\prime }\left( \tau ;\theta _{0}\right) \right] H\left( \tau
;\theta _{0}\right) ^{-1}, \\
&C_{\Psi \Psi } =\mathbb{E}\left[ \Psi \left( Z\right) \Psi ^{\prime
}\left( Z\right) \right] ,\quad C_{\Psi S}\left( \tau \right) =\mathbb{E}%
\left[\Psi \left( Z\right)  X^\prime\left( \theta _{0}\right) \left\{ 
\mathbb{I}\left[ Y\left( \theta _{0}\right) \leq X^{\prime }\left( \theta
_{0}\right) \beta \left( \tau ;\theta _{0}\right) \right] -\tau \right\} %
\right] .
\end{eqnarray*}
\end{theorem} 

\textbf{Proof of Theorem \ref{AN:th1}. }
Proposition \ref{P1} yields that%
\begin{eqnarray}\label{eq:AN1}
\sqrt{n}\left( \widehat{\beta }\left( \tau ;\widehat{\theta }\right) -\beta
\left( \tau; \theta_0 \right) \right)  &=&\sqrt{n}\left( \widehat{\beta }\left( \tau ;%
\widehat{\theta }\right) -\beta \left( \tau ,\widehat{\theta }\right)
\right) +\sqrt{n}\left( \beta \left( \tau ;\widehat{\theta }\right) -\beta
\left( \tau ;\theta _{0}\right) \right)   \nonumber \\
&=&\sqrt{n}\left( \widehat{\beta }\left( \tau ;\widehat{\theta }\right)
-\beta \left( \tau ,\widehat{\theta }\right) \right) +\left( \frac{\partial
\beta \left( \tau ;\theta _{0}\right) }{\partial \theta }+o_{\mathbb{P}%
}\left( 1\right) \right) \sqrt{n}\left( \widehat{\theta }-\theta _{0}\right) 
\nonumber \\
&=&\sqrt{n}\left( \widehat{\beta }\left( \tau ;\widehat{\theta }\right)
-\beta \left( \tau ,\widehat{\theta }\right) \right)   \nonumber \\
&&+\left( \frac{\partial \beta \left( \tau ;\theta _{0}\right) }{\partial
\theta }\right) ^{\prime }\frac{1}{\sqrt{n}}\sum_{i=1}^{n}\Psi \left(
Z_{i}\right) +o_{\mathbb{P}}\left( 1\right)   
\end{eqnarray}%
where the last line holds thanks to Assumption \ref{A3}. 
Equation \eqref{eq:AN1} and
Proposition \ref{P2} give%
\begin{equation}\label{eq:AN2}
\begin{aligned}
\sqrt{n}\left( \widehat{\beta }\left( \tau \right) -\beta
\left( \tau \right) \right)  &= H^{-1}\left( \tau ;\widehat{\theta }\right) 
\widehat{S}\left( \tau ;\widehat{\theta }\right) +\left( \frac{\partial
\beta \left( \tau ;\theta _{0}\right) }{\partial \theta }\right) ^{\prime }%
\frac{1}{\sqrt{n}}\sum_{i=1}^{n}\Psi \left( Z_{i}\right) +o_{\mathbb{P}%
}\left( 1\right)    \\
&=H^{-1}\left( \tau ;\theta _{0}\right) \widehat{S}\left( \tau ;\theta
_{0}\right) +\left( \frac{\partial \beta \left( \tau ;\theta _{0}\right) }{%
\partial \theta }\right) ^{\prime }\frac{1}{\sqrt{n}}\sum_{i=1}^{n}\Psi
\left( Z_{i}\right) +o_{\mathbb{P}}\left( 1\right),  
\end{aligned}
\end{equation}
where the last line results from \eqref{Stocheq} since Assumption \ref{A3} and taking $C$ large enough ensure that $\widehat{\theta}$ belongs to  $\mathcal{B}\left(
\theta _{0},C n^{-1/2} \right)$ with high probability. Since $ \frac{\partial \beta \left( \tau ;\theta _{0}\right) }{%
\partial \theta } = H(\tau;\theta_0)^{-1}D(\tau;\theta_0)$ from Proposition \ref{P1}, the Limit distribution of Theorem \ref{AN:th1} follows from  the Multivariate CLT.$\hfill \square $ 

\smallskip 

\textbf{Remark 1. }As Propositions \ref{P1} and \ref{P2} hold uniformly in $\tau $, the
expansion \eqref{eq:AN2} also does. Since Functional Central Limit Theorems
for $\widehat{S}\left( \tau ;\theta _{0}\right) $ can be applied, \eqref{eq:AN2} can be used to obtain a Functional Central Limit Theorem for the two
step quantile regression estimator.

\smallskip 

\textbf{Remark 2.} The order of the $o_{\mathbb{P}}\left( 1\right) $
remainder term in \eqref{eq:AN2} can be made more precise, strengthening the
smoothness Assumptions \ref{A1} and \ref{A2} to ensure that $\beta \left( \tau ;\theta
\right) $ is twice continuously differentiable using the Implicit Function
Theorem as in Proposition \ref{P1}. Indeed, if $\beta \left( \tau ;\theta \right) $ is twice
continuously differentiable with respect to $\theta $, the $o_{\mathbb{P}%
}\left( 1\right) $ remainder term in \eqref{eq:AN1} is an $O_{\mathbb{P}%
}\left( n^{-1/2}\right) $ and the order of the $o_{\mathbb{P}}\left(
1\right) $ remainder term in \eqref{eq:AN2} follows from (\ref{Baha}) and is $%
O_{\mathbb{P}}\left( n^{-1/4}\log ^{3/4}n\right) $.

\smallskip

\textbf{Remark 3.} The proof can be easily modified for the case where $\theta$ depends upon $\tau$.

\smallskip

\textbf{Remark 4.} For estimating the GQR asymptotic variance, a kernel-based approach can be employed with numerical derivatives. But bootstrap may be preferable and, indeed, is more suitable for quantile regression (see \cite{K05} and the references therein). The validity of bootstrap for obtaining asymptotic confidence interval of two-step semiparametric estimators with non-smooth objective function has been proven by \cite{CLV03}, implying its correctness for the GQR estimator. 

\section{Examples revisited} \label{sec:examples2}

In this section, we apply the asymptotic theory results of Section \ref{sec:asymptotics} to the motivating examples introduced in Section \ref{sec:examples1}.

\subsection{Quantile regression with constant slope }\label{subsec:regandQR2}

For the quantile regression model \eqref{eq:ols1}, recall that the constant paramater $\beta_1(\cdot)$ is estimated using least squares regression, and the quantile parameters $(\beta_0(\cdot), \beta_2(\cdot))$ are estimated using the generated dependent variable $Y_i(\widehat{\beta_1})=Y_i - \widehat{\beta_1}X_{1i}$ via the two-step quantile regression estimator of \eqref{eq:ols5}. Asymptotic normality of the first step OLS estimator is well established. Denote $X=[1,X_1,X_2]^{\prime}$. Assume that $\mathbb{E}[\varepsilon^2XX^{\prime}]$ is finite and $\mathbb{E}[XX^{\prime}]$ is full rank and finite.
The OLS estimator is asymptotically linear:
\[
\sqrt{n}\left(  \widehat{\beta}  -\beta  \right)= \sum_{i=1}^{n}\left[\mathbb{E}^{-1}\left[XX^{\prime} \right] X_i \varepsilon_i \right] / \sqrt{n} + o_{\mathbb{P}}(1).
\] 
 Denoting $i_{22}=[0,1,0]$, the asymptotic variance of $\widehat{\beta_1}$ is given by 
\begin{equation}\label{eq:ex1.1}
\mathcal{V}(\beta_1) =  i_{22} \left(\mathbb{E}^{-1}[XX^{\prime}]\mathbb{E}[\varepsilon^2XX^{\prime}] \mathbb{E}^{-1}[XX^{\prime}]\right) i_{22}^{\prime}.
\end{equation}

For the second step quantile regression, the dependent variable is generated as $Y(\widehat{\beta_1})=Y-\widehat{\beta_1}X_1$, and the regressors are denoted as $\widetilde{X}=[1,X_2]^{\prime}$. Asymptotic normality of the quantile parameters $\beta(\tau)=(\beta_0(\tau), \beta_2(\tau))^{\prime}$ follows directly from Theorem \ref{AN:th1}:

\begin{equation}\nonumber
\sqrt{n} \left[
\begin{array}
[c]{c}%
 \widehat{\beta_0}\left(  \tau \right)  -\beta_0\left(  \tau\right) \\
\widehat{\beta_2}\left(  \tau \right)  -\beta_2\left(  \tau\right)  
\end{array}
\right]  \stackrel{d}{\longrightarrow} \mathcal{N} \left(0, V(\tau) \right).
\end{equation}
The terms of $V$ are obtained from Theorem \ref{AN:th1}
by replacing $\theta_0 \equiv \beta_1$, $\beta(\tau) \equiv (\beta_{0}(\tau), \beta_{2}(\tau))^{\prime}$, $X(\theta_0) \equiv \widetilde{X}=[1,X_2]^{\prime}$and $Y(\theta_0) \equiv Y(\beta_1) = Y - \beta_1X_1$. Denoting the first $\tau$-derivative of $\beta(\tau)$ as $\beta^{(1)}(\tau)$, $V(\tau)$ comes as follows:
\begin{equation}\label{eq:ex1.2}
\begin{aligned}
&V(\tau) = H(\tau)^{-1} \left\{ J(\tau) +  D(\tau)\mathcal{V}(\beta_1)D(\tau)^\prime +  C(\tau)^\prime D(\tau) + C(\tau) D(\tau)^\prime  \right\}H(\tau)^{-1}, \\
\text{where } &H(\tau) = \mathbb{E}\left[\frac{\widetilde{X}\widetilde{X}^{\prime}}{\beta_0^{(1)}(\tau)+\beta_2^{(1)}(\tau)X_2} \right], \quad J(\tau) = \tau(1 - \tau)\mathbb{E}\left[\widetilde{X}\widetilde{X}^{\prime}\right],\\
&D(\tau) = - \mathbb{E}\left[\frac{X_1 \widetilde{X}}{\beta_0^{(1)}(\tau)+\beta_2^{(1)}(\tau)X_2} \right] \text{and} \\
&C(\tau) = \mathbb{E} \left[g(X) \left\{ \int_{0}^{\tau}\left(\beta_0(t) + \beta_2(t)X_2 \right)dt - \tau \left(\beta_0(\tau) + \beta_2(\tau)X_2   \right)               \right\}   \right], \\
\end{aligned}
\end{equation}
with $ g(X)= \widetilde{X}\left[0,1,0  \right]\mathbb{E}^{-1}\left[XX^\prime \right]X$.  

\subsection{Random coefficient model}\label{subsec:randomcoeff2}
For the random coefficient model in \eqref{eq:random1}, 
recall that for identification of the quantile specification in \eqref{eq:randomquant}, we normalise $\xi(1/2)=1$.  Denote the $\tau$-derivative of $\xi(\tau)$ by $\xi^{(1)}(\tau)$. The first step parameters $\theta \equiv (\mu,\Sigma)$ are estimated by \eqref{eq:randomfirststep}; denote $\mathcal{G}(\cdot) = X_i^\prime \mu + \left\vert\left\vert \Sigma^{1/2} X_i \right\vert\right\vert$. The $\theta$-derivative of $\mathcal{G}(\cdot)$ is given by
\begin{align*}
\mathcal{G}^{\theta} =\left[ \begin{matrix}
 X \\
 \frac{\partial\left\vert\left\vert \Sigma^{1/2} X \right\vert\right\vert}{\partial\sigma} 
\end{matrix}\right] 
\end{align*} 
where $\sigma$ is a $(K+1)^2 \times 1$ vector that stacks the columns of $\Sigma^{1/2}$. The non-linear median regression estimator of \eqref{eq:randomfirststep} is asymptotically linear (see Section 4.4 of \cite{K05}):
\begin{align*}
\sqrt{n}(\widehat{\theta}-\theta) = H_1^{-1} \sum_{i=1}^{n} \mathcal{G}_i^{\theta}\left[1/2 - \mathbb{I}\left(Y_i \leq \mathcal{G}_i(\cdot) \right) \right]/\sqrt{n} + o_{\mathbb{P}}(1)
\end{align*}
where $H_1 = \mathbb{E}\left[\frac{\mathcal{G}^{\theta}{\mathcal{G}^{\theta}}^\prime}{\left\vert\left\vert \Sigma^{1/2} X \right\vert\right\vert \xi^{(1)} (1/2)} \right]$. The asymptotic variance of $\widehat{\theta}$ is given by $\mathcal{V}(\theta)=H_1^{-1}\mathbb{E}\left[\mathcal{G}^{\theta}{\mathcal{G}^{\theta}}^\prime \right]H_1^{-1}/4$. 

The second stage involves finding empirical quantiles of the generated dependent variable $Y(\widehat{\theta}) \equiv Y(\widehat{\mu}, \widehat{\Sigma}) = \frac{Y_i - X_i^\prime \widehat{\mu}}{ \left\vert\left\vert \widehat{\Sigma}^{1/2}X \right\vert\right\vert }$ by \eqref{eq:randomsecondstep}. The asymptotic normality of $\widehat{\xi}(\tau)$ follows from Theorem \ref{AN:th1}: 
\[
\sqrt{n} \left[  \widehat{\xi}\left(  \tau \right)  -\xi\left(  \tau\right)
\right]  \stackrel{d}{\longrightarrow} \mathcal{N} \left(0, V(\tau) \right),
\]
where 
\[ V(\tau) = H(\tau)^{-1} \left\{ J(\tau) +  D(\tau)\mathcal{V}(\theta)D(\tau)^\prime +  C(\tau)^\prime D(\tau) + C(\tau) D(\tau)^\prime  \right\}H(\tau)^{-1}.
\]
The terms of $V(\tau)$ are:
\begin{eqnarray*}
&H(\tau) = \mathbb{E}\left[\frac{1}{\xi^{(1)} (\tau)} \right], \quad J(\tau) = \tau(1 - \tau), \quad D(\tau) = - \mathbb{E}\left[ \frac{1}{\xi^{(1)} (\tau)} \left[\begin{matrix} X \\
 \frac{\partial\left\vert\left\vert \Sigma^{1/2} X \right\vert\right\vert}{\partial\sigma} \xi(\tau) \end{matrix}\right] \right],\\
&C(\tau) = \mathbb{E} \left[ \Psi(\cdot) \left\{\mathbb{I}\left(\frac{Y - X^\prime {\mu}}{ \left\vert\left\vert {\Sigma}^{1/2}X \right\vert\right\vert } \leq \xi(\tau) \right) - \tau \right\} \right] \text{ where } \Psi(\cdot) = H_1^{-1} \mathcal{G}^{\theta}\left[1/2 - \mathbb{I}\left(Y \leq \mathcal{G}(\cdot) \right) \right]. 
\end{eqnarray*}

\subsection{Box-Cox power transformation}\label{subsec:boxcox2}

The box-cox transformation parameter of \eqref{eq:bc1} is estimated using the nonlinear IV (NIV) estimator of \eqref{eq:bc3}. The conditional quantile model for the generated dependent variable $Y(\widehat{\lambda})$ is assumed linear in parameters, which are estimated using the QR estimator of \eqref{eq:bc4}. \cite{A74} establishes the limiting behaviour of the NIV estimator. Assume that $\mathbb{E}\left[\left(Y(\lambda)-X^\prime \beta \right)^2 WW^\prime \right]$ is finite and $\Omega$ is full rank and finite.

Note that if $\beta$ is a $K$-dimension vector, then the NIV estimator estimates $(K+1)$ parameters, denoted by $\theta = [\lambda,\beta^{\prime}]^{\prime}$. Denote the $(K+1)$ order square matrix, 
\[
G = \mathbb{E}\left[W \frac{\partial Y(\lambda)}{\partial \lambda}, -WX^{\prime} \right].
\]
Then, the NIV estimator is asymptotically linear:
\[
\sqrt{n}\left(  \widehat{\theta}  -\theta  \right)= \sum_{i=1}^{n}\left[-\left(G^{\prime}\Omega G\right)^{-1}G^{\prime}\Omega W_i (Y_i(\lambda)-X_i^{\prime}\beta)\right] / \sqrt{n} + o_{\mathbb{P}}(1).
\]
The asymptotic variance of $\widehat{\lambda}$, denoted by $\mathcal{V}(\lambda)$, is the first term of the asymptotic variance-covariance matrix for $\widehat{\theta}$. Denoting $i_{11}=[1,\boldsymbol{0}_{K \times 1}]$, where $\boldsymbol{0}_{K \times 1}$ is a $K$-dimension row vector of zeros,
\begin{equation}\nonumber
\mathcal{V}(\lambda) =  i_{11} \left(\left(G^{\prime}\Omega G\right)^{-1}G^{\prime}\Omega \mathbb{E}\left[\left(Y(\lambda)-X^\prime \beta \right) WW^\prime \right]\Omega G \left(G^{\prime}\Omega G\right)^{-1} \right) i_{11}^{\prime}.
\end{equation}
Asymptotic normality for the quantile estimates obtained from QR of $Y(\widehat{\lambda})$ on $X$ follows directly from Theorem \ref{AN:th1}. 

\begin{equation}\nonumber
\sqrt{n} \left(\widehat{\beta}(\tau) - \beta(\tau) \right)  \stackrel{d}{\longrightarrow} \mathcal{N} \left(0, V(\tau) \right),
\end{equation}
where 
\[ V(\tau) = H(\tau)^{-1} \left\{ J(\tau) +  D(\tau)\mathcal{V}(\lambda)D(\tau)^\prime +  C(\tau)^\prime D(\tau) + C(\tau) D(\tau)^\prime  \right\}H(\tau)^{-1}.
\]
The terms of $V(\tau)$ are given by
\begin{eqnarray*}
&H(\tau) = \mathbb{E}\left[\frac{XX^{\prime}}{X^{\prime}\beta^{(1)}(\tau)} \right], \quad J(\tau) = \tau(1 - \tau)\mathbb{E}\left[XX^{\prime}\right], \quad D(\tau) = - \mathbb{E}\left[\frac{X}{X^\prime \beta^{(1)}(\tau)} \frac{\partial \left(X^\prime \beta(\tau) \lambda +1 \right)^{1/\lambda}}{\partial \lambda} \right]\\
&C(\tau) = \mathbb{E} \left[g(X) \left\{ \int_{0}^{\tau}X^{\prime} \beta(t) dt - \tau  X^{\prime} \beta(\tau)  \right\}   \right], 
\end{eqnarray*}
where $g(X)= X\left[1, \boldsymbol{0}_{K \times 1}  \right]\left(-\left(G^{\prime}\Omega G\right)^{-1}G^{\prime}\Omega W \right)$.

\subsection{Endogeneity in quantile regression - control variable approach}\label{subsec:IV2}

The quantile regression model in \eqref{eq:iv1} is estimated in two steps. The first step uses OLS estimator of \eqref{eq:iv2} to estimate $\widehat{\gamma}$. This is used to generate the control variable $\widehat{\eta}=\left(X_i-Z_i^\prime \widehat{\gamma}\right)$, which is included as a regressor in the quantile regression estimator of \eqref{eq:iv3} for estimating the quantile parameters $\delta(\tau)\equiv (\alpha(\tau)^{\prime},\beta(\tau), \lambda(\tau))^{\prime}$. Denote the generated regressors as $X(\gamma) = \left[W^\prime, X, \left(X-Z^\prime \gamma\right)\right]^{\prime}$. We assume that $\mathbb{E}\left[\eta ^2 \vert Z \right] = \sigma^2 $ and $\mathbb{E}\left(ZZ^{\prime}\right)$ is finite.
The OLS estimator is asymptotically linear:
\[
\sqrt{n}\left(  \widehat{\gamma}  -\gamma  \right)= \sum_{i=1}^{n}\left[\mathbb{E}^{-1}\left[ZZ^{\prime} \right] Z_i \eta_i \right] / \sqrt{n} + o_{\mathbb{P}}(1).
\]
The asymptotic normality of the quantile parameters $\delta(\tau)$ follows directly from Theorem \ref{AN:th1}, 
\[
\sqrt{n} \left[  \widehat{\delta}\left(  \tau \right)  -\delta\left(  \tau\right)
\right]  \stackrel{d}{\longrightarrow} \mathcal{N} \left(0, V(\tau) \right),
\]
where 
\[ V(\tau) = H(\tau)^{-1} \left\{ J(\tau) +  D(\tau)\sigma^2\mathbb{E}^{-1}[ZZ^{\prime}]D(\tau)^\prime +  C(\tau)^\prime D(\tau) + C(\tau) D(\tau)^\prime  \right\}H(\tau)^{-1}.
\]
The terms of $V(\tau)$ are given by
\begin{eqnarray*}
& J(\tau) = \tau(1 - \tau)\mathbb{E}\left[X(\gamma)X(\gamma)^{\prime}\right], \quad D(\tau) = -\left. \frac{\partial }{\partial
\gamma }\left[ \mathbb{E}\left[ \left\{ F\left( \left. X^{\prime }\left(
\gamma \right) \delta \right\vert W, X,\gamma \right) -\tau \right\} X\left(
\gamma \right) \right] \right] \right\vert _{\delta = \delta(\tau) } \\
&C(\tau) = \mathbb{E} \left[X(\gamma) \left(\mathbb{I}\left(Y \leq X(\gamma)^{\prime}\delta(\tau) \right) - \tau \right) \eta Z^\prime \mathbb{E}^{-1} [ZZ^\prime]\right], \quad H(\tau) = \mathbb{E}\left[\frac{X(\gamma)X(\gamma)^{\prime}}{X(\gamma)^{\prime}\delta^{(1)}(\tau)} \right].
\end{eqnarray*}

\section{Simulations}\label{sec:sims}
This section reports results of simulation exercises to illustrate the performance of the two-step GQR estimator and validate the asymptotic normality result of Theorem \ref{AN:th1}.  
The simulations are based on the quantile regression with constant slope model of Section \ref{subsec:regandQR1}\[
Q_{Y}\left( \tau |X\right) =\beta _{0}\left( \tau \right) +\beta
_{1}\left( \tau \right) X_{1}+\beta _{2}\left( \tau \right) X_{2},
\] with true parameters as, 
\begin{equation}\label{eq:DGP}
\beta_0(\tau)=e^{\tau}, \quad \beta_1(\tau) = \beta_1 = 1 \ \forall \tau, \quad \beta_2(\tau) = 2\tau^2. 
\end{equation}
Data are generated as $Y_i=\beta_0(U_i)+\beta_1X_{1i}+\beta_2(U_i)X_{2i}$, where $(X_{1i}, X_{2i})$ are uniform random variables  between $[1,5]$ and $[3,10]$ respectively, $U_i$ is a $[0,1]$-uniform random variable, $i= 1,\cdots, n$. Sample sizes of $n=100$ and $n=1000$ are considered. The number of simulation replications is set to $1000$. 

GQR estimation of the above model proceeds as in Section \ref{subsec:regandQR1}. We also compare GQR with standard quantile regression, where all parameters - both constant and quantile-varying ones - are estimated together by quantile regression of $Y$ on $X$'s. Also, to clearly see the effect of first stage estimation on overall variance, the GQR estimator is compared with an infeasible quantile regression (i-QR) estimator which uses the true value of the first step parameter instead of its estimate, that is, the unknown dependent variable $Y_i^*(\beta_1) = Y_i - \beta_1X_{1i}$, for QR based estimation of quantile parameters. 

Following Remark $4$, asymptotic variance estimation for validating the asymptotic normality result and for finding confidence intervals follows \cite{B94}'s design matrix bootstrap. Design matrix bootstrap is extensively used in empirical applications for quantile regression involving large samples, see, for instance, \cite{B94} and \cite{A02}\footnote{See \cite{B95} and \cite{KH01} for a comparison of various QR variance estimators; they conclude in favour of design matrix bootstrap.}. The approach is as follows.
For $B$ bootstrap replications, each of size of $m$ (drawn with replacement from an overall sample size of $n$), $b=1,\cdots,B$ bootstrap quantile estimates are obtained at each quantile level. 
This follows the so-called $m$-out-of-$n$ bootstrap technique which provides significant computational advantage when sample size is large. Following \cite{B94}, the sample covariance of these estimates, rescaled by $(m/n)$, constitutes a valid estimator of the covariance matrix of the QR estimator. Hence, the estimate for asymptotic covariance $V(\tau)$ with quantile parameters $\beta(\cdot)$ and the bootstrap estimates denoted by $\widehat{\beta}^b(\tau)$, $b=1,\cdots,B$, is given by
\begin{equation}\label{eq:boot}
\widehat{V}(\tau)=n \left(\frac{m}{n}\right)\frac{1}{B}\sum_{b=1}^{B}\left(\widehat{\beta}^b(\tau) - \widehat{\beta}^b_{\mathcal{A}}(\tau) \right)\left( \widehat{\beta}^b(\tau) - \widehat{\beta}^b_{\mathcal{A}}(\tau) \right)^{\prime},
\end{equation}
where $\widehat{\beta}^b_{\mathcal{A}}(\tau)$ is the average of the $B$ bootstrap estimates. We set $B=1000$; for $n=1000$, the bootstrap sample size is $m=300$, while for $n=100$, we have $m=n$. The choice of bootstrap replications and sample size are consistent with \cite{B95} and \cite{AB00}. We estimate $\widehat{V}(\tau)$ from \eqref{eq:boot} for each of the $1000$ simulations and report the average.

\subsection{Bias-RMSE and coverage rate}\label{sec:biasRMSEcov}

For GQR in Tables \ref{tab:RMSEbeta0}-\ref{tab:KS1000}, the first step least squares regression gives the mean  of $\hat{\beta_1}$ as $1.007$ (with average standard deviation $= 0.3953$) for a sample size of $100$, and $1.001$ (with average standard deviation $= 0.1242$) for a sample size of $1000$, respectively. The fact that OLS is unbiased is expected but the standard deviation is meaningful as it gives an idea of how much the first step impacts the overall variance.

Table \ref{tab:RMSEbeta0} reports the bias-root mean square error (RMSE) for  $\widehat{\beta}_0(\cdot)$ for GQR, standard QR and i-QR estimation methods, with varying $n$. All methods of estimation have low biases and the RMSE falls with increasing sample size. We note that while all estimation procedures have similar biases, the RMSE with GQR is greater than that of QR for the first quantile, and the opposite is true for the rest of the quantiles, an observation we investigate further in Section \ref{sec:GQRvsQR}. As expected, the RMSE with GQR is greater than that of i-QR at each quantile, with substantial difference in some, due to the added variance contribution from first step estimation in the former.  

Table \ref{tab:RMSEbeta2} reports the Bias-RMSE results for the slope parameter $\widehat{\beta}_2(\cdot)$. The bias and RMSE are similar for all three methods of estimation and the RMSE falls with increase in sample size. The following remark explains this. \vspace{-2mm}
\paragraph{Remark.}  In the GQR asymptotic variance for the QR with constant slope model as given by \eqref{eq:ex1.2}, if the covariates $X_1$ and $X_2$ are independent, as considered here, it holds that \vspace{-2mm}
\begin{enumerate}[(i)]
\item The covariance between first and second step estimates is zero: $C(\tau) = 0$.\vspace{-2mm}
\item The first step estimation has an effect on the second-step variance for the intercept, $\widehat{\beta_0}(\tau)$, but not for the slope parameter  $\widehat{\beta_2}(\tau)$, as $H(\tau)^{-1}D(\tau)$ in \eqref{eq:ex1.2} evaluates to $\left[-\mathbb{E}[X_1],0\right]^\prime$.
\end{enumerate}
Proofs are straightforward using basic matrix algebra and its outline is presented in Appendix \hyperref[sec:Ap3]{3}. 

\begin{table}[hptb]
\caption{Bias and RMSE of $\hat{\beta}_0(\cdot)$ for $n=100$ and $1000$}%
\label{tab:RMSEbeta0}
\begin{center}
\begin{tabularx}{0.7\textwidth}{cccccc}
      \hline
			\hline
			\multicolumn{1} {X} { } & \multicolumn{1} {X} { } & \multicolumn{2} {c} {$n=100$} & \multicolumn{2} {c} {$n=1000$} \\ [1 ex]
			\hline
			$\tau$&{}&Bias&RMSE&Bias&RMSE\\
			\hline
			$0.2$		&GQR &$0.0320$		&$1.4076$		&$-0.0004$		&$0.4357$		\\
							&QR &$0.0328$		&$0.9152$		&$0.0135$		&$0.2868$		\\
							&i-QR  &$0.0108$		&$0.6971$		&$-0.0124$		&$0.2256$		\\
			$0.4$		&GQR &$0.1374$		&$1.9558$		&$-0.0126$		&$0.6405$		\\
							&QR &$0.0125$		&$2.0384$		&$0.0272$		&$0.6662$		\\
							&i-QR  &$0.1546$		&$1.5300$		&$-0.0137$		&$0.4960$		\\
			$0.6$		&GQR &$0.1654$		&$2.5208$		&$-0.0486$		&$0.8433$		\\
							&QR &$0.0470$		&$2.8203$		&$0.0409$		&$0.9770$		\\
							&i-QR  &$0.0922$		&$2.2092$		&$0.0252$	&$0.7233$		\\
			$0.8$		&GQR &$0.0581$		&$2.7213$		&$-0.0481$		&$0.8440$		\\
							&QR &$-0.1241$		&$3.0094$		&$0.0129$		&$1.0336$		\\
							&i-QR  &$0.0412$		&$2.4851$		&$0.0006$	&$0.7875$		\\
			\hline
		\end{tabularx}
\end{center}
\end{table}

\begin{table}[hptb]
\caption{Bias and RMSE of $\hat{\beta}_2(\cdot)$ for $n=100$ and $1000$}%
\label{tab:RMSEbeta2}
\begin{center}
\begin{tabularx}{0.7\textwidth}{cccccc}
      \hline
			\hline
			\multicolumn{1} {X} { } & \multicolumn{1} {X} { } & \multicolumn{2} {c} {$n=100$} & \multicolumn{2} {c} {$n=1000$} \\ [1 ex]
			\hline
			$\tau$&{}&Bias&RMSE&Bias&RMSE\\
			\hline
			$0.2$		&GQR &$0.0148$		&$0.1362$		&$0.0017$	&$0.0397$		\\
							&QR &$0.0053$		&$0.1235$		&$-0.0034$	&$0.0375$		\\
							&i-QR  &$0.0009$		&$0.1242$		&$-0.0003$	&$0.0395$		\\
			$0.4$		&GQR &$-0.0089$		&$0.2756$		&$0.0014$	&$0.0918$		\\
							&QR &$-0.0028$		&$0.2759$		&$-0.0073$	&$0.0863$		\\		
							&i-QR  &$-0.0257$		&$0.2692$		&$-0.0003$	  &$0.0892$		\\
			$0.6$		&GQR &$-0.0371$		&$0.3972$		&$0.0066$	&$0.1348$		\\
							&QR &$-0.0346$		&$0.3948$		&$-0.0177$	&$0.1286$		\\
							&i-QR  &$-0.0314$		&$0.3897$		&$-0.0097$	  &$0.1288$		\\
			$0.8$		&GQR &$-0.0425$		&$0.4400$		&$0.0036$	&$0.1371$		\\
							&QR &$-0.0452$		&$0.4308$		&$-0.0148$	&$0.1433$		\\
							&i-QR  &$-0.0486$		&$0.4481$		&$-0.0082$		&$0.1408$		\\
			\hline
		\end{tabularx}
\end{center}
\end{table}

As a means for validating the asymptotic normality result, Tables \ref{tab:KS100}-\ref{tab:KS1000} compare the empirical $90\%, 95\%$ and $99\%$ GQR confidence intervals with that in theory for normal approximation, for $n=100$ and $n=1000$. For $\tau=\{0.2,0.4,0.6,0.8\}$, t-stat of the quantile parameters is computed using bootstrapped standard error (SE) from \eqref{eq:boot} and its absolute value is compared with the critical values for $(1-\alpha)$ confidence level of the normal approximation, $(1-\alpha)=0.9,0.95$ and $0.99$, to find if the true quantile parameter is inside the corresponding confidence interval. Repeating the exercise $1000$ times, we find the percentage of times when the true parameter lies inside the $(1-\alpha)$ confidence interval. Coverage rates for QR and i-QR are also reported. For the starting quantile in Table \ref{tab:KS100}, the coverage rate is higher as compared to the nominal level, suggesting variance overestimation for $\tau=0.2$, $n=100$, but it improves for $n=1000$ in Table \ref{tab:KS1000}. Overall, the empirical levels for confidence intervals are close to $(1-\alpha)$ and improves with increasing sample size, which suggests that the estimation procedure gives accurate central limit theorem based confidence intervals. 

\begin{table}[phtb]
\caption{Confidence intervals: nominal vs. empirical, $n=100$, $\text{simulations}=1000$}%
\label{tab:KS100}
\begin{center}
\begin{tabularx}{0.9\textwidth}{cccccccc}
			\hline
			\hline
			\multicolumn{1} {X} { } & \multicolumn{1} {X} { } & \multicolumn{3} {c} {CI for $\beta_0(\cdot)$} & \multicolumn{3} {c} {CI for $\beta_2(\cdot)$} \\ [1 ex]
			\hline
			Nominal level&  &$0.90$&$0.95$&$0.99$&$0.90$&$0.95$&$0.99$\\
			\hline
			Empirical level for $\tau=0.2$ &GQR	&$0.940$	&$0.977$	&$0.997$ &$0.960$	&$0.982$ &$0.995$	\\
			&QR	&$0.949$	&$0.978$	&$0.996$ &$0.923$	&$0.970$ &$0.991$	\\
			&i-QR	&$0.928$	&$0.972$	&$0.994$ &$0.907$	&$0.952$ &$0.986$	\\
			Empirical level for $\tau=0.4$	&GQR &$0.897$	&$0.951$	&$0.991$ &$0.888$	&$0.942$ &$0.980$	\\
			&QR &$0.899$	&$0.948$	&$0.987$ &$0.888$	&$0.940$ &$0.987$	\\
			&i-QR &$0.885$	&$0.940$	&$0.986$ &$0.878$	&$0.933$ &$0.981$	\\
			Empirical level for $\tau=0.6$	&GQR &$0.895$	&$0.944$	&$0.985$ &$0.882$	&$0.934$ &$0.984$	\\
			&QR &$0.893$	&$0.936$	&$0.981$ &$0.883$	&$0.940$ &$0.976$	\\
			&i-QR &$0.878$	&$0.929$	&$0.979$ &$0.877$	&$0.934$ &$0.981$	\\
			Empirical level for $\tau=0.8$	&GQR &$0.900$	&$0.940$	&$0.989$ &$0.895$	&$0.942$ &$0.981$	\\
			&QR &$0.905$	&$0.950$	&$0.984$ &$0.890$	&$0.942$ &$0.986$	\\
			&i-QR &$0.867$	&$0.928$	&$0.978$ &$0.875$	&$0.932$ &$0.975$	\\
			\hline
		\end{tabularx}
\end{center}
\end{table}

\begin{table}[phtb]
\caption{Confidence intervals: nominal vs. empirical, $n=1000$, $\text{simulations}=1000$}%
\label{tab:KS1000}
\begin{center}
\begin{tabularx}{0.9\textwidth}{cccccccc}
			\hline
			\hline
			\multicolumn{1} {X} { } & \multicolumn{1} {X} { } &\multicolumn{3} {c} {CI for $\beta_0(\cdot)$} & \multicolumn{3} {c} {CI for $\beta_2(\cdot)$} \\ [1 ex]
			\hline
			Nominal level& &$0.90$&$0.95$&$0.99$&$0.90$&$0.95$&$0.99$\\
			\hline
			Empirical level for $\tau=0.2$	&GQR &$0.897$	&$0.952$	&$0.992$ &$0.906$	&$0.954$ &$0.989$	\\
			&QR &$0.903$	&$0.953$	&$0.989$ &$0.911$	&$0.947$ &$0.981$	\\
			&i-QR &$0.884$	&$0.940$	&$0.985$ &$0.893$	&$0.939$ &$0.983$	\\
			Empirical level for $\tau=0.4$	&GQR &$0.885$	&$0.933$	&$0.987$ &$0.882$	&$0.944$ &$0.984$	\\
			&QR &$0.886$	&$0.942$	&$0.987$ &$0.907$	&$0.954$ &$0.987$	\\
			&i-QR &$0.901$	&$0.947$	&$0.988$ &$0.895$	&$0.946$ &$0.986$	\\
			Empirical level for $\tau=0.6$	&GQR &$0.886$	&$0.952$	&$0.985$ &$0.891$	&$0.937$ &$0.981$	\\
			&QR &$0.874$	&$0.945$	&$0.980$ &$0.887$	&$0.935$ &$0.984$	\\
			&i-QR &$0.893$	&$0.946$	&$0.985$ &$0.893$	&$0.942$ &$0.986$	\\
			Empirical level for $\tau=0.8$	&GQR &$0.902$	&$0.952$	&$0.991$ &$0.899$	&$0.941$ &$0.986$	\\
			&QR &$0.897$	&$0.946$	&$0.983$ &$0.876$	&$0.935$ &$0.984$	\\
			&i-QR &$0.892$	&$0.946$	&$0.990$ &$0.883$	&$0.942$ &$0.989$	\\
			\hline
		\end{tabularx}
\end{center}
\end{table}

\subsection{Asymptotic variance: Further analysis}\label{sec:GQRvsQR}
In this section, we compare the GQR and QR asymptotic variances both analytically and through simulation, following the RMSE pattern observed in Table \ref{tab:RMSEbeta0} which hints at their relative efficiency being quantile dependent. The data distribution and true parameter values, as assumed in the data generating process, allows comparison based on obtaining explicit asymptotic variance expressions for both GQR and QR. Although the discussion here is specific to the assumed QR with constant slope model, it provides interesting insights, in particular, to the role of first stage estimator in overall GQR variance. Note that while asymptotic variance of GQR, which estimates the constant parameter $\beta_1$ and quantile dependent ones $(\beta_0(\tau),\beta_1(\tau))$ separately, is given by \eqref{eq:ex1.2}, that for standard QR where all parameters are estimated together is given by 
\begin{equation}\label{eq:GQRvsQR1}
V(\tau)_{QR}= H(\tau)_{QR}^{-1}J(\tau)_{QR}H(\tau)_{QR}^{-1} 
\end{equation}
where denoting $X=[1,X_1,X_2]^{\prime}$,
\[
H(\tau)_{QR} = \mathbb{E}\left[\frac{XX^{\prime}}{\beta_0^{(1)}(\tau)+\beta_1^{(1)}(\tau)X_1+\beta_2^{(1)}(\tau)X_2} \right], \quad J(\tau)_{QR} = \tau(1 - \tau)\mathbb{E}\left[XX^{\prime}\right].
\]

\paragraph{Asymptotic variance for $\widehat{\beta}_0(\cdot)$.} Under the  remark noted in Section \ref{sec:biasRMSEcov}, the asymptotic variance of $\widehat{\beta}_0(\cdot)$ for GQR is obtained using \eqref{eq:ex1.2} as follows:
\begin{equation}\label{eq:truebeta0G}
V(\tau)_{GQR,0}= [1,0]H(\tau)^{-1}J(\tau)H(\tau)^{-1}[1,0]^{\prime} + \mathbb{E}^2[X_1]\mathcal{V}(\beta_1)
\end{equation}
where $H(\tau), J(\tau)$ are given by \eqref{eq:ex1.2}. For the true model parameters and distribution considered here, this evaluates to
\begin{equation}\label{eq:truebeta0GQR}
V(\tau)_{GQR,0}= \frac{\tau(1-\tau)}{\left(ac-b^2 \right)^2}\left(c^2 - 2bc \mathbb{E}[X_2] + b^2\mathbb{E}[X_2^2] \right) + \mathbb{E}^2[X_1]\mathcal{V}(\beta_1)
\end{equation}
where 
\begin{eqnarray*}
a &=& \mathbb{E}\left[\frac{1}{\beta_0^{(1)}(\tau)+\beta_2^{(1)}(\tau)X_2} \right] =\frac{1}{28 \tau}\ln{\left(\frac{e^{\tau} + 40 \tau}{e^{\tau} + 12 \tau} \right)}, \\
b &=& \mathbb{E}\left[\frac{X_2}{\beta_0^{(1)}(\tau)+\beta_2^{(1)}(\tau)X_2} \right] = \frac{1}{7 \times 16 \tau^2}\left(28 \tau - e^{\tau}\ln{\left(\frac{e^{\tau} + 40 \tau}{e^{\tau} + 12 \tau} \right)}\right), \\
c &=& \mathbb{E}\left[\frac{X_2^2}{\beta_0^{(1)}(\tau)+\beta_2^{(1)}(\tau)X_2} \right] = \frac{1}{448 \tau^3}\left(728 \tau^2 - 28 \tau e^{\tau} + e^{2\tau}\ln{\left(\frac{e^{\tau} + 40 \tau}{e^{\tau} + 12 \tau} \right)}\right)\\
\mathbb{E}[X_2] &=& 13/2,\quad \mathbb{E}[X_2^2]=139/3,\quad \mathbb{E}[X_1]=3.
\end{eqnarray*}
The first step asymptotic variance $\mathcal{V}(\beta_1)$ is given by \eqref{eq:ex1.1}, and for the model parameters considered here, evaluates to
\begin{eqnarray*}
\mathcal{V}(\beta_1) &=& i_{22}\left(\mathbb{E}[XX^{\prime}]\right)i_{22}^{\prime} \mathbb{E}\left[\varepsilon_0^2 \right]+ i_{22}\left(\mathbb{E}^{-1}[XX^{\prime}]\mathbb{E}[X_2^2 XX^{\prime}]\mathbb{E}^{-1}[XX^{\prime}]\right)i_{22}^{\prime} \mathbb{E}\left[ \varepsilon_2^2 \right] \\
&=& \frac{3}{4} \mathbb{E}\left[\varepsilon_0^2 \right]+ \frac{139}{4} \mathbb{E}\left[ \varepsilon_2^2 \right] =  \frac{3}{4} \times \left(\frac{1}{2}(e^2-1) - (e-1)^2 \right) + \frac{139}{4} \times \left(\frac{4}{5}-\left(\frac{2}{3} \right)^2 \right)
\end{eqnarray*}
where $\varepsilon_i= \left(\beta _{i}\left( U\right) -\overline{\beta }_{i}\right),\ i=0,2$.

\begin{figure}[hpt]
\centering
\includegraphics[width=.7\textwidth]{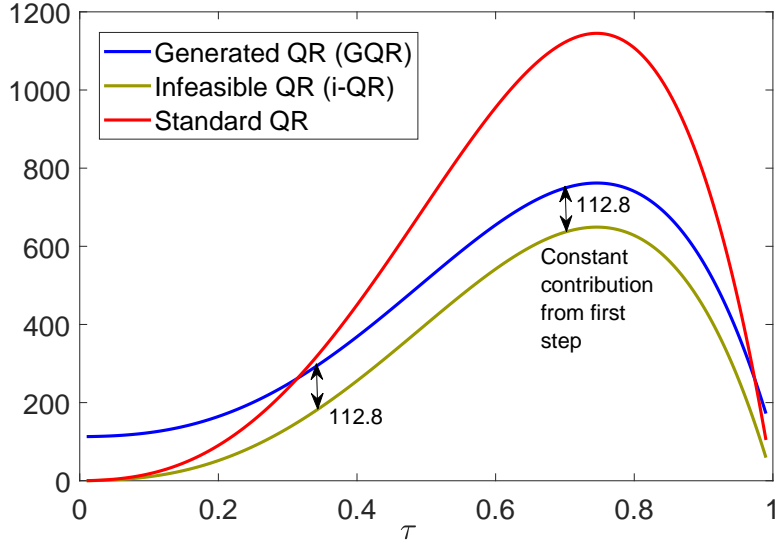}\hfill
\caption{Asymptotic variance: GQR vs i-QR vs QR}
\label{fig:GQR_QR_Var}
\end{figure}

The asymptotic variance of $\widehat{\beta}_0(\cdot)$ for the standard QR is given by the first element of \eqref{eq:GQRvsQR1}, which, for the model parameters and distribution assumed in this exercise, evaluates to

\begin{equation}\label{eq:truebeta0QR}
V(\tau)_{QR,0} = \frac{\tau(1-\tau)}{\left(ac-b^2 \right)^2}\left(c^2 - 2bc \mathbb{E}[X_2] + b^2\mathbb{E}[X_2^2] \right) + \frac{\tau(1-\tau)(bf-dc)^2}{\left(ac-b^2 \right)^2 a^2 \text{Var}(X_1)} 
\end{equation}
where $\text{Var}(X_1)= 4/3$, ($a$, $b$, $c$) are as in \eqref{eq:truebeta0GQR} and
\begin{eqnarray*}
d &=& \mathbb{E}\left[\frac{X_1}{\beta_0^{(1)}(\tau)+\beta_2^{(1)}(\tau)X_2} \right] = \frac{3}{28 \tau}\ln{\left(\frac{e^{\tau} + 40 \tau}{e^{\tau} + 12 \tau} \right)}\\
f &=& \mathbb{E}\left[\frac{X_1 X_2}{\beta_0^{(1)}(\tau)+\beta_2^{(1)}(\tau)X_2} \right] = \frac{3}{7 \times 16 \tau^2}\left(28 \tau - e^{\tau}\ln{\left(\frac{e^{\tau} + 40 \tau}{e^{\tau} + 12 \tau} \right)}\right). 
\end{eqnarray*}
It can be seen from \eqref{eq:truebeta0GQR} and \eqref{eq:truebeta0QR} that the GQR and QR asymptotic variances have a common quantile varying component; GQR has a constant additional component that depends on the first step asymptotic variance, while the additional part for QR is again quantile-dependent. The i-QR variance is given by \eqref{eq:truebeta0GQR} by setting $\mathcal{V}(\beta_1) = 0$. Figure \ref{fig:GQR_QR_Var} plots the asymptotic variance comparison for GQR and QR, as well as i-QR.

As can be seen from Figure \ref{fig:GQR_QR_Var}, the variance of the QR and i-QR estimators, being a function of $\tau(1-\tau)$, is close to $0$ at the very tails. The tail variance of both the QR and i-QR is less than that of GQR because the two step GQR procedure adds an additional constant variance contribution from the first step estimation irrespective of the quantile level. But the opposite is true for all other quantile levels, with GQR asymptotic variance being lesser than QR for most quantiles and especially prominent in the higher quantiles\footnote{The relatively high GQR variance at the boundaries is driven by Assumption \ref{A2} that the density of $X(\theta)$ is bounded away from 0. If it is relaxed, then the two-step estimator can become as good as one-step. When it is not relaxed, the support of $X(\theta)$ depends on $\theta$. In that case, we can have faster than OLS estimation, like \cite{S94}'s nonregular regression which converges at rate $1/n$, so that the two-step estimator is asymptotically unaffected by the first stage. The drawback of the latter is the requirement that error distribution is bounded away from 0 in its compact support. OLS, although slower, does not suffer from such restrictions.}. While this exercise considered $X_1$ and $X_2$ to be independent, the empirical application suggests that for the general case as well, the pattern for GQR versus QR asymptotic variances remains similar. However, we note that there isn't a clear efficiency gain of one method over the other - it depends on the quantile level and the choice of the first stage estimator which impacts the tail behaviour. 

\paragraph{Asymptotic variance for $\widehat{\beta}_2(\cdot)$.} Under the remark noted in Section \ref{sec:biasRMSEcov}, the asymptotic variance of $\widehat{\beta}_2(\cdot)$ is same for GQR, QR and i-QR, given by,	
\[
V(\tau)_{GQR,2} = V(\tau)_{QR,2} = [0,1]H(\tau)^{-1}J(\tau)H(\tau)^{-1}[0,1]^{\prime} = [0,0,1]H(\tau)_{QR}^{-1}J(\tau)_{QR}H(\tau)_{QR}^{-1}[0,0,1]^{\prime} 
\]
where $(H(\tau), J(\tau))$ and $(H(\tau)_{QR}, J(\tau)_{QR})$ are obtained from \eqref{eq:ex1.2} and \eqref{eq:GQRvsQR1}, respectively. For the true model parameters and distribution considered here, this evaluates to
\begin{equation}\label{eq:truebeta2}
V(\tau)_2= \frac{\tau(1-\tau)}{\left(ac-b^2 \right)^2}\left(b^2 - 2ab \mathbb{E}[X_2] + a^2\mathbb{E}[X_2^2] \right)
\end{equation}
where ($a$, $b$, $c$) are as in \eqref{eq:truebeta0GQR}.

Tables \ref{tab:asympSEbeta0}-\ref{tab:asympSEbeta2} compare the bootstrapped asymptotic SE of $\widehat{\beta}_0(\cdot)$, $\widehat{\beta}_2(\cdot)$ obtained from \eqref{eq:boot} (mean of $\widehat{V}(\tau)$ over the $1000$ simulations is reported) with the true values obtained from their analytical expressions as derived above. It can be seen in Table \ref{tab:asympSEbeta0} that the true asymptotic SE of $\widehat{\beta}_0(\cdot)$ is greater for GQR than QR for $\tau=0.2$ and the trend changes for all other quantile levels, while it is always greater than that of i-QR, which are as predicted by theory and discussed above. Bootstrap estimation of asymptotic standard error works well even for small sample size of $100$ (except for $\tau=0.2$ using GQR) and the estimation accuracy improves with samples size. Table \ref{tab:asympSEbeta0} also reports the coefficient of variation (CoV) for $\widehat{V}(\tau)$, which is the ratio of the standard deviation to the mean of $\widehat{V}(\tau)$ over the $1000$ simulations. CoV measures the precision in estimation of the asymptotic SE (or variability among the estimated values in each run of the simulation).  Looking at CoV, it is interesting to note that for GQR the estimates of asymptotic SE have lesser variation across simulations relative to their mean values, and CoV is very similar across quantiles, than that of i-QR or QR. This suggests that the GQR asymptotic SE estimates are less dispersed around the mean than that of i-QR or QR. The CoV falls for all methods with sample size; for $n=1000$, it is well within $10\%$ for GQR and slightly greater than $10\%$ for i-QR and QR. Table \ref{tab:asympSEbeta2} confirms that variance of $\hat{\beta}_2(\cdot)$ is unaffected by the two step procedure, as QR, i-QR and GQR yield identical true values, and similar bootstrapped estimates as well as CoV. Also, in Table \ref{tab:asympSEbeta0} and, to a lesser degree in Table \ref{tab:asympSEbeta2}, we find a slight overestimation of the GQR variance and underestimation of the QR one. 

\begin{table}[hptb]
\caption{Asymptotic standard error for $\hat{\beta_0}(\cdot)$, $B=1000$, $\text{simulations}=1000$}%
\label{tab:asympSEbeta0}
\centering
  \begin{threeparttable}
\begin{tabularx}{0.8\textwidth}{ccccccc}
      \hline
			\hline
			\multicolumn{1} {X} { } & \multicolumn{1} {X} { } & \multicolumn{1} {X} {} & \multicolumn{2} {c} {$n=100$} & \multicolumn{2} {c} {$n=1000$} \\ [1 ex]
			\hline
			$\tau$&{}&True&Estimated&CoV&Estimated&CoV\\
			\hline
			$0.2$		&GQR &$12.8272$		&$15.2959$	&$18.43\%$	&$13.8210$	&$7.63\%$		\\
							&QR &$9.5117$		&$10.5633$	&$33.22\%$	&$9.2487$	&$13.66\%$		\\
							&i-QR  &$7.1905$		&$7.8174$		&$31.51\%$  &$7.0263$	  &$12.52\%$		\\
			$0.4$		&GQR &$19.1989$		&$20.0756$	&$18.03\%$	&$19.6835$	&$7.80\%$		\\
							&QR &$21.2165$		&$20.6893$	&$24.49\%$	&$20.6343$	&$11.01\%$		\\
							&i-QR  &$15.9926$		&$15.8358$	&$24.46\%$  &$15.6959$	&$10.73\%$		\\
			$0.6$		&GQR &$25.5869$		&$25.7885$	&$18.03\%$	&$25.8453$	&$7.87\%$		\\
							&QR &$30.9088$		&$28.9673$	&$22\%$	&$29.7102$	&$10.25\%$		\\
							&i-QR  &$23.2778$		&$22.5501$	&$22.53\%$  &$22.8857$	&$10.37\%$		\\
			$0.8$		&GQR &$27.2149$		&$28.1044$	&$19.08\%$	&$27.4708$	&$8.55\%$		\\
							&QR &$33.2814$		&$32.3425$	&$22.83\%$	&$32.5182$	&$10.29\%$		\\
							&i-QR  &$25.0563$		&$25.0403$	&$22.82\%$ 	&$24.9103$  &$10.3\%$		\\
			\hline
		\end{tabularx}
\begin{tablenotes}
      \scriptsize
      \item The true asymptotic SE for GQR and QR are computed using \eqref{eq:truebeta0GQR} and \eqref{eq:truebeta0QR}, respectively, while for i-QR, the first step variance=0 in the formula for GQR. Mean over $1000$ simulations of the bootstrapped asymptotic SE \eqref{eq:boot} is reported. CoV denotes coefficient of variation and indicates the extent of variability in the estimates for each run of the simulation. 
    \end{tablenotes}
  \end{threeparttable}
\end{table}

\begin{table}[hptb]
\caption{Asymptotic standard error for $\hat{\beta_2}(\cdot)$, $B=1000$, $\text{simulations}=1000$}%
\label{tab:asympSEbeta2}
\centering
  \begin{threeparttable}
\begin{tabularx}{0.8\textwidth}{ccccccc}
      \hline
			\hline
			\multicolumn{1} {X} { } & \multicolumn{1} {X} { } & \multicolumn{1} {X} {} & \multicolumn{2} {c} {$n=100$} & \multicolumn{2} {c} {$n=1000$} \\ [1 ex]
			\hline
			$\tau$&{}&True&Estimated&CoV&Estimated&CoV\\
			\hline
			$0.2$		&GQR &$1.2450$		&$1.6212$	&$25.13\%$	&$1.2617$	  &$11.48\%$		\\
							&QR  &$1.2450$		&$1.3722$	&$33.06\%$	&$1.2084$	&$12.56\%$		\\
							&i-QR  &$1.2450$		&$1.3328$	&$32.32\%$	&$1.2183$	&$12.44\%$		\\
			$0.4$		&GQR &$2.8129$		&$2.8139$	&$23.43\%$	&$2.7845$	  &$10.16\%$		\\
							&QR  &$2.8129$		&$2.7485$	&$24.07\%$	&$2.7447$	&$10.03\%$		\\
							&i-QR  &$2.8129$		&$2.7656$	&$23.82\%$	&$2.7697$	&$10.51\%$		\\
			$0.6$		&GQR &$4.1156$		&$4.0434$	&$21.58\%$	&$4.0755$	  &$9.63\%$		\\
							&QR  &$4.1156$		&$3.9354$	&$21.50\%$	&$4.0237$	&$9.37\%$		\\
							&i-QR  &$4.1156$		&$4.0175$	&$22.66\%$	&$4.0526$	&$10.06$		\\
			$0.8$		&GQR &$4.4389$		&$4.4844$	&$22.36\%$	&$4.3899$	  &$10.07\%$		\\
							&QR  &$4.4389$		&$4.4241$	&$21.47\%$	&$4.3901$	&$9.92\%$		\\
							&i-QR  &$4.4389$		&$4.4241$	&$23.14\%$	&$4.4223$  &$10.65\%$		\\
			\hline
		\end{tabularx}
\begin{tablenotes}
      \scriptsize
      \item The true asymptotic SE for all methods is given by \eqref{eq:truebeta2}. The rest of explanation is as in Table \ref{tab:asympSEbeta0}.
    \end{tablenotes}
  \end{threeparttable}
\end{table}

As mentioned earlier, the GQR asymptotic variance is impacted by the choice of the first stage estimator. While OLS is a natural choice for estimating the constant first stage slope, as we have considered till now, linearity in OLS is a  restriction and choosing from a wider class including non-linear estimators is likely to produce variance improvement. In Table \ref{tab:DiffFirst}, we report bootstraped GQR asymptotic SE for $\hat{\beta}_0(\cdot)$ using two QR-based first stage estimators of $\beta_1$, apart from OLS: the mean of the quantile estimates $\hat{\beta}_1(\tau)$ over $19$ equidistant  quantile levels $\{0.05,0.10,\hdots,0.95\}$, and the weighted average of $\hat{\beta}_1(\tau)$ at $\tau = (1/3, 1/2,$, $2/3)$ with weights $(0.3,0.4,0.3)$, respectively (this corresponds to \cite{G66} estimator, to use the terminology in \cite{ABHHRT72}). For the sake of brevity, the results for $\hat{\beta}_2(\cdot)$ is not reported, since its variance is unaffected by first stage, as noted earlier. Comparison of bootstrapped GQR asymptotic SE using different first stage estimators shows that QR-based estimators yield more efficient results than OLS, which is not surprising as quantile regression can be more efficient than least squares in absence of i.i.d and Gaussian error assumption. QR mean, which assigns equal weights to all quantiles, results in poorer GQR efficiency than Gastwirth's weighted QR which gives more weight to the median and lesser to the tails - this estimator is known to have higher efficiency in a large class of distributions (see, for example, \cite{KB78}). Optimal first stage estimator for the constant slopes in QR models and semi-parametric efficiency of the two-step GQR is an interesting study left for future research.

\begin{table}[hptb]
\caption{GQR bootstrapped asymptotic SE for $\hat{\beta_0}(\cdot)$: Varying first stage estimators}%
\label{tab:DiffFirst}
\centering
  \begin{threeparttable}
\begin{tabularx}{0.47\textwidth}{clcc}
      \hline
      			\hline
			\multicolumn{1} {c} {$\tau$ } & \multicolumn{1} {l} {$\beta_1$ estimator} & \multicolumn{1} {c} {$n=100$} & \multicolumn{1} {c} {$n=1000$} \\ [1 ex]
					\multicolumn{1} {c} { } & \multicolumn{1} {l} {(First stage)}  & \multicolumn{1} {c} {} & \multicolumn{1} {c} {} \\ [1 ex]
			\hline
			\hline
			$0.2$		&OLS 		&$15.2959$		&$13.8210$		\\
						&QR mean 	&$14.2295$		&$12.9440$		\\
						&Weighted QR   &$13.8919$		&$12.5515$	\\
			$0.4$		&OLS 		&$20.0756$		&$19.6835$		\\
						&QR mean 	&$19.5125$		&$19.0885$		\\
						&Weighted QR   &$19.2904$		&$18.8691$	\\
			$0.6$		&OLS 		&$25.7885$		&$25.8453$		\\
						&QR mean 	&$25.5513$		&$25.3143$		\\
						&Weighted QR   &$25.2964$		&$25.3056$	\\
			$0.8$		&OLS 		&$28.1044$		&$27.4708$		\\
						&QR mean 	&$27.3706$		&$27.2047$		\\
						&Weighted QR   &$27.4267$		&$26.9379$	\\
			\hline
		\end{tabularx}
\begin{tablenotes}
      \scriptsize
      \item QR mean averages quantile estimates $\hat{\beta}_1(\tau)$ over $\tau=\{0.05,0.10,\hdots,0.95 \}$. 
      Weighted QR assigns the weights $(0.3,0.4,0.3)$ to $(\hat{\beta}_1(1/3), \hat{\beta}_1(1/2), \hat{\beta}_1(2/3))$, respectively. Bootstrapped SE is the mean of \eqref{eq:boot} over $1000$ simulations.
    \end{tablenotes}
  \end{threeparttable}
\end{table}

\section{Empirical Application} \label{sec:empirical}

The two step estimation procedure of Section \ref{subsec:regandQR1} can be useful in estimating auction models as in the quantile regression approach of \cite{G17}. In first price auctions, a quantile regression specification for the private value generates a quantile regression specification for the bid, see \cite{GG20}. The linear regression approach of Haile, Hong \& Shum (2003)\nocite{HHS03} for estimating first price auction models uses the `homogenized bid' technique, which implies constant slope parameters in the bid quantile regression model.
It is shown here that the two approaches can be combined, as in the example of Section \ref{subsec:regandQR1}. We apply the GQR estimator for the estimation of bid quantile specification containing both quantile-constant and quantile-dependent slope parameters. In the first step, following \cite{HHS03}, the constant slope parameter is estimated by regressing the bids on the observed covariates. This is then used to generate the dependent variable for the quantile regression for estimating the quantile parameters 
The aim of our empirical exercise is to see how imposing a constant slope for a given set of variables can improve the estimation of the other slope functions.

We illustrate our proposed methodology using data from first price timber auctions conducted by the US Forest Services (USFS) covering the western half of US in the year $1979$. This is the same data used by \cite{LP08}. The data consists of $214$ first price auctions with $2$ bidders, and the covariates listed are appraisal value and timber volume (in log). 

\begin{figure}[hpt]
\centering
\includegraphics[width=.33\textwidth]{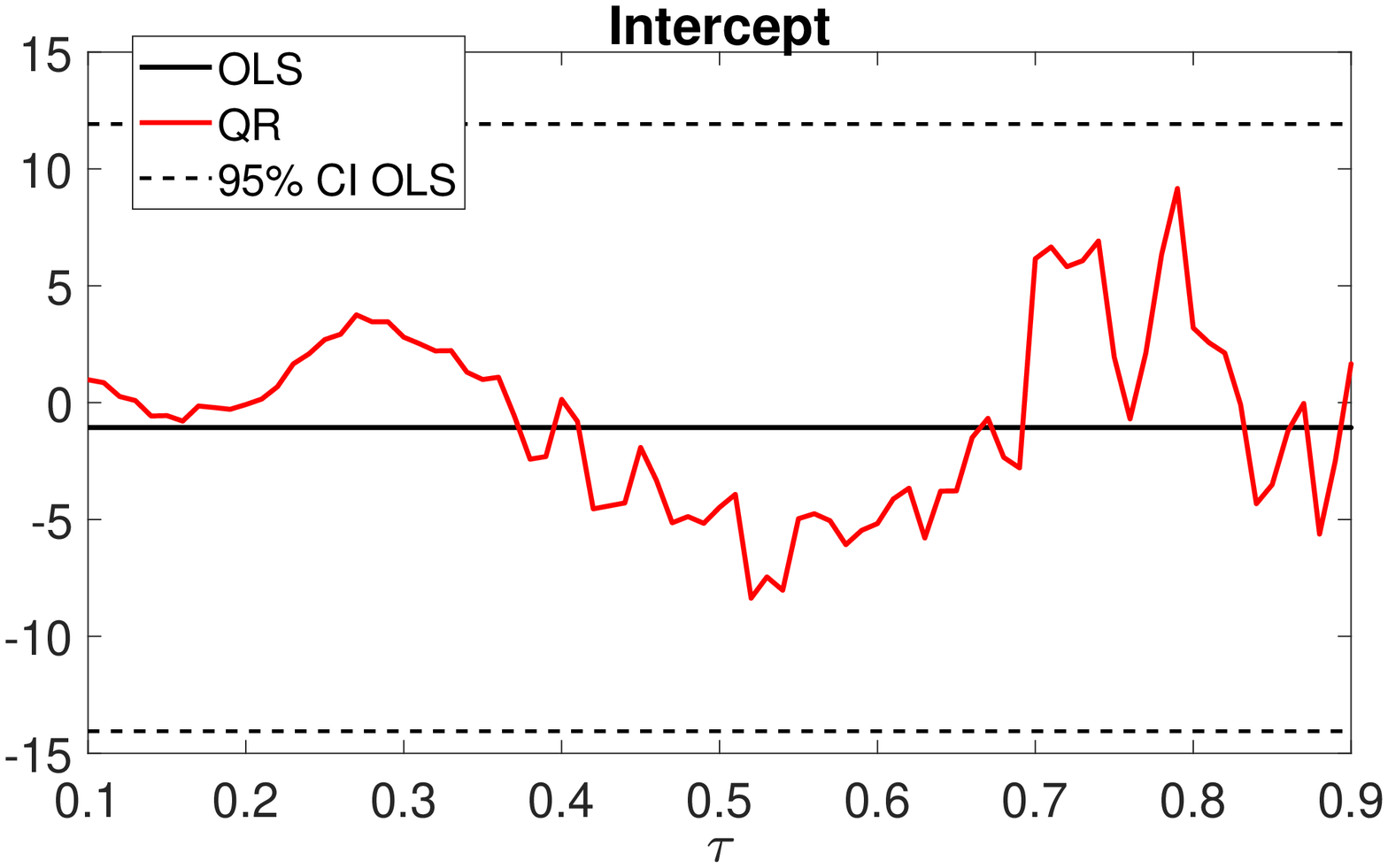}\hfill
\includegraphics[width=.33\textwidth]{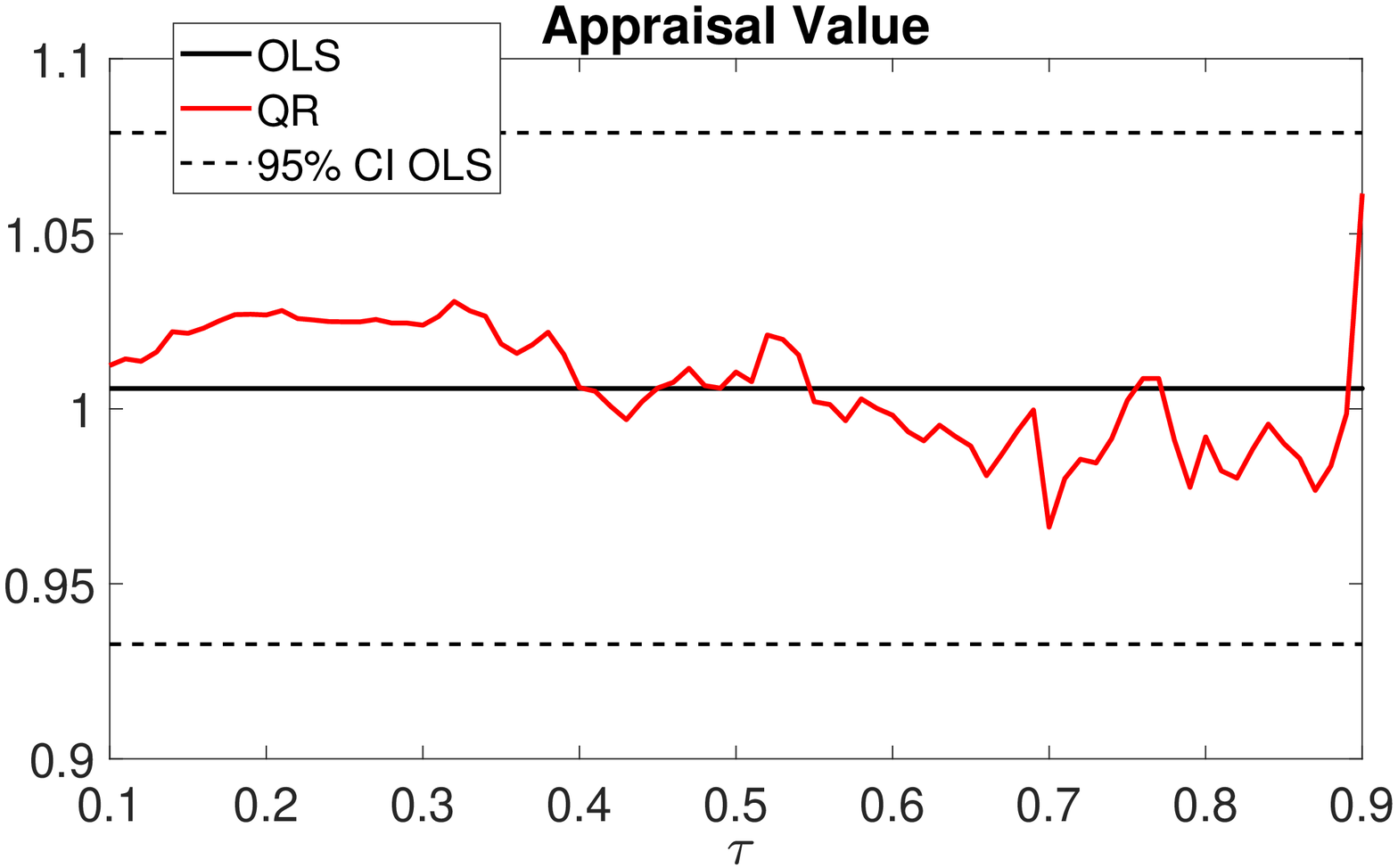}\hfill
\includegraphics[width=.33\textwidth]{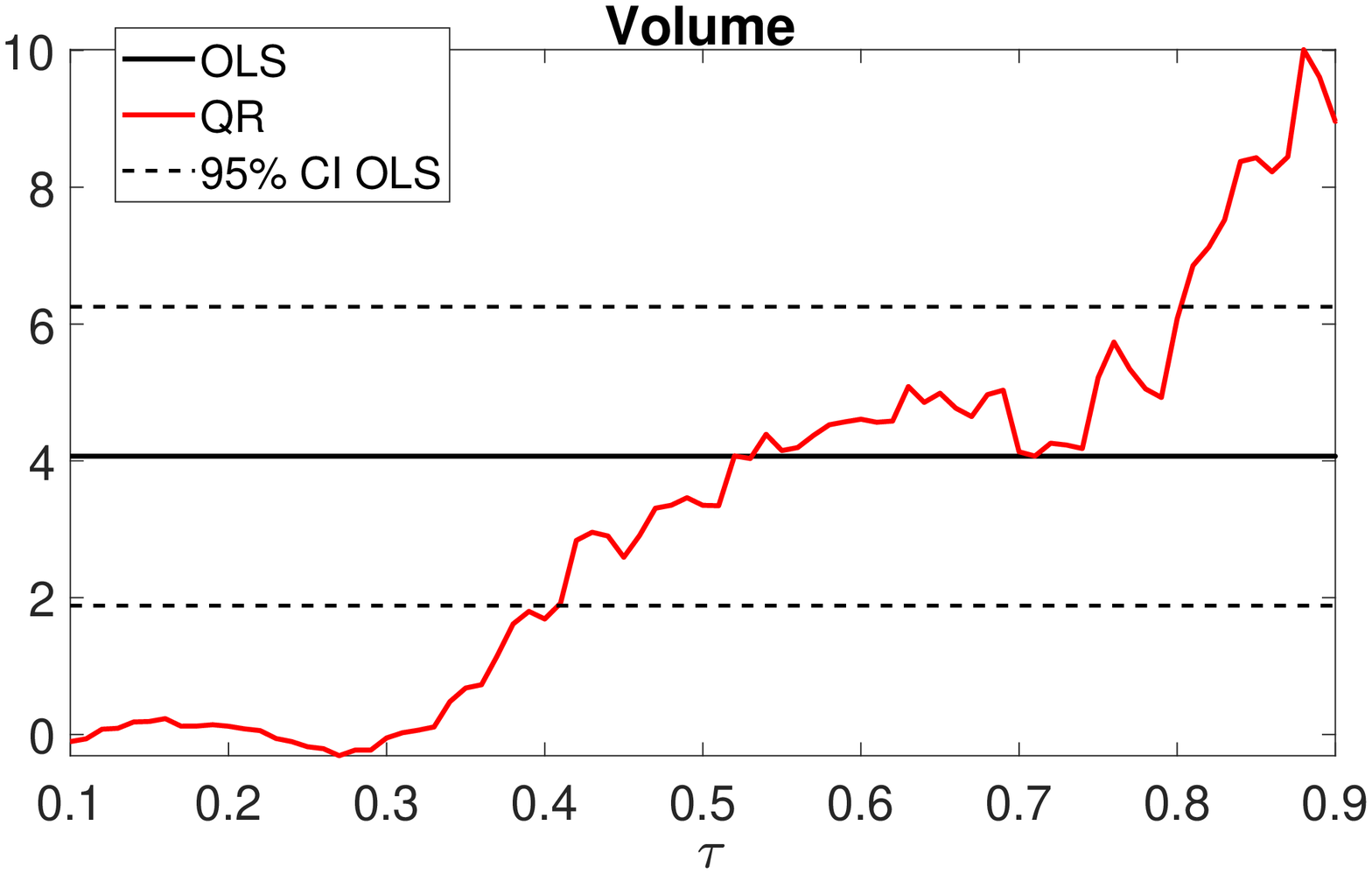}
\caption{Bid quantile parameter estimates}
\label{fig:OLS_QR_CI}
\end{figure}

\paragraph{Bid homogenization.} Figure \ref{fig:OLS_QR_CI} shows the bid quantile parameter estimates obtained from quantile regression of bids on the covariates along with the corresponding OLS estimates and its $95 \%$ confidence interval. 
Intercept and appraisal value quantile slope coefficients seem to satisfy the assumption of constancy across quantiles. However, the volume quantile parameter does not seem to be constant. 

\begin{figure}[hpt]
\centering
    \includegraphics[width=0.7\textwidth]{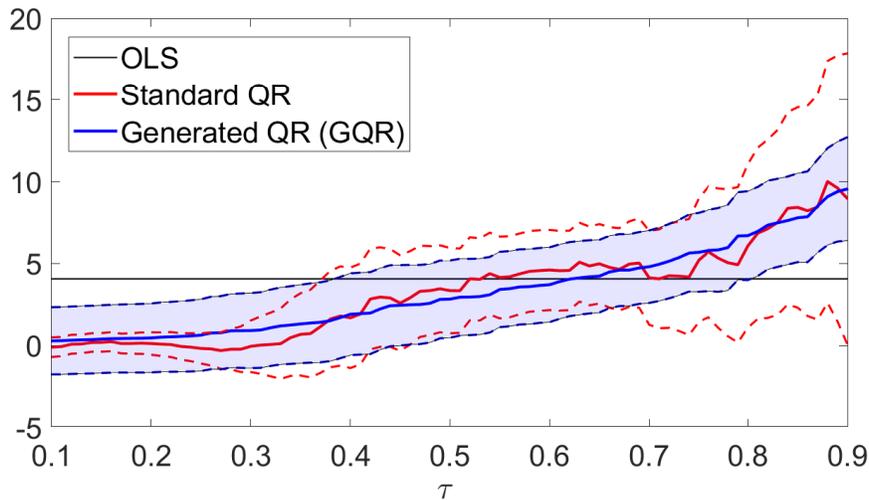}
    \caption{Second step - Bid QR estimates for volume}
    \label{fig:OLS_AQR_GQR_bid}
    	\small\textsuperscript{ 95\%  bootstrapped confidence interval for QR shown by red dotted lines, and for GQR by blue shaded region}
\end{figure}

\paragraph{Bid quantile estimation using GQR.} The GQR estimator involves constrained estimation assuming the intercept and appraisal value slope to be constant across quantiles, while the volume parameter is considered to be varying with quantile levels. Table \ref{tab:OLSApp} reports the result of linear regression of bids on the covariates. The first step estimates constitute the intercept and appraisal value slope regression estimates, while the quantile estimates for slope of volume is obtained through quantile regression of the generated dependent variable $(bids_i-(-1.07)-1.01 \times appraisal\ value_i)$ on $volume_i$. The second step GQR bid quantile estimate for slope of volume is shown in Figure \ref{fig:OLS_AQR_GQR_bid}. For comparison purpose, we also plot the results of unconstrained estimation of quantile parameters of volume. Table \ref{tab:GQRvsQR} also reports the bootstrapped standard error (SE) of the constrained and the unconstrained estimator obtained from $10,000$ bootstrap replications.
\begin{table}[hptb]
\caption{First step - bid regression}%
\label{tab:OLSApp}
\begin{center}
\begin{tabularx}{0.5\textwidth}{cccc}
      \hline
			\hline
			Intercept&Appraisal value&Volume&$R^2$\\
			\hline
			$-1.07$		&$1.01$		&$4.07$	  &$0.77$		\\
			$(6.72)$  &$(0.04)$ &$(1.12)$	& \\
			\hline
		\end{tabularx}
\end{center}
\end{table}

As can be seen in Figure \ref{fig:OLS_AQR_GQR_bid} and Table \ref{tab:GQRvsQR}, the GQR slope estimate is more regular than that of the unconstrained estimation; the GQR estimates increase with quantile level, which is consistent with an increasing bid conditional quantile function. The reason is likely that the first stage estimation removes some of the regressors in the GQR stage along with the associated variation, thereby improving monotonicity and smoothness of the quantile estimates. An accuracy improvement, especially in the higher quantiles, is also evident as the GQR bootstrap confidence interval is much smaller in higher quantiles. 
\begin{table}[hptb]
\caption{Bootstrapped SE for constrained vs unconstrained quantile estimation of volume}%
\label{tab:GQRvsQR}
\begin{center}
\begin{tabularx}{0.7\textwidth}{ccccc}
      \hline
			\hline
			\multicolumn{1} {X} { } & \multicolumn{2} {c} {Constrained (GQR)} & \multicolumn{2} {c} {Unconstrained} \\ [1 ex]
			\hline
			$\tau$&Estimate&SE&Estimate&SE\\
			\hline
			$0.1$		&$0.2837$ 	&$1.0457$		&$-0.1023$	&$0.3050$		\\
			$0.2$		&$0.4631$	  &$1.0735$		&$0.1219$ 	&$0.3492$		\\
			$0.3$		&$0.9131$		&$1.1677$		&$-0.0491$	&$0.8013$		\\
			$0.4$		&$1.9080$		&$1.2727$		&$1.6894$	  &$1.5634$		\\
			$0.5$		&$2.8312$		&$1.2024$		&$3.3503$	  &$1.3285$		\\
			$0.6$		&$3.7115$		&$1.1624$		&$4.6088$	  &$1.2478$		\\
			$0.7$		&$4.8065$		&$1.1281$		&$4.1329$ 	&$1.4609$		\\
			$0.8$		&$6.7040$		&$1.3835$		&$6.0770$		&$2.5452$		\\
			$0.9$		&$9.5601$		&$1.6093$		&$8.9608$		&$4.5484$		\\
			\hline
		\end{tabularx}
\end{center}
\end{table}
The SE pattern observed in Table \ref{tab:GQRvsQR} is as expected from the analysis in Section \ref{sec:GQRvsQR} although the covariates are no longer independent: constrained SE is lesser than that obtained by unconstrained estimation except for the first three quantile levels. Also note that the SE for the unconstrained estimator varies quite a lot across quantiles and is quite high for the higher quantiles, which are particularly important for auction models as winners reside here. An intuitive explanation for the SE pattern observed here is as follows. 
The asymptotic variance of the unconstrained estimator will have the form given by \eqref{eq:GQRvsQR1}: in the tails, while $\tau(1-\tau)$ tends to make the quantile estimate more precise, the derivative of an increasing quantile slope parameter has an opposite effect. In higher quantiles, as is typical with quantile regression, the latter effect dominates, reducing the precision of the quantile estimates in that region. The asymptotic variance of the GQR estimator will have the form of \eqref{eq:ex1.2}: there will be a constant effect of the first step variance at each quantile level, but in addition to the corresponding $H^{-1}JH^{-1}$ term which increases with $\tau$ for increasing slope parameter, there is a negative quantile effect due to the covariance term being negative in $\tau$ for an increasing slope parameter. So, the net quantile-dependent effect is reduced and the SE is more uniform across quantiles. Hence, at lower quantile levels, the SE of the GQR estimator is greater than that of the unconstrained one because of the constant contribution of first step variance. But at higher quantiles, SE of the unconstrained estimator is much greater. 

In general, the unconstrained quantile regression involves fitting the model at each quantile level, for estimating both the constant and the quantile-dependent model parameters, and thus loses out on the information that some covariate effects are common across quantiles. It is well noted in literature that for estimating quantile models that have some common covariate effects, efficiency gain can be achieved by aggregating the information across multiple quantiles, as in the combined quantile regression approach of \cite{ZY08}. The GQR estimator utilizes the commonality information and improves upon efficiency - the overall efficiency gain and tail behaviour will, as noted earlier, depend on the choice of first step estimator. 

\section{Conclusion} \label{sec:conclusion}

This paper studies two step estimation of quantile regression models with generated covariates and/or dependent variable.  The asymptotic normality of this generated quantile regression (GQR) estimator is derived using the Bahadur expansion approach. The results are verified using simulation and an application based on auctions is carried out. We mention some relevant areas of application. In particluar, analysis of QR models with some constant slopes suggests potential benefits of the two stage procedure in terms of improvements in monotonicity, smoothness and estimation accuracy. A key technical contribution of the paper is to provide Bahadur expansion which holds uniformly with respect to first step parameter and quantile levels, which can be utilised for developing specification tests (like those developed in \cite{KM99} and \cite{KX02}) as well as to obtain functional central limit theorem for the two step quantile regression estimator. 

A slightly different problem that can be studied using techniques developed here relates to quantile specifications where a first step estimation impacts the quantile level for the second stage quantile regression. Such specifications arise in \cite{AB17}'s method of quantile regression with ``rotated" check function to correct for sample selection in quantile regression models. A more challenging problem open for future research is to consider the case where the first stage converges at a slower rate, like in quantile regression models for panel data where the first step within estimator is usually $\sqrt{n}$-consistent and the quantile estimator is $\sqrt{nT}$-consistent.

\clearpage
\section*{Appendix 1. Proof section}\label{sec:Ap1}
\addcontentsline{toc}{section}{Appendix 1}
\renewcommand\theequation{A1.\arabic{equation}}
\paragraph{Notations.} The notation $\asymp$ is defined as: sequences $\{x_{n}\}$ and $\{y_{n}\}$ satisfy $x_{n} \asymp y_{n}$ if $\vert x_{n} \vert/C \leq \vert y_{n} \vert \leq C \vert x_{n} \vert$, for some $C>0$ and $n$ large enough. $\left\vert\left\vert \cdot \right\vert\right\vert$ is the Euclidean norm.
The largest eigenvalue in absolute value for a symmetric matrix $A$ is $\left\vert\left\vert A \right\vert\right\vert = \sup_{u \in \mathcal{B}(0,1)} \left\vert\left\vert Au \right\vert\right\vert = \sup_{u \in \mathcal{B}(0,1)}\left\vert u^{\prime}Au  \right\vert $. 
Also, for any matrix or vector $B$, $\left\vert\left\vert AB \right\vert\right\vert \leq \left\vert\left\vert A \right\vert\right\vert\ \left\vert\left\vert B \right\vert\right\vert$. We denote $\vert\vert f(\cdot \vert \cdot) \vert\vert_{\infty} \text{=} \sup_{y,x} \vert f(y \vert X) \vert$. And the notation $\succ$ denotes that, for two symmetric matrices $A_1, A_2$, $A_1 \succ A_2$ if and only if $A_1 - A_2$ is a positive definite symmetric matrix.

Define \vspace{-4mm}
\begin{align*}
Q\left( \beta ;\tau ,\theta \right) =\mathbb{E}\left[ \rho _{\tau }\left(
Y\left( \theta \right) -X^{\prime }\left( \theta \right) \beta \right) %
\right] -\mathbb{E}\left[ \rho _{\tau }\left( Y\left( \theta \right) \right) %
\right] .
\end{align*}
As $\rho _{\tau }\left( \cdot \right) $ is almost everywhere differentiable
with bounded derivatives, $\beta \mapsto Q\left( \beta ;\tau ,\theta \right) 
$ is differentiable with first derivative
$
Q^{\left( 1\right) }\left( \beta ;\tau ,\theta \right) = \mathbb{E}\left[
\left\{ \mathbb{I}\left( Y\left( \theta \right) \leq X^{\prime }\left(
\theta \right) \beta \right) -\tau \right\} X\left( \theta \right) \right] = \mathbb{E}\left[ \left\{ F\left( X^{\prime }\left( \theta \right) \beta
|X,\theta \right) -\tau \right\} X\left( \theta \right) \right] .
$
Hence $Q\left( \cdot ;\tau ,\theta \right) $ is twice continuously
differentiable with respect to $\beta $, with a second derivative \vspace{-2mm}
\begin{align*}
Q^{\left( 2\right) }\left( \beta ;\tau ,\theta \right)  =\mathbb{E}\left[
f\left( X^{\prime }\left( \theta \right) \beta |X,\theta \right) X\left(
\theta \right) X^{\prime }\left( \theta \right) \right]  = \int f\left( x^{\prime }\beta |x,\theta \right) xx^{\prime }f_X\left(
x|\theta \right) dx. 
\end{align*}
Let \emph{B}$\left( \theta \right) $ be the set of $\beta ^{\prime }$ such
that $0<F\left( x^{\prime }\left( \theta \right) \beta |x,\theta \right) <1$
for some inner $x$ of $\mathcal{X}\left( \theta \right) $, $
\emph{B}\left( \theta \right) =\left\{ \beta ;\text{there is an inner }x%
\text{ of }\mathcal{X}\left( \theta \right) \text{ such that }\underline{y}%
\left( \theta |x\right) <x^{\prime }\beta <\overline{y}\left( \theta
|x\right) \right\} 
$
where $\underline{y}\left( \theta |x\right) =F^{-1}\left( 0|x\right) $ and $%
\overline{y}\left( \theta |x\right) =F^{-1}\left( 1|x\right) $. The next
Lemma describes some key properties of $Q^{\left( 2\right) }\left( \beta
;\tau ,\theta \right) $ and $Q\left( \beta ;\tau ,\theta \right) $.
\begin{lemma}\label{AN:L2}
Under Assumption \ref{A2} it holds
\begin{enumerate}[(i)]
\item $Q^{\left( 2\right) }\left( \beta ;\tau ,\theta \right) $ is
continuous with respect to its three arguments, with \vspace{-2mm}
\begin{align*}
\left\Vert Q^{\left( 2\right) }\left( \beta _{1};\tau ,\theta \right)
-Q^{\left( 2\right) }\left( \beta _{0};\tau ,\theta \right) \right\Vert \leq
C\left\Vert \beta _{1}-\beta _{0}\right\Vert 
\end{align*}%
for all $\beta _{0}$ and $\beta _{1}$, $\theta \in \Theta $ and $\tau  \in \left[ 0,1\right] $.
\item $Q^{\left( 2\right) }\left( \beta ;\tau ,\theta \right) $ is strictly
positive for all $\beta \in $  \emph{B}$\left( \theta \right) $, $\theta \in %
\Theta $ and $\tau \in \left[ 0,1\right] $.
\item For $\theta \in \Theta $ and $\tau \in \left( 0,1\right) $, $%
Q\left( \beta ;\tau ,\theta \right) $ has a unique minimizer $\beta \left(
\tau ;\theta \right) $ which is continuously differentiable in $\theta $ and 
$\tau $ with \vspace{-2mm}
\begin{align*}
\frac{\partial \beta \left( \tau ;\theta \right) }{\partial \theta ^{\prime }%
} &=H\left( \tau ;\theta \right) ^{-1}D\left( \tau ;\theta \right) , \\
\frac{\partial \beta \left( \tau ;\theta \right) }{\partial \tau }
&=H\left( \tau ;\theta \right) ^{-1}\mathbb{E}\left[ X\left( \theta \right) %
\right] ,
\end{align*}%
where $H\left( \tau ;\theta \right) $ and $D\left( \tau ;\theta \right) $
are as in Proposition \ref{P1}. \vspace{-2mm}
\end{enumerate}
\end{lemma}

\subparagraph{Proof of Lemma \ref{AN:L2}.}

(i) directly follows from Assumption \ref{A2} and the Lebesgue Dominated
Convergence Theorem. For (ii), Assumption \ref{A2} gives that, for each $\beta $ in 
\emph{B}$\left( \theta \right) $, there is an open subset $\mathcal{O=O}%
_{\beta ,\theta }$ of $\mathcal{X}\left( \theta \right) $ such that
$
Q^{\left( 2\right) }\left( \beta ;\tau ,\theta \right) \succeq \bigintsss_{%
\mathcal{O}}xx^{\prime }dx.
$
Hence, $H(\tau; \theta)=Q^{\left( 2\right) }\left( \beta(\tau;\theta) ;\tau ,\theta \right)$ has an inverse.
For (iii), observe that $Q\left( \beta ;\tau ,\theta \right) $ is bounded
away from $-\infty $, so that it has local minimizers which must satisfy the
first order condition%
\begin{equation}
0=Q^{\left( 1\right) }\left( \beta ;\tau ,\theta \right) =\mathbb{E}\left[
\left\{ F\left( X^{\prime }\left( \theta \right) \beta |X,\theta \right)
-\tau \right\} X\left( \theta \right) \right] .  \label{FOC}
\end{equation}%
Hence these minimizers must lie in \emph{B}$\left( \theta \right) $ as
outside this set it holds $F\left( X^{\prime }\left( \theta \right) \beta
|X,\theta \right) =1$ a.s, or $F\left( X^{\prime }\left( \theta \right)
\beta |X,\theta \right) =0$ a.s. Now, if there are two such local minimizers 
$\beta _{0}\left( \tau ;\theta \right) $ and $\beta _{1}\left( \tau ;\theta
\right) $, convexity implies that all $\beta _{\pi }\left( \tau ;\theta
\right) =\left( 1-\pi \right) \beta _{0}\left( \tau ;\theta \right) +\pi
\beta _{1}\left( \tau ;\theta \right) $, $0\leq \pi \leq 1$, must be global
minimizers, contradicting that $Q^{\left( 2\right) }\left( \beta _{\pi
}\left( \tau ;\theta \right) ;\tau ,\theta \right) $ is strictly positive as 
$Q^{\left( 1\right) }\left( \beta _{\pi }\left( \tau ;\theta \right) ;\tau
,\theta \right) =0$ for all $\pi $ in $\left[ 0,1\right] $. The rest of
(iii) follows from (i), (ii) and the Implicit Function Theorem.$\hfill
\square $

\subparagraph{Proof of  Proposition \ref{P1}.}

Follows from Lemma \ref{AN:L2}-(iii).$\hfill \square $

\subparagraph{Proof of  Proposition \ref{P2}-(i).}

This proof conducts a uniform order study of the Bahadur error term \eqref{eq:def3}. Define the following \vspace{-2mm}
\begin{equation}\nonumber
\mathcal{L}_{n}(\gamma,\tau;\theta) = \sum_{i=1}^{n}\left\{\rho_{\tau}\left(Y_{i}(\theta)-X_{i}(\theta)^{\prime}\left(\frac{\gamma}{\sqrt{n}}+\beta(\tau;\theta)\right)\right) - \rho_{\tau}\left(Y_{i}(\theta)-X_{i}(\theta)^{\prime}\beta(\tau;\theta)\right)\right\},
\end{equation}
such that $\sqrt{n}\left(\widehat{\beta}(\tau;\theta)-\beta(\tau;\theta)\right) = \arg\min_{\gamma}\mathcal{L}_{n}(\gamma,\tau;\theta).
$
In what follows, we write \vspace{-2mm}
\begin{align}
\widehat{\alpha}(\tau;\theta) &\equiv -H^{-1}\left( \tau ;\theta \right) \widehat{S}\left( \tau ;\theta
\right) \label{AN:del}\\ 
\widehat{S}\left( \tau ;\theta
\right) &= \frac{1}{\sqrt{n}}\sum_{i=1}^{n}s_i(\tau;\theta). \vspace{-2mm} \label{AN:S}
\end{align}%
It follows from \eqref{eq:def3} that \vspace{-4mm}
\begin{align}\label{AN:eq7}
\widehat{\mathcal{E}}(\tau;\theta) &= \arg\min_{\epsilon}\mathbb{L}_{n}(\widehat{\alpha}(\tau;\theta), \epsilon, \tau;\theta),\ \text{where}\nonumber \\
\mathbb{L}_{n}(\gamma,\epsilon,\tau;\theta) &= \mathcal{L}_{n}(\gamma+\epsilon,\tau;\theta) -  \mathcal{L}_{n}(\gamma,\tau;\theta). \vspace{-2mm}
\end{align}
Consider the following decomposition of $\mathbb{L}_{n}(\gamma,\epsilon,\tau;\theta)$. \vspace{-2mm}
\begin{align}\label{AN:eq8}
\mathbb{L}_{n}(\gamma,\epsilon,\tau;\theta) &= \mathbb{L}^{0}_{n}(\gamma,\epsilon,\tau;\theta) + \mathbb{R}_{n}(\gamma,\epsilon,\tau;\theta), \ \text{where}\nonumber \\
\mathbb{L}^{0}_{n}(\gamma,\epsilon,\tau;\theta) &= \widehat{S}(\tau;\theta)^{\prime} (\gamma + \epsilon) + \frac{1}{2}(\gamma + \epsilon)^{\prime}H(\tau;\theta)(\gamma + \epsilon) -  \widehat{S}(\tau;\theta)^{\prime} \gamma - \frac{1}{2}\gamma^{\prime}H(\tau;\theta)\gamma \nonumber\\
&= \widehat{S}(\tau;\theta)^{\prime}\epsilon + \frac{1}{2}\epsilon^{\prime}H(\tau;\theta)(\epsilon + 2\gamma). \vspace{-2mm}
\end{align}
$\mathbb{L}^{0}_{n}(\gamma,\epsilon,\tau;\theta)$ is the quadratic approximation of $\mathbb{L}_{n}(\gamma,\epsilon,\tau;\theta)$ and $\mathbb{R}_{n}(\gamma,\epsilon,\tau;\theta)$ is the remainder term. As mentioned under `Heuristics' in Section \ref{sec:asymptotics}, a uniform order for $\widehat{\mathcal{E}}(\tau;\theta)$ relies on a uniform order study for the remainder term $\mathbb{R}_{n}(\gamma,\epsilon,\tau;\theta)$, using concepts of maximal inequality under bracketing conditions given in \cite{M07}, and on linearization techniques to study $\widehat{\mathcal{E}}(\tau;\theta)$ given in \cite{HP11}. 
\paragraph{Uniform order for $\mathbb{R}_{n}(\gamma,\epsilon,\tau;\theta)$.} The remainder term is $\mathbb{R}_{n}(\gamma,\epsilon,\tau;\theta) = \mathbb{L}_{n}(\gamma,\epsilon,\tau;\theta) - \mathbb{L}^{0}_{n}(\gamma,\epsilon,\tau;\theta) = \sum_{i=1}^{n}\mathbb{R}_{i}(\gamma,\epsilon,\tau;\theta)$, where
\begin{equation}\label{AN:eq9}
\begin{aligned}
\mathbb{R}_{i}(\gamma,\epsilon,\tau;\theta) &= \left\{\rho_{\tau
}\left(Y_{i}(\theta)-X_{i}(\theta)^{\prime}\left(\frac{\gamma \text{+} \epsilon}{\sqrt{n}}+\beta(\tau;\theta)\right)\right) - \rho_{\tau
}\left(Y_{i}(\theta)-X_{i}(\theta)^{\prime}\left(\frac{\gamma}{\sqrt{n}}\text{+}\beta(\tau;\theta)\right)\right)\right\} \\
&- \frac{{s}_{i}(\tau;\theta)}{\sqrt{n}}^{\prime}\epsilon - \frac{1}{2}\epsilon^{\prime}\frac{H(\tau;\theta)}{n}(\epsilon + 2\gamma).
\end{aligned}
\end{equation}
Define also \vspace{-2mm}
\begin{equation}\label{AN:eq10}
R_{i}(\gamma,\epsilon,\tau;\theta)= \mathbb{R}_{i}(\gamma,\epsilon,\tau;\theta) + \frac{1}{2}\epsilon^{\prime}\frac{H(\tau;\theta)}{n}(\epsilon + 2\gamma),
\end{equation}
\begin{equation}\label{AN:eq11}
\mathbb{R}^{1}_{i}(\gamma,\epsilon,\tau;\theta)= R_{i}(\gamma,\epsilon,\tau;\theta)-\mathbb{E}[R_{i}(\gamma,\epsilon,\tau;\theta)\vert X_{i}(\theta)],
\end{equation}
\begin{equation}\label{AN:eq12}
\mathbb{R}^{2}_{i}(\gamma,\epsilon,\tau;\theta)= \mathbb{E}[R_{i}(\gamma,\epsilon,\tau;\theta)\vert X_{i}(\theta)] - \frac{1}{2}\epsilon^{\prime}\frac{H(\tau;\theta)}{n}(\epsilon + 2\gamma),
\end{equation}
such that \vspace{-1mm}
\begin {equation}\label{AN:eq13}
\begin{aligned}
&\mathbb{R}_{n}(\gamma,\epsilon,\tau;\theta) = \mathbb{R}^{1}_{n}(\gamma,\epsilon,\tau;\theta) + \mathbb{R}^{2}_{n}(\gamma,\epsilon,\tau;\theta),\ \text{with}, \\
&\mathbb{R}^{j}_{n}(\gamma,\epsilon,\tau;\theta) = \sum_{i=1}^{n}\mathbb{R}^{j}_{i}(\gamma,\epsilon,\tau;\theta),\ j=1,2.
\end{aligned}
\end{equation}
The following Lemmas provide uniform bounds for the suprema of the constituents of the remainder term $\mathbb{R}_{n}$ and for $\hat{S}$ (see Appendix \hyperref[sec:Ap2]{2} for their proofs).
\begin{lemma}\label{AN:L4}
Under Assumption \ref{A2}, for real numbers $t_{\gamma}$,$t_{\epsilon} > 0$ with $t_{\gamma} \asymp \log^{1/2}{n}$, $t_{\gamma} \geq 1$, $t_{\epsilon} = \left(t \log^{3/4}{n}\right)/n^{1/4} $ for some $t>0$, such that $(t_{\gamma}+t_{\epsilon})^{1/2}/t_{\epsilon} \leq O\left(n^{1/4}/\log^{1/2}n\right)$, for large $n$, \vspace{-1mm}
\begin{equation}\nonumber
\mathbb{E}\left[\sup_{(\gamma,\epsilon,\tau;\theta) \in \mathcal{B}(0,t_{\gamma})\times \mathcal{B}(0,t_{\epsilon})\times [\underline{\tau },\overline{\tau}] \times \Theta} \vert \mathbb{R}^{1}_{n}(\gamma,\epsilon,\tau;\theta) \vert\right] \leq C \frac{\log^{1/2}{n}}{n^{1/4}}t_{\epsilon}(t_{\gamma}+t_{\epsilon})^{1/2}.
\end{equation}
\end{lemma}
\begin{lemma}\label{AN:L5}
Under Assumption \ref{A2}, for real numbers $t_{\gamma}$,$t_{\epsilon} > 0$ defined as in Lemma \ref{AN:L4}, such that $t_{\gamma}/t_{\epsilon} = O\left(n/\log^{1/2}n \right)$, for large $n$, \vspace{-1.5mm}
\begin{equation}\nonumber
\mathbb{E}\left[\sup_{(\gamma,\epsilon,\tau;\theta) \in \mathcal{B}(0,t_{\gamma})\times \mathcal{B}(0,t_{\epsilon}) \times[\underline{\tau },\overline{\tau}] \times \Theta} \vert \mathbb{R}^{2}_{n}(\gamma,\epsilon,\tau;\theta) \vert\right] \leq C \frac{t_{\epsilon}(t_{\gamma}+t_{\epsilon})^{2}}{n^{1/2}}.
\end{equation}
\end{lemma}
\begin{lemma}\label{AN:L6}
Under Assumption \ref{A2}, \vspace{-2mm}
\[
\sup_{(\tau ,\theta) \in [\underline{\tau },\overline{\tau}] \times \Theta} \left\vert\left\vert \widehat{S}(\tau;\theta) \right\vert\right\vert = O_{\mathbb{P}}(\log^{1/2}{n}).
\]
\end{lemma}
In what follows, 
$t_n = t \frac{\log^{3/4}{n}}{n^{1/4}},\ t>0\ ,
$ such that $t_n = o\left(\log^{1/2}{n}\right)$. $t_n $ plays the role of $t_\epsilon$ in the Lemmas, while $t_\gamma$ is chosen such that $t_\gamma \asymp \log^{1/2}{n}$. Hence,
\begin{align*}
\frac{\left(t_\gamma + t_\epsilon \right)^{1/2}}{t_\epsilon} &\asymp \frac{n^{1/4} \log^{1/4}{n}}{t \log^{3/4}{n}} = \frac{1}{t}O\left(\frac{n^{1/4}}{\log^{1/2}{n}} \right) \\
\frac{t_\gamma}{t_\epsilon} &\leq C\frac{n^{1/4}\log^{1/2}{n}}{t \log^{3/4}{n}} \leq C\frac{n^{1/2}}{\log^{1/2}{n}} \leq C\frac{n^{1/2}n^{1/2}}{\log^{1/2}{n}} = O\left(\frac{n}{\log^{1/2}{n}}\right),
\end{align*}
for large $n$. These choices for $t_\gamma$ and $t_\epsilon$ satisfy the requirements for the Lemmas. Lemma \ref{AN:L2}-(ii), which proves existence of $H^{-1}$ for all $\tau \in [\underline{\tau },\overline{\tau}]$ and $\theta \in \Theta$, implies that  $\widehat{\alpha}(\tau;\theta)$ is well defined with a probability tending to $1$. Lemma \ref{AN:L6} implies \vspace{-2mm}
\begin{equation}\label{AN:eq14}
\sup_{(\tau, \theta) \in [\underline{\tau },\overline{\tau}] \times \Theta} \left\vert\left\vert \widehat{\alpha}(\tau;\theta) \right\vert\right\vert = O_{\mathbb{P}}\left(\log^{1/2}{n}\right).
\end{equation}
Consider $\xi>0$ arbitrarily small. Then there exists a $C_{\xi}$ such that, for large $n$ and some $\underline{\varphi} > 0$, \vspace{-2mm}
\begin{align*}
&\mathbb{P}\left(\sup_{\left(\epsilon,\tau,\theta \right) \in \mathcal{B}(0,t_n) \times [\underline{\tau },\overline{\tau}] \times \Theta} \left\vert \mathbb{R}_{n}(\widehat{\alpha}(\tau;\theta),\epsilon,\tau;\theta) \right\vert \geq \frac{\underline{\varphi}t^2_n}{4} \right) \\
&\leq \mathbb{P}\left(\sup_{\left(\epsilon,\tau,\theta \right) \in \mathcal{B}(0,t_n) \times [\underline{\tau },\overline{\tau}] \times \Theta} \left\vert \mathbb{R}_{n}(\widehat{\alpha}(\tau;\theta),\epsilon,\tau;\theta) \right\vert \geq \frac{\underline{\varphi}t^2_n}{4},\  \sup_{\tau,\theta \in [\underline{\tau },\overline{\tau}] \times \Theta} \left\vert\left\vert \widehat{\alpha}(\tau;\theta) \right\vert\right\vert \leq C_{\xi}\log^{1/2}{n} \right) \\
&\ \ \ + \mathbb{P}\left(\sup_{\tau,\theta \in [\underline{\tau },\overline{\tau}] \times \Theta} \left\vert\left\vert \widehat{\alpha}(\tau;\theta) \right\vert\right\vert > C_{\xi}\log^{1/2}{n} \right) \leq \mathbb{P}\left(\sup_{\substack{\left(\gamma, \epsilon,\tau,\theta \right) \in \mathcal{B}(0,C_{\xi}\log^{1/2}{n}) \\ \times \mathcal{B}(0,t_n) \times [\underline{\tau },\overline{\tau}] \times \Theta}}  \left\vert \mathbb{R}_{n}(\gamma,\epsilon,\tau;\theta) \right\vert \geq \frac{\underline{\varphi}t^2_n}{4} \right) + \xi.
\end{align*}
Since $\mathbb{R}_{n} = \mathbb{R}^{1}_{n} + \mathbb{R}^{2}_{n}$, Lemmas \ref{AN:L4}-\ref{AN:L5}, and Markov inequality give \vspace{-2mm}
\begin{align*}
&\mathbb{P}\left(\sup_{\left(\gamma, \epsilon, \tau, \theta \right) \in \mathcal{B}(0,C_{\xi}\log^{1/2}{n}) \times \mathcal{B}(0,t_n) \times [\underline{\tau },\overline{\tau}] \times \Theta} \left\vert \mathbb{R}_{n}(\gamma,\epsilon,\tau;\theta) \right\vert \geq \frac{\underline{\varphi}t^2_n}{4} \right)\\
&\leq \frac{C}{t^2_n}\left(\mathbb{E}\left[
\sup_{\substack{(\gamma,\epsilon,\tau,\theta) \in \mathcal{B}(0,C_{\xi}\log^{1/2}{n})\\ \times \mathcal{B}(0,t_{\epsilon}) \times [\underline{\tau },\overline{\tau}] \times \Theta}}
 \left\vert \mathbb{R}^{1}_{n}(\gamma,\epsilon,\tau;\theta) \right\vert
 \right] + 
\mathbb{E}\left[
\sup_{\substack{(\gamma,\epsilon,\tau,\theta) \in \mathcal{B}(0,C_{\xi}\log^{1/2}{n})\\ \times \mathcal{B}(0,t_{\epsilon}) \times [\underline{\tau },\overline{\tau}] \times \Theta}}
 \left\vert \mathbb{R}^{2}_{n}(\gamma,\epsilon,\tau;\theta) \right\vert
 \right] \right) \notag \\ 
&\leq \frac{C}{t_n}\frac{\log^{3/4}{n}}{n^{1/4}}\left(\left( C_{\xi} + \frac{t_n}{\log^{1/2}{n}} \right)^{1/2} + \left(\frac{\log{n}}{n} \right)^{1/4} \left(C_{\xi} + \frac{t_n}{\log^{1/2}{n}} \right)^2 \right).  
\end{align*}
Using $t_n= \left(t \log^{3/4}{n} \right)/n^{1/4}$ and since $(\log{n})/n = o(1)$, we get \vspace{-2mm}
\begin{align}\label{AN:eq15}
\lim_{n\rightarrow \infty} \mathbb{P}\left(\sup_{\left(\epsilon, \tau, \theta \right) \in \mathcal{B}(0,t_n) \times [\underline{\tau },\overline{\tau}] \times \Theta} \left\vert \mathbb{R}_{n}(\widehat{\alpha}(\tau;\theta),\epsilon,\tau;\theta) \right\vert \geq \frac{\underline{\varphi}t^2_n}{4} \right) = \xi + O\left(\frac{{C_{\xi}}^{1/2}}{t} \right).
\end{align}
\paragraph{Uniform order for $\widehat{\mathcal{E}}(\tau;\theta)$.} Consider ${\mathcal{T}}_n \geq t_n$ and $\epsilon = {\mathcal{T}}_n e$, $\left\vert\left\vert e \right\vert\right\vert = 1$ so that $\left\vert\left\vert \epsilon \right\vert\right\vert \geq t_n$. Since $\rho_{\tau}(\cdot)$ is convex, ${\mathbb{L}}_n\left(\beta(\tau; \theta),\epsilon,\tau;\theta \right)$ is convex.  Recall that from \eqref{AN:eq7} and \eqref{AN:eq8}, ${\mathbb{L}}_n\left(\beta(\tau; \theta),0,\tau; \theta \right) = 0$ and $\mathbb{L}_{n} = \mathbb{L}^{0}_{n} + \mathbb{R}_{n}$. Then, using convexity property, \vspace{-2mm}
\begin{align*} 
\frac{t_n}{\mathcal{T}_n}{\mathbb{L}}_n\left( \widehat{\alpha}(\tau; \theta),\epsilon,\tau;\theta \right) &= \frac{t_n}{\mathcal{T}_n}{\mathbb{L}}_n\left( \widehat{\alpha}(\tau; \theta),\epsilon,\tau;\theta \right) + \left(1- \frac{t_n}{\mathcal{T}_n} \right){\mathbb{L}}_n\left(\widehat{\alpha}(\tau; \theta),0,\tau; \theta \right) \geq {\mathbb{L}}_n\left(\widehat{\alpha}(\tau; \theta), \frac{t_n \epsilon}{\mathcal{T}_n}  ,\tau; \theta \right) \\
&= \mathbb{L}_n\left(\widehat{\alpha}(\tau; \theta), t_n e  ,\tau; \theta \right) = \mathbb{L}^{0}_{n}\left(\widehat{\alpha}(\tau; \theta), t_n e  ,\tau; \theta \right) + \mathbb{R}_{n}\left(\widehat{\alpha}(\tau; \theta), t_n e ,\tau; \theta \right).
\end{align*}
Since $\widehat{\mathcal{E}} (\tau; \theta) = \arg\min_{\epsilon} {\mathbb{L}}_n\left( \widehat{\alpha}(\tau; \theta),\epsilon,\tau;\theta \right)$, we have \vspace{-2mm}
\begin{align*} 
\left\{\left\vert\left\vert \widehat{\mathcal{E}} (\tau; \theta)  \right\vert\right\vert \geq t_n \right\} &\subset \left\{\inf_{\epsilon; \left\vert\left\vert \epsilon  \right\vert\right\vert \geq t_n} {\mathbb{L}}_n\left( \widehat{\alpha}(\tau; \theta),\epsilon,\tau;\theta \right) \leq  \inf_{\epsilon; \left\vert\left\vert \epsilon  \right\vert\right\vert < t_n} {\mathbb{L}}_n\left( \widehat{\alpha}(\tau; \theta),\epsilon,\tau;\theta \right) \right\} \\
&\subset \left\{\inf_{\epsilon; \left\vert\left\vert \epsilon  \right\vert\right\vert \geq t_n} {\mathbb{L}}_n\left( \widehat{\alpha}(\tau; \theta),\epsilon,\tau;\theta \right) \leq {\mathbb{L}}_n\left( \widehat{\alpha}(\tau; \theta),0,\tau; \theta \right) = 0 \right\}\\
&\subset \left\{\inf_{e; \left\vert\left\vert e  \right\vert\right\vert = 1} \left[\mathbb{L}^{0}_{n}\left(\widehat{\alpha}(\tau; \theta), t_n e  ,\tau; \theta \right) + \mathbb{R}_{n}\left(\widehat{\alpha}(\tau; \theta), t_n e ,\tau; \theta \right)  \right] \leq 0 \right\}\\
&\subset \left\{\inf_{\left\vert\left\vert \epsilon  \right\vert\right\vert = t_n} \mathbb{L}^{0}_{n}\left(\widehat{\alpha}(\tau; \theta), \epsilon  ,\tau; \theta \right) - \sup_{\left\vert\left\vert \epsilon  \right\vert\right\vert = t_n} \left\vert \mathbb{R}_{n}\left(\widehat{\delta}(\tau; \theta), \epsilon ,\tau; \theta \right)   \right\vert \leq 0  \right\}.
\end{align*}
Then, it follows for supremum of $\left\vert\left\vert \widehat{\mathcal{E}} (\tau; \theta)  \right\vert\right\vert$ that \vspace{-2mm}
\begin{align}
&\left\{\sup_{(\tau, \theta) \in [\underline{\tau },\overline{\tau}]\times \Theta} \left\vert\left\vert \widehat{\mathcal{E}} (\tau;\theta)  \right\vert\right\vert \geq t_n  \right\} = \bigcup_{(\tau, \theta) \in [\underline{\tau },\overline{\tau}]\times \Theta} \left\{\left\vert\left\vert \widehat{\mathcal{E}} (\tau; \theta)  \right\vert\right\vert \geq t_n \right\}\nonumber \\
&\subset \bigcup_{(\tau, \theta) \in [\underline{\tau },\overline{\tau}]\times \Theta} \left\{ \inf_{\left\vert\left\vert \epsilon  \right\vert\right\vert = t_n} \mathbb{L}^{0}_{n}\left(\widehat{\alpha}(\tau;\theta), \epsilon,\tau  ;\theta \right) - \sup_{\left\vert\left\vert \epsilon  \right\vert\right\vert = t_n} \left\vert \mathbb{R}_{n}\left(\widehat{\alpha}(\tau  ;\theta), \epsilon,\tau ;\theta \right)   \right\vert \leq 0  \right\}\nonumber \\
&\subset \left\{ \inf_{(\tau, \theta) \in [\underline{\tau },\overline{\tau}]\times \Theta} \inf_{\left\vert\left\vert \epsilon  \right\vert\right\vert = t_n} \mathbb{L}^{0}_{n}\left(\widehat{\alpha}(\tau  ;\theta), \epsilon,\tau  ;\theta \right) \leq 
\sup_{\left\vert\left\vert \epsilon  \right\vert\right\vert = t_n} \left\vert \mathbb{R}_{n}\left(\widehat{\alpha}(\tau  ;\theta), \epsilon,\tau ;\theta \right) \right\vert
  \right\}\label{AN:eq16}.
\end{align}
Under Assumption \ref{A2}, there exists a $C>0$ such that for all $\tau \in [\underline{\tau },\overline{\tau}]$ and $\theta \in \Theta$,
$
H(\tau;\theta) \succ CM$ where $M=\mathbb{E}\left[X(\theta)X(\theta)^{\prime}\right],
$
and for all $u$ in $\mathbb{R}^{P}$, \vspace{-2mm}
\begin{equation}\nonumber
\begin{aligned}
u^{\prime}Mu &=\mathbb{E}\left[u^{\prime} X(\theta)X(\theta)^{\prime} u \right] = \mathbb{E}\left[ \left(u^{\prime} X(\theta) \right)^2 \right] \\
&= \int{\left(u^{\prime} x \right)^2 f_{X}(x \vert \theta) dx} \geq C\int_{\mathcal{H}}{\left(u^{\prime} x \right)^2 dx} \geq C\left\vert\left\vert u \right\vert\right\vert^2,
\end{aligned}
\end{equation}
where the last bound uses the fact that $u \mapsto \left(\int{\left(u^{\prime} x \right)^2 dx} \right)^{1/2}$ is a norm and norm over $\mathbb{R}^{P}$ are equivalent. Hence, for any non-zero $u \in \mathbb{R}^{P}$, $M$ is a positive definite matrix. This implies that if $\underline{\phi}_M$ is the smallest eigenvalue of $M$, then, $\underline{\phi}_M > 0$. Since $H(\tau;\theta) \succ CM$, it follows for the smallest eigenvalue of the positive definite symmetric matrix $H(\tau;\theta)$, denoted by $\underline{\phi}(\tau;\theta)$, that \vspace{-2mm}
\begin{equation}\label{eq:L3.1}
\inf_{(\tau, \theta) \in [\underline{\tau },\overline{\tau}] \times \Theta}{\underline{\phi}(\tau;\theta)} \geq C\underline{\phi}_M + o_{\mathbb{P}}(1); \text{ for some } \underline{\phi}_M > 0.
\end{equation}
Consider $\inf_{(\tau, \theta) \in [\underline{\tau },\overline{\tau}]\times \Theta} \inf_{\left\vert\left\vert \epsilon  \right\vert\right\vert = t_n} \mathbb{L}^{0}_{n}\left(\widehat{\alpha}(\tau  ;\theta), \epsilon,\tau  ;\theta \right)$. The above result gives, for any $\epsilon$ with $\left\vert\left\vert \epsilon  \right\vert\right\vert \geq t_n$, 
$
\mathbb{L}^{0}_{n}\left(\widehat{\alpha}(\tau  ;\theta), \epsilon,\tau  ;\theta \right) = \frac{1}{2} \epsilon^{\prime} H(\tau;\theta)\epsilon \geq \frac{1}{2} {\underline{\phi}} t^2_n.
$
Hence, from \eqref{AN:eq15} and \eqref{AN:eq16}, we have 
\begin{align*} 
&\lim_{n \rightarrow \infty}\mathbb{P}\left(\sup_{\substack{(\tau, \theta) \in [\underline{\tau },\overline{\tau}] \\ \times \Theta}} \left\vert\left\vert \widehat{\mathcal{E}} (\tau;\theta)  \right\vert\right\vert \geq t_n \right) \leq \lim_{n \rightarrow \infty}\mathbb{P}\left(\sup_{\substack{(\epsilon,\tau ,\theta) \in \mathcal{B}(0,t_n)\\ \times[\underline{\tau },\overline{\tau}]\times \Theta}} \left\vert \mathbb{R}_{n}\left(\widehat{\alpha}(\tau  ;\theta), \epsilon,\tau ;\theta \right) \right\vert \geq \frac{1}{2} {\underline{\phi}} t^2_n \right)\\
&\leq \lim_{n \rightarrow \infty}\mathbb{P}\left(\sup_{\substack{(\epsilon,\tau ,\theta) \in \mathcal{B}(0,t_n) \\ \times[\underline{\tau },\overline{\tau}] \times \Theta} }\left\vert \mathbb{R}_{n}\left(\widehat{\alpha}(\tau  ;\theta), \epsilon,\tau ;\theta \right) \right\vert \geq \frac{1}{4} {\underline{\phi}} t^2_n   \right) = \xi + O\left(\frac{{C_{\xi}}^{1/2}}{t} \right). 
\end{align*}
The latter can be made arbitrarily small by choosing $\xi$ arbitrarily small and $t$ large enough. Recalling $t_n=(t \log^{3/4}{n})/n^{1/4}$ proves Proposition \ref{P2}-(i). Note that $O_{\mathbb{P}}\left( \frac{\log^{3/4}{n}}{n^{1/4}} \right) = \left( \frac{\log^{3/4}{n}}{n^{1/4}} \right)O_{\mathbb{P}}(1)=o_{\mathbb{P}}(1)$. $\hfill \square $

\subparagraph{Proof of  Proposition \ref{P2}-(ii).}
Setting $Z_{i}\left( \theta
\right) =Y_{i}\left( \theta \right) -X_{i}^{\prime }\left( \theta \right)
\beta \left( \tau ;\theta \right) $, \vspace{-2mm}
\begin{align*}
\widehat{S}\left( \tau; \theta \right) -\widehat{S}\left(
\tau; \theta _{0}\right) &= \frac{1}{\sqrt{n}}\sum_{i=1}^{n} \widetilde{s}_i(\tau;\theta), \text{ where} \\
\widetilde{s}_i(\tau;\theta) &= \left[{X}_i(\theta)\left\{\mathbb{I}(Z_i(\theta) \leq 0) - \tau\right\} - {X}_i(\theta_0)\left\{\mathbb{I}(Z_i(\theta_0) \leq 0) - \tau\right\}\right] \\
&\leq 2({X}_i(\theta) + {X}_i(\theta_0))
\end{align*}%
Denoting ${\widetilde{s}_{i\ell}(\tau;\theta)}$ as the $\ell$-th coordinate of the vector ${\widetilde{s}_i(\tau;\theta)}$ implies \vspace{-4mm}
\begin{align*}
\left\vert\frac{\widetilde{s}_{i\ell}(\tau;\theta)}{\sqrt{n}}\right\vert &\leq \frac{C}{\sqrt{n}} \asymp n^{-1/2} \equiv {\overline{\nu}}^{\prime\prime\prime}  
\end{align*}
By Assumption \ref{A1}, for $C^{\left(
1\right) }<\infty $ such that $\sup_{\theta \in \Theta }\left\Vert \frac{%
\partial }{\partial \theta ^{\prime }}\left[ Z_i(\theta) %
\right] \right\Vert \leq C^{\left( 1\right) }$, Taylor inequality gives $\left\vert Z_i(\theta) - Z_i(\theta_0) \right\vert \leq C^{\left( 1\right) } \left\vert\left\vert \theta - \theta_0 \right\vert\right\vert $. Then, under Assumptions \ref{A3} and \ref{A2}, and removing the subscript $i$ to denote random variables, we have
\begin{align*}
&{Var}\left(\frac{\widetilde{s}_\ell(\tau;\theta)}{\sqrt{n}} \right) =\frac{1}{n}\mathbb{E}\left[ \left( \left(X_{\ell}\left( \theta_0 \right)\tau -   X_{\ell}\left( \theta \right)\tau \right) +  \left( X_{\ell}\left( \theta \right) \mathbb{I}\left[ Z\left( \theta \right) \leq 0\right] -  X_{\ell}\left( \theta_0 \right) \mathbb{I}\left[ Z\left( \theta_0 \right) \leq 0\right] \right) \right) ^{2}\right]  \\
& \leq \frac{2}{n} \mathbb{E} \left[ \left(X_{\ell}\left( \theta_0 \right)\tau -   X_{\ell}\left( \theta \right)\tau \right)^2 + \left(X_{\ell}\left( \theta \right) \left(\mathbb{I}\left[ Z\left( \theta \right) \leq 0 \right] - \mathbb{I}\left[ Z\left( \theta_0 \right) \leq 0 \right]\right)+ \mathbb{I}\left[ Z\left( \theta_0 \right) \leq 0 \right] \left(X_{\ell}\left( \theta \right) -   X_{\ell}\left( \theta_0 \right) \right) \right)^2    \right]   
\\
& \leq \frac{C}{n}\left\Vert \theta -\theta _{0}\right\Vert ^{2}+\frac{C}{n}\mathbb{E}%
\left[ \mathbb{I}\left( -\frac{C }{\sqrt{n}}\leq Z\left(
\theta _{0}\right) \leq \frac{C}{\sqrt{n}}\right) \right]
\leq \frac{C}{n}\left\Vert \theta -\theta _{0}\right\Vert =O\left( n^{-3/2}\right).
\end{align*}%
Hence, the standard deviation of ${\widetilde{s}_{\ell}(\tau;\theta)}/{\sqrt{n}} $ is ${\overline{\sigma}}^{\prime\prime\prime} \asymp n^{-3/4}$. Then arguing as in Steps \ref{L2.1S2}-\ref{L2.1S3} of Lemma \ref{AN:L4} (see Appendix 2), 
\begin{align*}
\mathbb{E}\left[ \sup_{\tau \in [\underline{\tau },\overline{\tau}],\vert\vert \theta - \theta_0 \vert\vert \leq C/\sqrt{n}}  \left\vert \left\vert \widehat{S}_{\ell}\left(\tau; \theta \right) -\widehat{S}_{\ell}\left(\tau;
\theta _{0}\right) \right\vert \right\vert \right] &= O\left(n^{1/2}{\overline{\sigma}}^{\prime\prime\prime} \log^{1/2}{n} + \left({\overline{\sigma}}^{\prime\prime\prime} + {\overline{\nu}}^{\prime\prime\prime} \right)  \log{n} \right) 
= O\left(\frac{\log^{1/2}{n}}{n^{1/4} }\right).
\end{align*} 
Note that by Lemma \ref{AN:L2} we have $\sup_{\left( \tau ,\theta \right) \in \left[ \underline{\tau },\overline{%
\tau }\right] \times \mathcal{B}\left( \theta _{0},Cn^{-1/2}\right)
}\left\Vert H^{-1}\left( \theta ;\tau \right) -H^{-1}\left( \theta _{0};\tau
\right) \right\Vert = O\left( n^{-1/2}\right)$ and $ 
\sup_{\left( \tau ,\theta \right) \in \left[ \underline{\tau },\overline{%
\tau }\right] \times \mathcal{B}\left( \theta _{0},Cn^{-1/2}\right)
}\left\Vert H^{-1}\left( \theta ;\tau \right) \right\Vert \leq C$. Markov inequality and Lemma \ref{AN:L6}, then, explain the order in (\ref{Stocheq}). $\hfill \square $

\section*{Appendix 2. Proofs of intermediary Lemmas for Proposition \ref{P2}} \label{sec:Ap2}
\addcontentsline{toc}{section}{Appendix 2}
\renewcommand\theequation{A2.\arabic{equation}}
\setcounter{equation}{0}

\subparagraph{Proof of Lemma \ref{AN:L4}.} Bound for $\mathbb{R}^{1}_{n}(\gamma,\epsilon,\tau;\theta)$ is based on Massart's maximal inequality under bracketing entropy Theorem $6.8$, the conditions for which are proven in Step \hyperref[L2.1S2]{1}. This first requires studying variance of $R(\gamma,\epsilon,\tau;\theta)$. \\
\smallskip
\textit{Variance of $R(\gamma,\epsilon,\tau;\theta)$.}
Note that $\rho_{a}(b) = (a-\mathbb{I}(b<0))b = \int_{0}^{b}(a-\mathbb{I}(t<0))dt$. 
Denoting \vspace{-2mm}
\begin{align}\label{eq:L6.1}
\delta(\gamma;\theta) = X(\theta)^{\prime}\gamma/\sqrt{n}, \text{ and } Z(\tau;\theta)=Y(\theta)-X(\theta)^{\prime}\beta(\tau;\theta), \vspace{-2mm}
\end{align} 
and using definitions in \eqref{AN:eq9} and \eqref{AN:eq10}, for a given $\theta \in \Theta$, \vspace{-2mm}
\begin{align}
R(\gamma,\epsilon,\tau;\theta) &= \rho_{\tau}\left(Z(\tau;\theta)-\delta(\gamma + \epsilon;\theta)\right) - \rho_{\tau}\left(Z(\tau;\theta)-\delta(\gamma;\theta)\right) - \delta(\epsilon;\theta)\left(\mathbb{I}\left(Z(\tau;\theta) \leq 0 \right) - \tau \right) \nonumber \\
&= \int_{\delta(\gamma;\theta)}^{\delta(\gamma;\theta)+\delta(\epsilon;\theta)} \left(\mathbb{I}\left(Z(\tau;\theta) \leq t\right) - \mathbb{I} \left(Z(\tau;\theta) \leq 0\right) \right) dt. \label{eq:L6.2} \vspace{-2mm}
\end{align} 
Using Cauchy-Schwarz inequality, $R(\gamma,\epsilon,\tau;\theta)^{2} \leq \left\vert{\delta(\epsilon;\theta)}\right\vert \left\vert{\bigintsss_{\delta(\gamma;\theta)}^{\delta(\gamma;\theta)+\delta(\epsilon;\theta)} \left(\mathbb{I}\left(Z(\tau;\theta) \leq t\right) - \mathbb{I} \left(Z(\tau;\theta) \leq 0\right) \right)^{2} dt}\right\vert \leq \left\vert{\delta(\epsilon;\theta)}\right\vert \left\vert{\bigintsss_{\delta(\gamma;\theta)}^{\delta(\gamma;\theta)+\delta(\epsilon;\theta)} \mathbb{I}\left(\vert Z(\tau;\theta) \vert \leq \vert t \vert\right) dt}\right\vert$. Under Assumption \ref{A2},\vspace{-3mm}
\begin{align*}
&\mathbb{E}[R^{2}(\gamma,\epsilon,\tau;\theta) \vert X(\theta)] \leq \left\vert{\delta(\epsilon;\theta)}\right\vert \left\vert{\int_{\delta(\gamma;\theta)}^{\delta(\gamma;\theta)+\delta(\epsilon;\theta)} \left\{\int \mathbb{I} \left(\vert y - X(\theta)^{\prime}\beta(\tau;\theta) \vert \leq \vert t \vert \right) f(y \vert X, \theta) dy \right\}} dt \right\vert \\
&\leq  \vert\vert f(\cdot \vert \cdot, \cdot) \vert\vert_{\infty} \left\vert{\delta(\epsilon;\theta)}\right\vert \left\vert{2 \int_{\delta(\gamma;\theta)}^{\delta(\gamma;\theta)+\delta(\epsilon;\theta)}  \vert t \vert} dt \right\vert = \vert\vert f(\cdot \vert \cdot, \cdot) \vert\vert_{\infty} \left\vert{\delta(\epsilon;\theta)}\right\vert \left\vert{2 \int_{0}^{\delta(\epsilon;\theta)}  \vert \delta(\gamma;\theta) + u \vert} du \right\vert 
\\
&\leq \vert\vert f(\cdot \vert \cdot, \cdot) \vert\vert_{\infty} \left\vert{\delta(\epsilon;\theta)}\right\vert \left\vert{2 \int_{0}^{\vert \delta(\epsilon;\theta) \vert}  (\vert \delta(\gamma;\theta) \vert + \vert u \vert} )du \right\vert
\leq \frac{C \vert\vert X(\theta) \vert\vert^{3}}{n^{3/2}} \vert\vert \epsilon \vert\vert^{2} (\vert\vert \gamma \vert\vert + \vert\vert \epsilon \vert\vert).\vspace{-2mm}
\end{align*}
Therefore, $\text{Var}(R(\gamma,\epsilon,\tau;\theta)) \leq \mathbb{E}[R^{2}(\gamma,\epsilon,\tau;\theta)] = \mathbb{E}[\mathbb{E}[R^{2}(\gamma,\epsilon,\tau;\theta) \vert X(\theta)]] \leq \mathbb{E} \left[\frac{C \vert\vert X(\theta) \vert\vert^{3}}{n^{3/2}} \vert\vert \epsilon \vert\vert^{2} (\vert\vert \gamma \vert\vert + \vert\vert \epsilon \vert\vert)\right] = \frac{C \vert\vert \epsilon \vert\vert^{2} (\vert\vert \gamma \vert\vert + \vert\vert \epsilon \vert\vert) }{n^{3/2}} \bigintsss \vert\vert x \vert\vert^{3} f_{X}(x\vert \theta)dx \leq \frac{C \vert\vert \epsilon \vert\vert^{2} (\vert\vert \gamma \vert\vert + \vert\vert \epsilon \vert\vert) }{n^{3/2}}. 
$

\begin{step}[Brackets of $\{R(\gamma,\epsilon,\tau;\theta)\}$.]\label{L2.1S2}
Let $\mathcal{F} = \{R(\gamma,\epsilon,\tau;\theta); (\gamma,\epsilon,\tau;\theta) \in \mathcal{B}(0,t_{\gamma}) \times \mathcal{B}(0,t_{\epsilon}) \times[\underline{\tau },\overline{\tau}] \times  \Theta \}$. This step finds coverings of $\mathcal{F}$ with brackets $[\underline{R}, \overline{R}]$, where the bracket $[\underline{R}, \overline{R}]$ is the set of all $R_{j}$ such that $\underline{R} \leq R_{j} \leq \overline{R}$ almost surely. Define for $\gamma$ in $\mathbb{R}^{P}$
$\widetilde{R}(\gamma,\tau ; \theta) = \int_{0}^{\delta(\gamma;\theta)} \left(\mathbb{I}\left(Z(\tau;\theta) \leq t \right) - \mathbb{I}\left(Z(\tau;\theta) \leq 0 \right) \right)dt,
$ which is such that, from \eqref{eq:L6.2}, \vspace{-3mm}
\begin{align}\label{eq:L6.4}
R(\gamma,\epsilon,\tau;\theta) = \widetilde{R}(\gamma + \epsilon,\tau;\theta) - \widetilde{R}(\gamma,\tau ; \theta)
\end{align}
Let $\text{sgn}(t) = \mathbb{I}(t \geq 0) - \mathbb{I}(t < 0)$, such that with a change of variable $u = t/\text{sgn}(\delta(\gamma;\theta)) $, we have \vspace{-2mm}
\begin{align}\label{eq:L6.5}
\widetilde{R}(\gamma,\tau ; \theta) &= \int_{0}^{\vert \delta(\gamma;\theta) \vert} (\mathbb{I}(Z(\tau;\theta) \leq \text{sgn}(\delta(\gamma;\theta))u ) - \mathbb{I}(Z(\tau;\theta) \leq 0 )) \text{sgn}(\delta(\gamma;\theta)) du \nonumber \\
&= \int_{0}^{\vert \delta(\gamma;\theta) \vert} \vert \mathbb{I}(Z(\tau;\theta) \leq \text{sgn}(\delta(\gamma;\theta))u ) - \mathbb{I}(Z(\tau;\theta) \leq 0 )\vert du \nonumber \\
&= \vert \delta(\gamma;\theta) \vert \int_{0}^{1} \vert \mathbb{I}(Z(\tau;\theta) \leq \delta(\gamma;\theta)v ) - \mathbb{I}(Z(\tau;\theta) \leq 0 ) \vert dv, \nonumber \\
&= \vert \delta(\gamma;\theta) \vert \int_{0}^{1} \vert \mathbb{I}( Z(\tau;\theta) \text{ lies between $0$ and $\delta(\gamma;\theta)v$})  \vert dv,  
\end{align}
where the second last line is obatined using change of variable $v = u/\vert \delta(\gamma;\theta) \vert$. Hence, $0 \leq \widetilde{R}(\gamma,\tau ; \theta) \leq \vert \delta(\gamma;\theta) \vert$. Then, using the definition of $\delta(\gamma;\theta)$ in \eqref{eq:L6.1}, we get for all $\gamma \in \mathcal{B}(0, t_{\gamma} + t_{\epsilon})$, \vspace{-2mm}
\begin{equation}\label{eq:L6.6}
\vert \widetilde{R}(\gamma,\tau ; \theta) \vert \leq \vert\vert X(\theta) \vert\vert \frac{\vert\vert \gamma \vert\vert}{\sqrt{n}} \leq \frac{\overline{\nu}}{4}, \text{ where } \overline{\nu} \asymp \frac{t_{\gamma} + t_{\epsilon}}{\sqrt{n}}.
\end{equation}
It follows from \eqref{eq:L6.4} and the variance bound obtained earlier that \vspace{-4mm}
\begin{align}\label{eq:L6.7}
&\mathbb{E}\left[\left\vert R(\gamma,\epsilon,\tau;\theta) - \mathbb{E}[R(\gamma,\epsilon,\tau;\theta)] \right\vert^{k}\right] \nonumber \\
&= \mathbb{E}\left[ \left\vert \widetilde{R}(\gamma \text{+} \epsilon,\tau;\theta) - \mathbb{E}\left[\widetilde{R}(\gamma \text{+} \epsilon,\tau;\theta)\right] - 
\left\{\widetilde{R}(\gamma,\tau ; \theta) - \mathbb{E}\left[\widetilde{R}(\gamma,\tau ; \theta)\right]\right\} \right\vert^{k-2} 
 \left\vert R(\gamma,\epsilon,\tau;\theta) - \mathbb{E}\left[R(\gamma,\epsilon,\tau;\theta)\right] \right\vert^{2} \right] \nonumber \\
&\leq \left(4 \times \frac{\overline{\nu}}{4}\right)^{k-2} \text{Var}(R(\gamma,\epsilon,\tau;\theta)) \leq \frac{k!}{2} {\overline{\nu}}^{k-2}{\overline{\sigma}}^{2}, 
\text{ where } {\overline{\sigma}}^{2} \asymp \frac{t^{2}_{\epsilon} (t_{\epsilon} + t_{\gamma})}{n^{3/2}}. 
\end{align}
In order to find covering for $\mathcal{F}$, we first define $\widetilde{\mathcal{F}}_{t} = \{\widetilde{R}(\gamma,\tau ; \theta); (\gamma, \tau, \theta) \in \mathcal{B}(0,t) \times[\underline{\tau },\overline{\tau}] \times \Theta \}$ and show that it is sufficient to find covering of $\widetilde{\mathcal{F}}_{t}$, with set of brackets $\{[\underline{R}_j, \overline{R}_j], 1 \leq j \leq e^{h(t_b;t)} \}$, where $t_b \in (0,1)$ denotes length of a bracket, satisfying, \vspace{-2mm}
\begin{equation}\label{eq:L6.8}
\mathbb{E}\left[\left\vert \overline{R}_j - \underline{R}_j \right\vert^{k} \right] \leq \frac{k!}{8} \left(\frac{\overline{\nu}}{2}\right)^{k-2} t^2_b,  
\end{equation}
\begin{equation}\label{eq:L6.9}
h(t_b;t) \leq C\log{\left(\frac{nt}{t_b} \right)}.
\end{equation}
Consider the following two coverings of $\widetilde{\mathcal{F}}_{t_\gamma}$ and $\widetilde{\mathcal{F}}_{t_\gamma + t_\epsilon}$ \vspace{-2mm}
\begin{equation}\nonumber
\widetilde{\mathcal{F}}_{t_\gamma} \subset \bigcup_{1 \leq j \leq e^{h(t_b;t_\gamma)}}\left[ \underline{R}^{1}_j, \overline{R}^{1}_j \right],\ \ \widetilde{\mathcal{F}}_{t_\gamma + t_\epsilon} \subset \bigcup_{1 \leq j \leq e^{h(t_b;t_\gamma + t_\epsilon)}}\left[ \underline{R}^{2}_j, \overline{R}^{2}_j \right]
\end{equation}
If such coverings of $\widetilde{\mathcal{F}}_{t_\gamma}$ and $\widetilde{\mathcal{F}}_{t_\gamma + t_\epsilon}$ exist, then for every $(\gamma,\epsilon,\tau;\theta)$, $\widetilde{R}(\gamma,\tau ; \theta) \in \left[\underline{R}^{1}_{j_1}, \overline{R}^{1}_{j_1} \right]$, $\widetilde{R}(\gamma + \epsilon,\tau;\theta) \in \left[\underline{R}^{2}_{j_2}, \overline{R}^{2}_{j_2} \right]$, for some $j_1$ and $j_2$, and from \eqref{eq:L6.4}, we have $R(\gamma,\epsilon,\tau;\theta) \in \left[ \underline{R}^{2}_{j_2} - \overline{R}^{1}_{j_1}, \overline{R}^{2}_{j_2} - \underline{R}^{1}_{j_1} \right]$. Hence, $\mathcal{F}$ can be covered by $e^{h^{\prime}(t_b;t)}$ brackets such that, using \eqref{eq:L6.8} and \eqref{eq:L6.9},\vspace{-2mm}
\begin{align*}
h^{\prime}(t_b;t) = h(t_b;t_\gamma) + h(t_b;t_\gamma + t_\epsilon ) &\leq C\log{\left(\frac{n(t_\gamma + t_\epsilon)}{t_b} \right)}, \text{ and} \\ 
\mathbb{E}\left[\left\vert \overline{R}^{2}_{j_2} - \underline{R}^{1}_{j_1} - \left( \underline{R}^{2}_{j_2} - \overline{R}^{1}_{j_1}\right) \right\vert^{k} \right] &= \mathbb{E}\left[\left\vert \left(\overline{R}^{2}_{j_2} - \underline{R}^{2}_{j_2}\right) + \left( \overline{R}^{1}_{j_1} - \underline{R}^{1}_{j_1}\right) \right\vert^{k} \right]\\
&\leq 2^{k-1}\left(\mathbb{E}\left[\left\vert \overline{R}^{2}_{j_2} - \underline{R}^{2}_{j_2} \right\vert^{k} \right] + \mathbb{E}\left[ \left\vert \overline{R}^{1}_{j_1} - \underline{R}^{1}_{j_1} \right\vert^{k} \right] \right) \\
&\leq 2^{k-1} \frac{k!}{8} \left(\frac{\overline{\nu}}{2}\right)^{k-2}t^2_b = \frac{k!}{2} {\overline{\nu}}^{k-2}t^2_b.,
\end{align*}
where the inequality in the second line of the above equation follows because, for $a>0$, $b > 0$, $(a+b)^k \leq 2^{k-1} (a^k + b^k)$. 

We now construct covering for $\widetilde{\mathcal{F}}_{t}$. Lemma \ref{AN:L2} proves that $\beta(\tau;\theta)$ is continuously differentiable in $\mu = (\tau,\theta)$ over $[\underline{\tau },\overline{\tau}] \times \Theta$ with bounded derivative. Then from from Taylor's inequality we get, for all $\mu_{1}, \mu_{2}$ in $[\underline{\tau },\overline{\tau}] \times \Theta$, \vspace{-2.5mm}
\begin{equation}\label{eq:L6.10}
\left\vert x(\theta)^{\prime}\beta(\mu_{1}) - x(\theta)^{\prime}\beta(\mu_{2}) \right\vert \leq C \left\vert\left\vert \mu_{1}-\mu_{2} \right\vert\right\vert.
\end{equation}
Also, given $\theta \in \Theta$, for all $\gamma_1, \gamma_2$ in $\mathbb{R}^P$, we have \vspace{-3mm}
\begin{equation}\label{eq:L6.11}
\left\vert \delta(\gamma_1;\theta_1) - \delta(\gamma_2;\theta_2) \right\vert \leq \frac{C}{\sqrt{n}} \left\vert\left\vert \gamma_1-\gamma_2 \right\vert\right\vert.
\end{equation}
Define $r(q,\delta) = \int_{0}^{1} \rho(q,\delta v)dv$,
$ \rho(q,\delta) = \left\vert \mathbb{I}(q \leq \delta) - \mathbb{I}(q \leq 0) \right\vert = \mathbb{I}\left(q \in (0, \delta] \right)\mathbb{I}\left( \delta \geq 0 \right) + \mathbb{I}\left(q \in [\delta, 0) \right)\mathbb{I}\left( \delta <0 \right).
$
From \eqref{eq:L6.5}, $
\widetilde{R}(\gamma, \tau;\theta)= \left\vert \delta(\gamma;\theta) \right\vert r \left(Z(\tau;\theta),\delta(\gamma;\theta) \right).
$
Note that $\rho(q,\delta)$ is a step function which is $1$ for $q$ between $0$ and $\delta$, and $0$ elsewhere, for a given $\delta$. Let $\underline{\rho}(q,\delta)$ and $\overline{\rho}(q,\delta)$ be smooth approximations of $\rho(q,\delta)$, constructed using Friedrichs mollifier of the form \vspace{-2.5mm}
\begin{align*}
 \Phi(x)= C
\begin{cases}
e^{-1/\left(1-|x|^2 \right)},& \text{if } |x|< 1\\
    0,              & \text{if } |x|\geq 1
\end{cases},
\end{align*} 
where $C>0$ and chosen so that $\int_{-1}^{1}\Phi(x)dx = 1$ (see \cite{S11}, chapter $6$ for details). As such, for $\eta>0$, the convolution procedure yields that there exist smooth approximation functions $\underline{\rho}(q,\delta)$, $\overline{\rho}(q,\delta)$, and an open set $D_{\eta} \subset \mathbb{R}^2$ such that: \vspace{-2mm}
\begin{enumerate}[(i)]
  \item $0 \leq \underline{\rho}(q,\delta) \leq \rho(q,\delta) \leq \overline{\rho}(q,\delta) \leq 1$ for all $(q,\delta) \in D_{\eta}$, with $\underline{\rho}(q,\delta) = \rho(q,\delta) = \overline{\rho}(q,\delta)$ if $(q,\delta) \in \mathbb{R}^2 \backslash D_{\eta}$,\vspace{-1mm}
  \item $\sup_{(q,\delta) \in D_{\eta}} \left(\left\vert \frac{\partial \underline{\rho}(q,\delta)}{\partial q} \right\vert + \left\vert \frac{\partial \underline{\rho}(q,\delta)}{\partial \delta} \right\vert + \left\vert \frac{\partial \overline{\rho}(q,\delta)}{\partial q} \right\vert + \left\vert \frac{\partial \overline{\rho}(q,\delta)}{\partial \delta} \right\vert     \right) \leq C\eta^{-1/2}$, and, $\frac{\partial \underline{\rho}(q,\delta)}{\partial q} = \frac{\partial \underline{\rho}(q,\delta)}{\partial \delta} = \frac{\partial \overline{\rho}(q,\delta)}{\partial q} =  \frac{\partial \overline{\rho}(q,\delta)}{\partial \delta} = \frac{\partial{\rho}(q,\delta)}{\partial q} = \frac{\partial{\rho}(q,\delta)}{\partial \delta} = 0$, when $(q,\delta) \in \mathbb{R}^2 \backslash D_{\eta}$,\vspace{-1mm}
  \item $D_{\eta} \subset D^{\prime}_{\eta}= \left\{(q,\delta) \in \mathbb{R}^2; \vert q \vert \leq C\eta^{1/2} \text{ or } \vert q-\delta \vert \leq C\eta^{1/2} \right\}$ 
\end{enumerate}
Define $\underline{r}(q,\delta) = \int_{0}^{1}\underline{\rho}(q,v\delta)dv$, $\overline{r}(q,\delta) = \int_{0}^{1}\overline{\rho}(q,v\delta)dv$, and
$\underline{R}(\gamma,\tau; \theta)= \left\vert \delta(\gamma;\theta) \right\vert \underline{r} \left(Z(\tau;\theta),\delta(\gamma;\theta) \right)$, $
\overline{R}(\gamma,\tau; \theta)= \left\vert \delta(\gamma;\theta) \right\vert \overline{r} \left(Z(\tau;\theta),\delta(\gamma;\theta) \right) 
$ such that condition (i) implies \vspace{-2mm}
\begin{equation}\label{eq:L6.12}
\underline{R}(\gamma,\tau; \theta) \leq \widetilde{R}(\gamma,\tau; \theta) \leq \overline{R}(\gamma,\tau; \theta).
\end{equation}
We now bound $\underline{R}(\gamma_1, \mu_{1}) - \underline{R}(\gamma_2, \mu_{2})$ and $\overline{R}(\gamma_1, \mu_{1}) - \overline{R}(\gamma_2, \mu_{2})$. \vspace{-2mm}
\begin{align*}
&\left\vert \underline{R}(\gamma_1, \mu_{1}) - \underline{R}(\gamma_2, \mu_{2}) \right\vert 
= \left\vert \left\vert \delta(\gamma_1;\theta_1) \right\vert \underline{r} \left(Z(\mu_{1}),\delta(\gamma_1;\theta_1) \right) - \left\vert \delta(\gamma_2;\theta_2) \right\vert \underline{r} \left(Z(\mu_{2}),\delta(\gamma_2;\theta_2) \right) \right\vert \\
&= \left\vert \left\vert \delta(\gamma_1;\theta_1) \right\vert \underline{r} \left(Z(\mu_{1}),\delta(\gamma_1;\theta_1) \right) - \left\vert \delta(\gamma_2;\theta_2) \right\vert \underline{r} \left(Z(\mu_{2}),\delta(\gamma_2;\theta_2) \right) \right. \\
&\left.\quad + \left\vert \delta(\gamma_2;\theta_2) \right\vert \underline{r} \left(Z(\mu_{1}),\delta(\gamma_1;\theta_1) \right) - \left\vert \delta(\gamma_2;\theta_2) \right\vert \underline{r} \left(Z(\mu_{1}),\delta(\gamma_1;\theta_1) \right) \right\vert \\
&\leq \vert \left\vert \delta(\gamma_1;\theta_1) - \delta(\gamma_2;\theta_2) \right\vert \underline{r} \left(Z(\mu_{1}),\delta(\gamma_1;\theta_1) \right) + \left\vert \delta(\gamma_2;\theta_2) \right\vert \left\vert  \underline{r} \left(Z(\mu_{1}),\delta(\gamma_1;\theta_1) \right) - \underline{r} \left(Z(\mu_{2}),\delta(\gamma_2;\theta_2) \right)\right\vert \vert.
\end{align*}
Using the definitions of $Z(\tau;\theta)$ and $\delta(\gamma;\theta)$ given in \eqref{eq:L6.1}, the bounds on increments of $x(\theta)^{\prime}\beta(\tau;\theta)$ and $\delta(\gamma;\theta)$ obtained in \eqref{eq:L6.10} and \eqref{eq:L6.11}, respectively, conditions (i, ii) and Taylor's inequality, we have, for all $(\gamma_1,\mu_{1})$, $(\gamma_2,\mu_{2})$ in $\mathcal{B}(0,t) \times[\underline{\tau },\overline{\tau}] \times \Theta$, where $t=t_{\gamma} + t_{\epsilon} \geq 1$, \vspace{-2mm}
\begin{align*}
\left\vert \underline{R}(\gamma_1, \mu_{1}) - \underline{R}(\gamma_2, \mu_{2}) \right\vert &\leq
\frac{C \left\vert\left\vert  \gamma_1-\gamma_2 \right\vert\right\vert}{\sqrt{n}} + C\frac{t\eta^{-1/2}}{\sqrt{n}}\left(\left\vert\left\vert \mu_{1}-\mu_{2}  \right\vert\right\vert + \frac{\left\vert\left\vert  \gamma_1-\gamma_2 \right\vert\right\vert}{\sqrt{n}}\right)\\
&\leq \frac{C }{\sqrt{n}} \left(1+ t\eta^{-1/2}\right)\left(\left\vert\left\vert \mu_{1}-\mu_{2}  \right\vert\right\vert + {\left\vert\left\vert  \gamma_1-\gamma_2 \right\vert\right\vert}\right).
\end{align*}
Arguing similarly gives \vspace{-4mm}
\begin{align*}
\left\vert \overline{R}(\gamma_1, \mu_{1}) - \overline{R}(\gamma_2, \mu_{2}) \right\vert \leq
\frac{C }{\sqrt{n}} \left(1+ t\eta^{-1/2}\right)\left(\left\vert\left\vert \mu_{1}-\mu_{2}  \right\vert\right\vert + {\left\vert\left\vert  \gamma_1-\gamma_2 \right\vert\right\vert}\right).
\end{align*}
From \cite{VdG00} there exists a covering of $\mathcal{B}(0,t) \times[\underline{\tau },\overline{\tau}] \times \Theta$ by $L$ balls $\mathcal{B}((\gamma_j,\mu_{j}),\eta)$ with centre $(\gamma_j,\mu_{j})$ and radius $\eta$ such that \vspace{-4mm}
\begin{equation}\label{eq:L6.13}
L \leq \max{\left(1, \frac{Ct^P}{\eta^{P+d+1}}\right)}, \text{ where } \gamma \in \mathbb{R}^P, \mu = (\tau;\theta) \in [\underline{\tau },\overline{\tau}] \times \mathbb{R}^d.
\end{equation}
Note that for a ball of radius $\eta$ with centre $(\gamma_j,\mu_{j})$ and $(\gamma_2,\mu_{2})$ inside this ball,
$
\left\vert \underline{R}(\gamma_j, \mu_{j}) - \underline{R}(\gamma_2, \mu_{2}) \right\vert \leq \frac{C }{\sqrt{n}} \left(1+ t\eta^{-1/2}\right)\eta$, $\left\vert \overline{R}(\gamma_j, \mu_{j}) - \overline{R}(\gamma_2, \mu_{2}) \right\vert \leq \frac{C }{\sqrt{n}} \left(1+ t\eta^{-1/2}\right)\eta
$. Define \vspace{-4mm}
$ {\underline{R}}^{\prime}_j = \underline{R}(\gamma_j, \mu_{j}) - \frac{C }{\sqrt{n}} \left(1+ t\eta^{-1/2}\right)\eta$, ${\overline{R}}^{\prime}_j = \overline{R}(\gamma_j, \mu_{j}) + \frac{C }{\sqrt{n}} \left(1+ t\eta^{-1/2}\right)\eta,
$ and \vspace{-4mm}
\begin{align}\label{eq:L6.14}
{\underline{R}}_j = \max{(0, {\underline{R}}^{\prime}_j )},\ \ {\overline{R}}_j = \min{\left(\frac{\overline{\nu}}{2}, {\overline{R}}^{\prime}_j \right)}.
\end{align}
Then, from \eqref{eq:L6.12}, for $(\gamma,\theta)$ in $\mathcal{B}((\gamma_j,\mu_{j}),\eta)$, we have \vspace{-2.5mm}
\begin{align}\label{eq:L6.15}
{\underline{R}}^{\prime}_j \leq {\underline{R}}_j \leq \widetilde{R}(\gamma,\theta) \leq {\overline{R}}_j \leq  {\overline{R}}^{\prime}_j
\end{align}
This implies that $\{\left[{\underline{R}}_j, {\overline{R}}_j \right], j = 1,\cdots, L\}$ is a covering of ${\widetilde{\mathcal{F}}}_t$, with, \vspace{-2.5mm}
\begin{align}\label{eq:L6.16}
\left\vert {\underline{R}}_j - {\overline{R}}_j \right\vert \leq \frac{\overline{\nu}}{2} \asymp \frac{t}{\sqrt{n}},
\end{align}
since $0 \leq {\underline{R}}_j \leq {\overline{R}}_j \leq {\overline{\nu}}/{2}$.
We now bound $\mathbb{E}\left[\left({\overline{R}}_j - {\underline{R}}_j \right)^2\right]$ and $\mathbb{E}\left[\left\vert {\underline{R}}_j - {\overline{R}}_j \right\vert^{k} \right]$. The definitions of $\delta(\gamma;\theta)$, $Z(\tau;\theta)$ in \eqref{eq:L6.1}, conditions (i, iii), Assumption \ref{A2}, \eqref{eq:L6.15} and the inequality $(a+b)^2 \leq 2(a^2 + b^2)$ give
\begin{align*}
&\mathbb{E}\left[\left({\overline{R}}_j - {\underline{R}}_j \right)^2\right] \leq \mathbb{E}\left[\left({\overline{R}}^{\prime}_j - {\underline{R}}^{\prime}_j \right)^2\right] 
=\mathbb{E}\left[\left(\left(\overline{R}(\gamma_j, \mu_{j}) - \underline{R}(\gamma_j, \mu_{j})\right) \text{+} \frac{2C }{\sqrt{n}} \left(1+ t\eta^{-1/2}\right)\eta \right)^2 \right] \\
&\leq 2 \mathbb{E}\left[\left(\overline{R}(\gamma_j, \mu_{j}) - \underline{R}(\gamma_j, \mu_{j}) \right)^2 \right] + \frac{C }{n} \left(1+ t\eta^{-1/2}\right)^2\eta^2 \leq 2\mathbb{E}\left[\left(\overline{R}(\gamma_j, \mu_{j}) - \underline{R}(\gamma_j, \mu_{j}) \right)^2 \right] + \frac{C(1+t)^2(\eta+\eta^2)}{n} \\
&= 2\mathbb{E}\left[\delta^2(\gamma_j;\theta_j) \left(\overline{r}(Z(\mu_{j}), \delta(\gamma_j;\theta_j)) - \underline{r}(Z(\mu_{j}), \delta(\gamma_j;\theta_j)) \right)^2 \right] + \frac{C(1+t)^2(\eta+\eta^2)}{n}   \\
&\leq \frac{2 \left\vert\left\vert \gamma_j \right\vert\right\vert^2 }{n} \int{\left\vert\left\vert x \right\vert\right\vert^2 
\left\{\int
{\left\{\int_{0}^{1} {
\mathbb{I}\left( \left(Z(\mu_{j}), \delta(\gamma_j;\theta_j)v \right) \in D_\eta \right) 
 dv}  \right\}}
f(y \vert x, \theta)dy \right\} 
f_{X}(x\vert \theta)dx} + \frac{C(1+t)^2(\eta+\eta^2)}{n}   \\
&\leq C\frac{(1+t)^2}{n}(\eta+\eta^{2}+\eta^{1/2}),
\end{align*}
where the last inequality follows from Assumption \ref{A2} and condition (iii), since \vspace{-3mm}
\begin{align*}
&\int{\int_{0}^{1}{\mathbb{I} \left(\left( Z(\mu_{j}),\delta(\gamma_j;\theta_j)v \right) \in D_\eta \right) dv  }f(y \vert x, \theta) dy} \leq \int{\mathbb{I} \left(y \in D_\eta+ x(\theta)^{\prime}\beta(\mu_{j}) \right)  f(y \vert x, \theta) dy}\\
&\leq C\int{\mathbb{I} \left(y \in D_\eta+ x(\theta)^{\prime}\beta(\mu_{j}) \right) dy} 
= C \text{(length of $D_\eta$)} \leq C\eta^{1/2}.
\end{align*}
The above bound, together with \eqref{eq:L6.16}, gives for any integer $k \geq 2$,
\small
\begin{align*}
\mathbb{E}\left[\left\vert {\overline{R}}_j - {\underline{R}}_j \right\vert^{k} \right] &= \mathbb{E}\left[\left\vert {\overline{R}}_j - {\underline{R}}_j \right\vert^2 \left\vert {\overline{R}}_j - {\underline{R}}_j \right\vert^{k-2} \right] \leq \left(\frac{\overline{\nu}}{2}\right)^{k-2} \mathbb{E}\left[ \left({\overline{R}}_j - {\underline{R}}_j \right)^2 \right] \leq \frac{k!}{8} \left(\frac{\overline{\nu}}{2}\right)^{k-2}C\frac{(1+t)^2}{n}(\eta+\eta^{2}+\eta^{1/2}).
\end{align*}
\normalsize
Hence, \eqref{eq:L6.8} holds if $
\eta = \frac{1}{3C} \min{\left(\left(\frac{n}{(1+t)^2} \right)^{1/2}t_b, \left(\frac{n}{(1+t)^2} \right)t^2_b, \left(\frac{n}{(1+t)^2} \right)^{2}t^4_b \right)}.
$
Recall that $t \geq 1$ and $t_b \in (0,1)$. Then it follows from \eqref{eq:L6.13}, $
L = e^{h(t_b;t)} \leq \max{\left(1, \frac{Ct^P}{\min{\left(\left(\frac{n}{(1+t)^2} \right)^{1/2}t_b, \left(\frac{n}{(1+t)^2} \right)t^2_b, \left(\frac{n}{(1+t)^2} \right)^{2}t^4_b \right)}^{P+d+1}}  \right)} 
\leq \max{\left(1, \frac{Cnt^5}{t^4_b} \right)^{P+d+1}}
$, such that for large n, $
{h(t_b;t)} \leq \max{\left(0, (P+d+1) \log{\left(\frac{Cnt^5}{t^4_b}\right)}  \right)} = C(\log{n} + 5\log{t} - 4\log{t_b}) 
 \leq 4C (\log{n} + \log{t} - \log{t_b}) \text{+} C \log{t} \leq 4C \log{\left(\frac{nt}{t_b} \right)} \text{+} C \log{\left(\frac{nt}{t_b} \right)} \leq C \log{\left(\frac{nt}{t_b} \right)},
$ which proves \eqref{eq:L6.9}. This completes our task of constructing covering for $\widetilde{\mathcal{F}}_t$.
\end{step}

\begin{step}[Bound for $\mathbb{E}\left(\sup_{(\gamma,\epsilon,\tau;\theta)}\left\vert \mathbb{R}^{1}_{n}(\gamma,\epsilon,\tau;\theta) \right\vert \right)$.]\label{L2.1S3} \vspace{-2mm}
\begin{align*}
&\mathbb{E}\left[\sup_{\substack{(\gamma,\epsilon,\tau;\theta) \in \mathcal{B}(0,t_{\gamma})\\ \times \mathcal{B}(0,t_{\epsilon}) \times [\underline{\tau },\overline{\tau}] \times \Theta}} \left\vert \mathbb{R}^{1}_{n}(\gamma,\epsilon,\tau;\theta) \right\vert\right] =\mathbb{E}\left[\sup_{\substack{(\gamma,\epsilon,\tau;\theta) \in \mathcal{B}(0,t_{\gamma})\\ \times \mathcal{B}(0,t_{\epsilon}) \times [\underline{\tau },\overline{\tau}] \times \Theta}} \left\vert \sum_{i=1}^{n} \left(R_{i}(\gamma,\epsilon,\tau;\theta)-\mathbb{E}[R_{i}(\gamma,\epsilon,\tau;\theta)\vert X(\theta)] \right)  \right\vert\right] \\
&\leq \mathbb{E}\left[\sup_{(\gamma,\epsilon,\tau;\theta)} \left\vert \sum_{i=1}^{n} \left(R_{i}(\gamma,\epsilon,\tau;\theta)-\mathbb{E}[R_{i}(\gamma,\epsilon,\tau;\theta)] \right)  \right\vert\right] \text{+} \mathbb{E}\left[\sup_{(\gamma,\epsilon,\tau;\theta)} \left\vert  \mathbb{E}\left[\sum_{i=1}^{n}\left(R_{i}(\gamma,\epsilon,\tau;\theta) -\mathbb{E}[R_{i}(\gamma,\epsilon,\tau;\theta)] \right) \vert X(\theta) \right] \right\vert\right]\\
&\leq 2 \mathbb{E}\left[\sup_{(\gamma,\epsilon,\tau;\theta) \in \mathcal{B}(0,t_{\gamma})\times \mathcal{B}(0,t_{\epsilon}) \times [\underline{\tau },\overline{\tau}] \times \Theta} \left\vert \sum_{i=1}^{n} \left(R_{i}(\gamma,\epsilon,\tau;\theta)-\mathbb{E}[R_{i}(\gamma,\epsilon,\tau;\theta)] \right)  \right\vert\right].
\end{align*}
Let $\overline{\nu}$, $\overline{\sigma}$, and $h(\cdot;\cdot)$ be as defined in Step \ref{L2.1S2} by equations \eqref{eq:L6.6}, \eqref{eq:L6.7} and \eqref{eq:L6.9}. Recall that $t = t_\gamma + t_\epsilon \geq 1$ and $\overline{\sigma} < 1 \leq n(t_\gamma + t_\epsilon)$. Let us use the notation $h(u;t) = h(u)$. Applying Theorem $6.8$ of \cite{M07}, we get \vspace{-2mm}
\begin{align*}
\mathbb{E}\left[\sup_{\substack{(\gamma,\epsilon,\tau;\theta) 
\in \mathcal{B}(0,t_{\gamma})\\ \times \mathcal{B}(0,t_{\epsilon}) \times [\underline{\tau },\overline{\tau}] \times \Theta}} \left\vert \sum_{i=1}^{n} \left(R_{i}(\gamma,\epsilon,\tau;\theta)-\mathbb{E}[R_{i}(\gamma,\epsilon,\tau;\theta)] \right)  \right\vert\right] \leq C\left(n^{\text{1/2}}\int_{0}^{\overline{\sigma}} h^{1/2}(u) du \text{+} \left(\overline{\nu} \text{+} \overline{\sigma}\right)  h(\overline{\sigma})\right).
\end{align*}
From the discussion in Step \ref{L2.1S2} equation \eqref{eq:L6.9}, since $\overline{\sigma} < 1$, for all $u \in (0, \overline{\sigma}]$, $h(u;t) = h(u) \leq C\log{(n(t_\gamma + t_\epsilon)/u)}$. Therefore, by Cauchy-Schwarz inequality, we have \vspace{-2mm}
\begin{align*}
n^{1/2}\int_{0}^{\overline{\sigma}} h^{1/2}(u) du &\leq {\left(n\overline{\sigma}\right)}^{1/2} \left(\int_{0}^{\overline{\sigma}} h(u) du \right)^{1/2} \leq C {\left(n\overline{\sigma}\right)}^{1/2}\left(\int_{0}^{\overline{\sigma}} \log{\left(\frac{n(t_\gamma + t_\epsilon)}{u}\right)} du \right)^{1/2} \\
&= C {\left(n\overline{\sigma}\right)}^{1/2} \left(\overline{\sigma} \left( \log{\left(\frac{n(t_\gamma + t_\epsilon)}{\overline{\sigma}}\right)} + 1 \right) \right)^{1/2} \leq Cn^{1/2}\overline{\sigma}\log^{1/2}{\left(\frac{n(t_\gamma + t_\epsilon)}{\overline{\sigma}}\right)}.
\end{align*}
With the assumptions on the order of $t_\gamma$ and $t_\epsilon$ as stated in the statement of Lemma \ref{AN:L4} and the order of $\overline{\sigma}$ obtained in \eqref{eq:L6.7}, it follows \vspace{-2.5mm}
\begin{align*}
\log{\left(\frac{n(t_\gamma + t_\epsilon)}{\overline{\sigma}}\right)} &\leq C\log{\left(\frac{n^{7/4}(t_\gamma + t_\epsilon)^{1/2}}{t_\epsilon}\right)} \leq C\log{\left(\frac{n^{7/4}n^{1/4}}{\log^{1/2}{n}}\right)} \leq C\log{n}.
\end{align*}
Hence, on substituting, we get \vspace{-2.5mm}
\begin{align*}
&\mathbb{E}\left[\sup_{(\gamma,\epsilon,\tau;\theta) \in \mathcal{B}(0,t_{\gamma})\times \mathcal{B}(0,t_{\epsilon}) \times [\underline{\tau },\overline{\tau}] \times \Theta} \left\vert \mathbb{R}^{1}_{n}(\gamma,\epsilon,\tau;\theta) \right\vert\right] \leq C\left(n^{1/2}\overline{\sigma}\log^{1/2}{n} + \left(\overline{\nu} + \overline{\sigma}\right)\log{n} \right) \\
&\leq C \frac{t_\epsilon \left(t_\epsilon + t_\gamma\right)^{1/2}\log^{1/2}{n} }{n^{1/4}}\left(1\text{+} \log^{1/2}{n}\left(\frac{1}{n^{1/2}} + \frac{\left(t_\epsilon + t_\gamma\right)^{1/2}}{t_\epsilon n^{1/4}} \right)  \right) \leq C \frac{\log^{1/2}{n}}{n^{1/4}}t_\epsilon \left(t_\epsilon \text{+} t_\gamma\right)^{1/2}, 
\end{align*}
which proves Lemma \ref{AN:L4}. $\hfill \square $
\end{step} 
\subparagraph{Proof of Lemma \ref{AN:L5}.}
The proof of Lemma \ref{AN:L5} follows the same steps as in Lemma \ref{AN:L4} and, hence, a sketch of the proof is provided here. Treating quantities varying with $i$ as random variables, the expressions for $R(\gamma,\epsilon,\tau;\theta)$ given in \eqref{eq:L6.2}, $\mathbb{R}^{2}(\gamma,\epsilon,\tau;\theta)$ from \eqref{AN:eq12} and $H(\tau;\theta)$ gives \vspace{-2.5mm}
\begin{align*}
&\mathbb{R}^{2}(\gamma,\epsilon,\tau;\theta) = \int\limits_{\mathclap{\delta(\gamma;\theta)}}^{\mathclap{\delta(\gamma;\theta) \text{+} \delta(\epsilon;\theta)}}\left(F\left( X(\theta)^{\prime}\beta(\tau;\theta) \text{+} t \vert X,\theta \right) - F\left( X(\theta)^{\prime}\beta(\tau;\theta) \vert X,\theta \right)  \right)dt - \frac{1}{2} \epsilon^{\prime} H(\tau;\theta) (\epsilon + 2 \gamma) \\
&=\int\limits_{\mathclap{\delta(\gamma;\theta)}}^{\mathclap{\delta(\gamma;\theta) + \delta(\epsilon;\theta)}} \left(F\left( X(\theta)^{\prime}\beta(\tau;\theta) \text{+} t \vert X,\theta \right) \text{-} F\left( X(\theta)^{\prime}\beta(\tau;\theta) \vert X,\theta \right)\text{-}tf\left(X(\theta)^{\prime}\beta(\tau;\theta) \vert X,\theta \right) \right)dt\\
&=\int\limits_{\mathclap{\delta(\gamma;\theta)}}^{\mathclap{\delta(\gamma;\theta) + \delta(\epsilon;\theta)}} t \left\{\int_{0}^{1}\left(f\left(X(\theta)^{\prime}\beta(\tau;\theta) + vt \vert X,\theta \right) - f\left(X(\theta)^{\prime}\beta(\tau;\theta) \vert X, \theta \right)\right)dv  \right\} dt.
\end{align*}
Define $
r(\gamma,\tau;\theta) = \bigintsss_{0}^{\delta(\gamma;\theta)} t \left\{\int_{0}^{1}\left(f\left(X(\theta)^{\prime}\beta(\tau;\theta) + vt \vert X,\theta \right) - f\left(X(\theta)^{\prime}\beta(\tau;\theta) \vert X,\theta \right)\right)dv  \right\} dt$ which implies that $\mathbb{R}^{2}(\gamma,\epsilon,\tau;\theta) = r(\gamma + \epsilon,\tau;\theta) - r(\gamma,\tau;\theta)$. Using the definition of $\delta(\gamma;\theta)$ in \eqref{eq:L6.1} and because under Assumption \ref{A2} we have $n_0 > 0$ such that $\left\vert f(a+b \vert x,\theta) - f(a\vert x,\theta) \right\vert \leq n_0 \left\vert b \right\vert$, from Lemma \ref{AN:L2}, we have \vspace{-2.5mm}
\begin{equation}\label{eq:L7.1}
\begin{split}
&\left\vert \mathbb{R}^{2}(\gamma,\epsilon,\tau;\theta) \right\vert \leq \frac{n_0}{2} \left\vert \int_{\delta(\gamma;\theta)}^{\delta(\gamma;\theta) + \delta(\epsilon;\theta)} t^2 dt \right\vert = C \left\vert \delta(\epsilon;\theta) \left(3\delta(\gamma;\theta)^2 + 3\delta(\gamma;\theta)\delta(\epsilon;\theta) + \delta(\epsilon;\theta)^2 \right)  \right\vert \\
&\leq C \left\vert \delta(\epsilon;\theta) \right\vert \left(3\left\vert \delta(\gamma;\theta) \right\vert^2 \text{+} 3\left\vert \delta(\gamma;\theta) \right\vert \left\vert \delta(\epsilon;\theta) \right\vert \text{+} \left\vert \delta(\epsilon;\theta) \right\vert^2 \right) \leq C \left\vert \delta(\epsilon;\theta) \right\vert \left( \left\vert \delta(\gamma;\theta) \right\vert \text{+}  \left\vert \delta(\epsilon;\theta) \right\vert \right)^2 \\
&\leq C \frac{\left\vert\left\vert X(\theta) \right\vert\right\vert^3 \left\vert\left\vert \epsilon \right\vert\right\vert \left(\left\vert\left\vert \gamma \right\vert\right\vert \text{+} \left\vert\left\vert \epsilon \right\vert\right\vert \right)^2 }{n^{3/2}}.\\
\end{split}
\end{equation}
$\left\vert r(\gamma,\tau;\theta) \right\vert \leq C \left\vert  \delta(\gamma;\theta) \right\vert^3 \leq C \frac{\left\vert\left\vert X(\theta) \right\vert\right\vert^3 \left\vert\left\vert \gamma \right\vert\right\vert^3}{n^{3/2}}$. Thus, for all $\gamma \in \mathcal{B} (0, t_\gamma + t_\epsilon)$ and all $(\tau,\theta) \in [\underline{\tau },\overline{\tau}]\times \Theta$, \vspace{-2.5mm}
\[
\left\vert r(\gamma,\tau;\theta) \right\vert \leq \frac{{\overline{\nu}}^{\prime}}{2};\ \ {\overline{\nu}}^{\prime} \asymp \frac{\left(  t_\gamma + t_\epsilon \right)^3}{n^{3/2}}.
\]
From \eqref{eq:L7.1}, under Assumption \ref{A2}, \vspace{-4mm}
\begin{align*}
\text{Var}\left(\mathbb{R}^{2}(\gamma,\epsilon,\tau;\theta) \right) &\leq \mathbb{E}\left[\mathbb{R}^{2}(\gamma,\epsilon,\tau;\theta)^2 \right] \leq \left(C \frac{ \left\vert\left\vert \epsilon \right\vert\right\vert \left(\left\vert\left\vert \gamma \right\vert\right\vert + \left\vert\left\vert \epsilon \right\vert\right\vert \right)^2 }{n^{3/2}} \right)^2 \int{\left\vert\left\vert x(\theta) \right\vert\right\vert^3 f_{X}(x \vert \theta)dx} \\
&\leq C \frac{ \left\vert\left\vert \epsilon \right\vert\right\vert^2 \left(\left\vert\left\vert \gamma \right\vert\right\vert + \left\vert\left\vert \epsilon \right\vert\right\vert \right)^4 }{n^{3}} \leq \left({\overline{\sigma}}^{\prime}\right)^2;\ \ {\overline{\sigma}}^{\prime} \asymp \frac{t_\epsilon \left(t_\gamma + t_\epsilon \right)^2}{n^{3/2}}. \end{align*}
Then arguing as in step \ref{L2.1S2} of Lemma \ref{AN:L4} to construct brackets, \vspace{-3mm}
\small
\begin{align*}
&\mathbb{E}\left[\sup_{\substack{(\gamma,\epsilon,\tau,\theta) \in \mathcal{B}(0,t_{\gamma})\\ \times \mathcal{B}(0,t_{\epsilon}) \times[\underline{\tau },\overline{\tau}] \times \Theta}} \left\vert \mathbb{R}^{2}_{n}(\gamma,\epsilon,\tau;\theta)  - \mathbb{E}\left[\mathbb{R}^{2}_{n}(\gamma,\epsilon,\tau;\theta) \right]   \right\vert \right] \leq Cn^{1/2}{\overline{\sigma}}^{\prime} \log^{1/2}{\left(\frac{n(t_\gamma + t_\epsilon)}{{\overline{\sigma}}^{\prime}}\right)} + \left({\overline{\sigma}}^{\prime} + {\overline{\nu}}^{\prime} \right) \log{\left(\frac{n(t_\gamma + t_\epsilon)}{{\overline{\sigma}}^{\prime}}\right)}
\end{align*}
\normalsize
It follows from \eqref{eq:L7.1} and Assumption \ref{A2} that for all $(\gamma, \epsilon, \tau, \theta)$ in $\mathcal{B}(0,t_\gamma) \times \mathcal{B}(0,t_\epsilon) \times[\underline{\tau },\overline{\tau}] \times \Theta$, \vspace{-4mm}

\begin{align*}
\left\vert \mathbb{E}\left[\mathbb{R}^{2}_{n}(\gamma,\epsilon,\tau;\theta)  \right]  \right\vert &= \left\vert n \mathbb{E}\left[\mathbb{R}^{2}_{i}(\gamma,\epsilon,\tau;\theta)  \right]  \right\vert \leq n \mathbb{E}\left[\left\vert\mathbb{R}^{2}_{i}(\gamma,\epsilon,\tau;\theta) \right\vert \right] \leq \frac{C}{n^{1/2}} \mathbb{E}\left[\left\vert\left\vert {X(\theta)} \right\vert\right\vert^3 \left\vert\left\vert \epsilon \right\vert\right\vert \left(\left\vert\left\vert \gamma \right\vert\right\vert \text{+} \left\vert\left\vert \epsilon \right\vert\right\vert \right)^2 \right] \\
&= \frac{C}{n^{1/2}}  \left\vert\left\vert \epsilon \right\vert\right\vert \left(\left\vert\left\vert \gamma \right\vert\right\vert \text{+} \left\vert\left\vert \epsilon \right\vert\right\vert \right)^2 \int{\left\vert\left\vert x(\theta) \right\vert\right\vert^3 f_{X}(x \vert \theta) dx} \leq \frac{C}{n^{1/2}}  t_\epsilon \left(t_\gamma + t_\epsilon \right)^2,
\end{align*}
and using the conditions on orders of $t_\gamma$ and $t_\epsilon$ as specified in Lemma \ref{AN:L4}, such that $t_\gamma \geq 1$ and $t_{\gamma}/t_{\epsilon} = O\left(n/\log^{1/2}n \right)$, we have \vspace{-3mm}
\begin{equation}\nonumber
\begin{split}
&\mathbb{E}\left[\sup_{(\gamma,\epsilon,\tau,\theta) \in \mathcal{B}(0,t_{\gamma})\times \mathcal{B}(0,t_{\epsilon}) \times[\underline{\tau },\overline{\tau}] \times \Theta} \vert \mathbb{R}^{2}_{n}(\gamma,\epsilon,\tau;\theta) \vert\right] \\
&\leq \mathbb{E}\left[\sup_{(\gamma,\epsilon,\tau,\theta) \in \mathcal{B}(0,t_{\gamma})\times \mathcal{B}(0,t_{\epsilon}) \times[\underline{\tau },\overline{\tau}]\times \Theta} \left(\left\vert \mathbb{R}^{2}_{n}(\gamma,\epsilon,\tau;\theta) - \mathbb{E}\left[\mathbb{R}^{2}_{n}(\gamma,\epsilon,\tau;\theta)\right] \right\vert + \left\vert  \mathbb{E}\left[\mathbb{R}^{2}_{n}(\gamma,\epsilon,\tau;\theta)\right]  \right\vert  \right) \right] \\
&\leq Cn^{1/2}{\overline{\sigma}}^{\prime} \log^{1/2}{\left(\frac{n(t_\gamma + t_\epsilon)}{{\overline{\sigma}}^{\prime}}\right)} + \left({\overline{\sigma}}^{\prime} + {\overline{\nu}}^{\prime} \right) \log{\left(\frac{n(t_\gamma + t_\epsilon)}{{\overline{\sigma}}^{\prime}}\right)} + \frac{C}{n^{1/2}}  t_\epsilon \left(t_\gamma + t_\epsilon \right)^2 \\
&\leq C\frac{t_\epsilon \left(t_\gamma + t_\epsilon \right)^2}{n}\log^\text{{1/2}}{\left(\frac{n^\text{{5/2}}}{t_\epsilon \left(t_\gamma + t_\epsilon \right)}\right)} \left(1 + \frac{\left(t_\gamma + t_\epsilon \right)}{t_\epsilon n^{1/2}} \log^{1/2}{\left(\frac{n^{5/2}}{t_\epsilon \left(t_\gamma + t_\epsilon \right)}\right)}   \right) + \frac{C}{n^{1/2}}  t_\epsilon \left(t_\gamma \text{+} t_\epsilon \right)^\text{2}. \\
\end{split}
\end{equation}
Recall $t_\epsilon = {t \log^{3/4}{n}}/{n^{1/4}}= o(\log^{1/2}{n})$ and $t_\gamma \asymp \log^{1/2}{n}$, such that $t_\gamma + t_\epsilon \asymp \log^{1/2}{n}$. It follows, for large $n$, ${n^\text{{5/2}}}/\left({t_\epsilon \left(t_\gamma + t_\epsilon \right)}\right) \leq ({C}/{t})({n^{1/4}}/{\log^{5/4}{n}} ) \leq {C{n^{1/4}}}/{t}$, such that $\log{({n^{5/2}}/\left({t_\epsilon \left(t_\gamma + t_\epsilon \right)}\right))} \leq C\log{n}$. Similarly, ${\left(t_\gamma + t_\epsilon \right)}/({t_\epsilon n^{1/2}}) \leq C/\left(n \log{n}\right)^{1/4}$. \\ Thus, $({\left(t_\gamma + t_\epsilon \right)}/({t_\epsilon n^{1/2}})) \times \log^{1/2}{({n^{5/2}}/\left({t_\epsilon \left(t_\gamma + t_\epsilon \right)}\right))} \leq C (\log{n}/n)^{1/4} = o(1)$ and \\
$(1 + {\left(t_\gamma + t_\epsilon \right)}/({t_\epsilon n^{1/2}}) \times \log^{1/2}{({n^{5/2}}/({t_\epsilon (t_\gamma + t_\epsilon )}))}   ) = 1+ o(1) = 1$. Therefore, it follows, \vspace{-2.5mm}
\begin{align*}
\mathbb{E}\left[\sup_{\substack{(\gamma,\epsilon,\tau,\theta) \in \mathcal{B}(0,t_{\gamma})\\ \times \mathcal{B}(0,t_{\epsilon}) \times[\underline{\tau },\overline{\tau}] \times \Theta}} \vert \mathbb{R}^{2}_{n}(\gamma,\epsilon,\tau;\theta) \vert\right] 
&\leq C\frac{t_\epsilon \left(t_\gamma + t_\epsilon \right)^2}{n}\log^{1/2}{n} + \frac{C}{n^{1/2}}  t_\epsilon \left(t_\gamma \text{+} t_\epsilon \right)^\text{2} \leq  C\frac{t_\epsilon \left(t_\gamma \text{+} t_\epsilon \right)^\text{2}}{n^{1/2}}  ,  
\end{align*}
for large $n$, which proves Lemma \ref{AN:L5}. $\hfill
\square $

\subparagraph{Proof of Lemma \ref{AN:L6}.}
\renewcommand\theequation{L5.\arabic{equation}}
\setcounter{equation}{0}
The first order condition for $Q(\beta,\tau;\theta)$ gives 
$
\mathbb{E}\left[X(\theta)\left\{F(X^{\prime}(\theta)\beta(\tau;\theta) \vert X, \theta) - \tau \right\}  \right] = 0.
$ 
Let $s_{ i \ell}(\tau;\theta)$ denote the $\ell^{th}$ entry of the vector $\frac{{s}_{i}(\tau;\theta)}{\sqrt{n}}$ in \eqref{AN:S}. Assumption \ref{A2} gives, uniformly in $(\tau,\theta) \in [\underline{\tau },\overline{\tau}] \times \Theta$ for all $i$, \vspace{-4mm}
\begin{align*}
&\left\vert {s}_{ i \ell}(\tau;\theta) \right\vert \leq {\overline{\nu}}^{\prime\prime}, \text{ where } {\overline{\nu}}^{\prime\prime} \asymp n^{-1/2}\\
&\text{Var}({s}_{\ell }(\tau;\theta)) \leq \mathbb{E}\left[\left( {s}_{\ell }(\tau;\theta) \right)^2 \right] \leq \mathbb{E}\left[\frac{1}{n} X_{\ell }(\theta)^2 \right] = \frac{1}{n}\int{x^2f_{X}(x \vert \theta)dx} \leq \left({\overline{\sigma}}^{\prime\prime}\right)^2, \text{ where }{\overline{\sigma}}^{\prime\prime} \asymp n^{-1/2}.
\end{align*} 
Hence, arguing as in Steps \ref{L2.1S2}-\ref{L2.1S3} of Lemma \ref{AN:L4}, \vspace{-4mm}
\begin{align*}
\mathbb{E}\left[ \sup_{(\tau,\theta) \in [\underline{\tau },\overline{\tau}]\times \Theta}  \left\vert \left\vert  \widehat{S}_{\ell}(\tau;\theta) \right\vert \right\vert \right] &= O\left(n^{1/2}{\overline{\sigma}}^{\prime\prime} \log^{1/2}{n} + \left({\overline{\sigma}}^{\prime\prime} + {\overline{\nu}}^{\prime\prime} \right)  \log{n} \right) \\
&\leq C\left(\log^{1/2}{n} + \left(\frac{\log{n}}{n} \right)^{1/2} \log^{1/2}{n}  \right) = O\left(\log^{1/2}{n} \right).
\end{align*} 
Markov inequality, then, proves Lemma \ref{AN:L6}. $\hfill
\square $

\section*{Appendix 3. Proof of remark in Section \ref{sec:biasRMSEcov}} \label{sec:Ap3}
\addcontentsline{toc}{section}{Appendix 3}
\renewcommand\theequation{A3.\arabic{equation}}
\setcounter{equation}{0}

\begin{enumerate}[(i)]
\item In the expression for $C(\tau)$ in \eqref{eq:ex1.2}, $\left\{ \int_{0}^{\tau}\left(\beta_0(t) + \beta_2(t)X_2 \right)dt - \tau \left(\beta_0(\tau) + \beta_2(\tau)X_2   \right)\right\}$ will have the form $p(\tau)+q(\tau) X_2$. Recall that $\widetilde{X}=[1,X_2]^\prime$, $X=[1,X_1,X_2]^\prime$ and  $ g(X)= \widetilde{X}\left[0,1,0  \right]\mathbb{E}^{-1}\left[XX^\prime \right]X$, then, \vspace{-2mm}
\begin{equation}\nonumber
C(\tau)= \mathbb{E}\left[
\begin{array}
[c]{c}%
\left(\left[0,1,0  \right]\mathbb{E}^{-1}\left[XX^\prime \right]X \right)\left(p(\tau)+q(\tau) X_2\right)\\
\left(\left[0,1,0  \right]\mathbb{E}^{-1}\left[XX^\prime \right]X X_2\right)\left(p(\tau)+q(\tau) X_2 \right).
\end{array}
\right] 
\end{equation}
If $X_1$ and $X_2$ are independent, elementary matrix algebra gives that
\[
\left[0,1,0  \right]\mathbb{E}^{-1}\left[XX^\prime \right]X = \frac{1}{D}\left\{\left(\mathbb{E}[X_1]\mathbb{E}^2[X_2] - \mathbb{E}[X_1]\mathbb{E}[X_2^2]\right) + \left(\mathbb{E}[X_2^2] - \mathbb{E}^2[X_2] \right)X_1 \right\},
\]
where $D$ is the determinant of the matrix $\mathbb{E}[XX^\prime]$. Plugging in this expression in $C(\tau)$ and simplifying using independence of $X_1$ and $X_2$ proves the result.
\item  Given the result in (i), the increase in variance of the quantile estimates due to first step estimation, over standard quantile regression had the first step been known, is given by \eqref{eq:ex1.2} as $H(\tau)^{-1}D(\tau)\mathcal{V}(\beta_1)D(\tau)^\prime H(\tau)^{-1}$. Using $H(\tau)$ and $D(\tau)$ as given in \eqref{eq:ex1.2}, under independence of $X_1$ and $X_2$, the vector $H(\tau)^{-1}D(\tau)$ evaluates to $\left[-\mathbb{E}[X_1],0\right]^\prime$. Therefore, the additional variance due to two-step estimation is given by \vspace{-2mm}
\[
H(\tau)^{-1}D(\tau)\mathcal{V}(\beta_1)D(\tau)^\prime H(\tau)^{-1} = \left[\begin{matrix}
\mathbb{E}^2[X_1]\mathcal{V}(\beta_1) & 0\\
0 & 0\\
\end{matrix}
\right]. 
\] 
This proves (ii). $\hfill \square $
\end{enumerate}

\clearpage

\bibliographystyle{apalike2} 
{\small\bibliography{mybib}}

\end{document}